\documentclass[letterpaper,12pt]{article}
\usepackage[utf8]{inputenc}
\usepackage[margin=2cm]{geometry}
\usepackage{tcolorbox}
\usepackage{lipsum}
\usepackage{booktabs}
\usepackage{titling}
\usepackage[export]{adjustbox}
 \usepackage{comment}

\usepackage{graphicx}
\usepackage[affil-it]{authblk} 
\usepackage[labelfont=bf]{caption}
\usepackage{textcomp}
\usepackage[colorlinks=true, allcolors=blue]{hyperref}
\usepackage{float}
\usepackage{amssymb}
\usepackage{amsmath}
\usepackage{subfig}
\usepackage[arrowdel]{physics}
\usepackage{circuitikz}
\usepackage{fancyhdr}
\usepackage{multicol}
\usepackage{makecell}
\usepackage{setspace}
\usepackage[normalem]{ulem}
\usepackage[backend=biber,style=nature]{biblatex}
\newcommand{\CJ}{C_{J_1}}
\newcommand{\CJtwo}{C_{J_2}}
\newcommand{\phizpf}{\phi_{\rm zpf}}
\newcommand{\Qzpf}{Q_{\rm zpf}}
\newcommand{\GT}[2]{G^{#1,#2}}       
\newcommand{\Ucan}{\mathcal{U}_{\rm can}}
\newcommand{\Uphys}{\mathcal{U}_{\rm phys}}
\providecommand{\keywords}[1]
{
  \small	
  \textbf{\textit{Keywords---}} #1
}

\title{{\it \textbf{\large{Effective Study of Superconducting Quantum Circuits}}}}

\author[1]{C. R. Javier Valdez\footnote{Corresponding author: a310861@uach.mx}}
\author[1,2]{H. Hernandez-Hernandez\footnote{hhernandez@uach.mx}}
\author[2]{G. Chacon-Acosta\footnote{gchacon@cua.uam.mx}}
 
\affil[1]{Universidad Autonoma de Chihuahua, Facultad de Ingenieria, Nuevo Campus Universitario, Chihuahua 31125, Mexico}

\affil[2]{Departmento de Matematicas Aplicadas y Sistemas, Universidad Autonoma Metropolitana-Cuajimalpa, Av. Vasco de Quiroga 4871, Ciudad de Mexico, 05348, Mexico}

\date{\today}

\begin{document}
\maketitle
\begin{abstract}
We apply the momentous quantum mechanics formalism to the
non-perturbative study of superconducting circuits in the transmon
regime, deriving effective equations of motion for the bare Josephson
junction (JJ), the cavity--JJ system, and a dissipative resonator.
A Gaussian closure on the hierarchy of quantum moments resums the full
cosine nonlinearity into the closed-form effective Hamiltonian
$\tfrac{1}{2}[V(\phi+\phi_s)+V(\phi-\phi_s)]$, which is non-perturbative and preserves the
periodicity and boundedness of the Josephson potential at all phase
amplitudes; the standard Kerr (Duffing) approximation is recovered as
a special case. We derive the quantum-dressed frequency
$\omega_{\rm eff}=\Omega_p\sqrt{\cos(\phi_{\rm zpf}/\phi_0)}$, and
benchmark the effective dynamics against exact Mathieu-function
diagonalization, with the quantum width $G^{2,0}(t)$ providing the
most sensitive diagnostic of the closure's validity and marks the boundary of
the Gaussian approximation more sharply than $\langle\hat\phi\rangle(t)$ does.
We compare the Caldirola--Kanai, Bateman, and Lindblad descriptions:
the Bateman dynamics, quantized with a switched symplectic structure,
preserves the Heisenberg bound, admits exact closed-form moment
solutions, and reproduces the Lindblad benchmark for weak damping, i.e., within the analytic
error bound $(\lambda/\omega_1)^2\,\phi_{\rm zpf}^2$.
\end{abstract}


\newpage
\section{Introduction}
\label{sec:intro}
Recently, there has been growing interest in the development of
quantum computers, as they have been shown to tackle problems that
classical computation cannot solve efficiently.  However, the
physical platform on which quantum computers will ultimately operate
remains an open question.  Solid-state qubits are a promising
candidate, as they leverage existing nanofabrication techniques, and
the underlying physics, known as circuit quantum electrodynamics
(cQED), is well understood through its deep analogies with cavity
quantum electrodynamics and quantum
optics~\cite{Quantum_Machines,RevModPhys.93.025005}.
In this respect, the study of cQED is not only important because it
provides insight into the dynamics of superconducting qubits, but
also because it constitutes a platform where quantum-optical systems
can be replicated and engineered.  The fine control available over
the circuit parameters enables both fundamental research and the
exploration of otherwise inaccessible quantum regimes.

The Josephson tunnel junction (JJ) is a central superconducting
circuit element: its nonlinear inductance gives rise to
artificial atoms with unequally spaced energy levels, and its
coupling to a resonator reproduces the atom--cavity dynamics of
quantum optics.  Due to its fundamental importance, it is essential
to understand the JJ fully and across diverse operating regimes.
In particular, its cosine nonlinearity makes the standard quantum
treatment rely on second- and fourth-order Taylor approximations of
the cosine potential.  Although adequate in the deep transmon limit
$E_J/E_C \gg 1$, these approximations discard the periodic structure
of the potential and fail at moderate $E_J/E_C \sim 10$--$50$ or at
large flux amplitudes $\phi$.  Moreover, despite their
superconducting nature, these circuits are not perfectly isolated
from the environment, and realistic modeling must account for
dissipative losses, e.g.\ dielectric losses, equilibrium
quasiparticles~\cite{PhysRevX.13.041005}, and their interaction with
an environment such as a transmission
line~\cite{RevModPhys.93.025005}.

From a theoretical standpoint, the JJ Hamiltonian is characterised
by a single dimensionless parameter: the ratio $E_J/E_C$ of the
Josephson coupling energy to the charging energy.  This ratio governs
the depth of the cosine potential well relative to the quantum charge
fluctuations and determines which approximation schemes are
applicable.  In the transmon regime ($E_J/E_C \gg 1$), realised in
the widely studied transmon qubit~\cite{Koch2007}, the standard
approach expands the cosine to fourth order, arriving at a Duffing
(Kerr) oscillator.  While this approximation suffices for many
purposes, it discards the periodicity of the potential and fails at
large phase amplitudes or at moderate
$E_J/E_C$~\cite{RevModPhys.93.025005}.  A central goal of the present
work is to go beyond this limitation using the non-perturbative tools
provided by momentous quantum mechanics.

When it comes to complex quantum phenomena, effective theories have
always stood out as a powerful approach: they allow not only to
approximate the behaviour of the systems under study, but also to
develop mathematical intuition that helps interpret the results.
In this vein, momentous quantum mechanics
(MQM)~\cite{doi:10.1142/S0129055X06002772}, an effective method
grounded in the geometric formulation of quantum mechanics, provides
an alternative to standard perturbative approaches for studying
quantum dynamics.  The MQM framework approximates the
Schr\"odinger equation by a coupled system of ordinary differential
equations, producing a natural separation between classical and
quantum effective degrees of
freedom~\cite{doi:10.1142/S0219887807001941}.  Under certain
conditions, it is possible to arrive at a finite closed
system of equations of motion by expressing the infinite hierarchy of
statistical moments through a Gaussian
closure~\cite{bojowald2014canonical,PhysRevA.99.042114}.  The
formalism has been applied to the double-slit experiment, quantum
cosmological scenarios, tunnelling-time
problems, and dissipative quantum
mechanics~\cite{HernandezHernandez_2023,PhysRevD.110.043506,
PhysRevA.109.032209,JavierValdez_2025,PhysRevA.98.063417}.

The versatility of MQM makes its application to superconducting
quantum circuits both natural and timely.  This work constitutes a
first step towards extending the effective method to quantum optics
and quantum information, and towards establishing a common
mathematical framework connecting those fields with, for instance, quantum
cosmological models.  Several related semiclassical frameworks exist
in the literature: cumulant-expansion
methods~\cite{Kubo1962,scholl2016control,bellac1992quantum},
quantized Hamilton dynamics
(QHD)~\cite{Prezhdo2000,prezhdo2006quantized}, the black-box
quantization (BBQ) approach of Nigg
\textit{et al.}~\cite{Nigg2012}, and the Gaussian variational methods
of Shi, Demler, and Cirac~\cite{ShiDemlerCirac2018}.  As we discuss
in Sec.~\ref{sec:comparison_methods}, MQM is distinguished from these
alternatives by its unambiguous Weyl-ordered operator prescription
and, crucially for the Josephson problem, by the fact that its
Hamiltonian formulation in the extended phase space allows the
all-orders resummation of the cosine potential into the closed form
$\tfrac{1}{2}[V(\varphi+s)+V(\varphi-s)]$ without truncating the
Taylor series at any finite order---a feature that neither QHD nor
cumulant methods achieve at second order, and that BBQ does not
target.

This paper is organised as follows.
Section~\ref{section: SQC} reviews the quantization of the relevant
superconducting circuits: the bare JJ, the cavity--JJ system, and the
coupled cavity--cavity system used to model dissipation.
Section~\ref{section:effective_formalism} introduces the MQM formalism
and derives the all-orders effective Hamiltonian under the Gaussian
closure.
Section~\ref{Sec:EffectiveJJ} applies the effective formalism to each
circuit, derives the all-orders equations of motion, contrasts them
against exact Mathieu-function diagonalization, and obtains analytical
expressions for the quantum-dressed frequency and anharmonicity.
Section~\ref{Discussion} presents the quantitative comparison of the
three dissipation models (Caldirola--Kanai, Bateman, and Lindblad),
derives exact closed-form solutions for the Bateman quantum moments
together with an analytic error bound relative to the Lindblad
standard, establishes the temporal validity window of the Bateman
effective description, and situates MQM within the landscape of
related semiclassical methods (cumulant expansions, QHD, BBQ, and
Gaussian variational approaches).
Section~\ref{sec:conclusions} summarises the central results and
outlines directions for future work.
Appendices collect technical details on the expectation values of
trigonometric operators, the second-order effective dynamics of the
cavity--cavity system, the analytical solutions of the
Caldirola--Kanai model, and the quantitative numerical comparison
between the Bateman and Lindblad moment dynamics.

\section{Superconducting Quantum Circuits}
\label{section: SQC}%
In this work, we study the dynamics of superconducting circuits,
including the Josephson junction~(JJ), its coupling to a resonant
cavity, and the capacitive coupling of a resonant cavity to a~JJ.

In this section, we provide a brief review of the physics behind
these systems in terms of standard quantum mechanics. Comprehensive
reviews of the quantization of superconducting quantum circuits exist
elsewhere; we recommend in particular the excellent
introductions~\cite{Quantum_Machines,RevModPhys.93.025005}.

\subsection{Quantization of the Josephson Junction}
\label{subsection: QJJ}
The first circuit element we review is the parallel $LC$ (inductor-capacitor) circuit. It
is the simplest such element and serves as the foundation upon which
more complex circuits are built, and constitutes a preamble for the
quantization of the JJ.

\begin{figure}[h]
    \centering
    \ctikzset{bipoles/length=1cm}
    \begin{circuitikz}
     \draw 
     (4,3.75) -- (5,3.75)
     (5,3) to[barrier=$L_{J}$](7,3)
     (5,3) -- (5,4.5)
     (5,4.5) to[capacitor=$C_{J}$](7,4.5)
     (7,3) -- (7,4.5)
     (7,3.75) -- (8, 3.75);
     \filldraw[black] (4,3.75) circle (1pt) node[anchor=north]{$\phi_{1}$};
     \filldraw[black] (8,3.75) circle (1pt) node[anchor=north]{$\phi_{2}$};
    \end{circuitikz}
    \caption{Diagram of a Josephson Junction circuit.}
    \label{fig:JJ_circuit_diagram}
\end{figure}
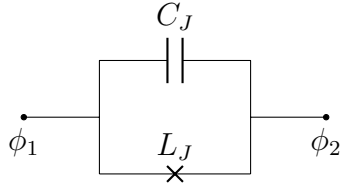

The dynamics of a parallel $LC$ superconducting circuit is expressed
in terms of a single degree of freedom $\phi$, the magnetic flux
through the inductor~\cite{Quantum_Machines}. The Lagrangian of the
circuit is
\begin{equation}
    \mathcal{L} = \frac{1}{2}C\dot{\phi}^{2}-\frac{1}{2L}\phi^{2}.
\end{equation}
The flux is defined as $\phi(t) = \int^{t}_{t_{0}} d\tau\, V(\tau)$,
where $V$ is the voltage across the inductor.
$\tfrac{1}{2}C\dot{\phi}^{2}$ acts as a kinetic term while
$\phi^{2}/(2L)$ is the potential energy. The conjugate momentum $Q$
is
\begin{equation*}
    Q = \frac{\partial\mathcal{L}}{\partial \dot{\phi}} = C\dot{\phi},
\end{equation*}
and from the Legendre transformation
$H = \sum_{i}Q_{i}\dot{\phi}_{i}-\mathcal{L}$
the corresponding Hamiltonian for the $LC$ circuit is
\begin{equation}
\label{eq:LC_Hamiltonian}
    H = \frac{Q^{2}}{2C}+\frac{\phi^{2}}{2L} = \frac{Q^{2}}{2C}+\frac{1}{2}C\omega^{2}\phi^{2},
\end{equation}
with $\omega^{2} = 1/LC$. This Hamiltonian has the structure of a
harmonic oscillator. Taking advantage of this, we can proceed with
the usual canonical quantization procedure. The canonical structure
is given by
\begin{equation}
\label{eq:canonical_relations}
    \{\phi,Q\} = 1 \longrightarrow [\hat{\phi},\hat{Q}] = i\hbar,
\end{equation}
and
\begin{equation}
\label{eq:creation_annihilation_operators}
    \hat{\phi} = \phizpf\,(\hat{a}+\hat{a}^{\dagger}), \quad
    \hat{Q} = -iQ_{\rm zpf}(\hat{a}-\hat{a}^{\dagger}),
\end{equation}
where $\phizpf = \sqrt{\hbar/2C\omega}$ and
$Q_{\rm zpf} = \sqrt{\hbar\omega C/2}$ are the zero-point
fluctuations, and $(\hat{a}^{\dagger},\hat{a})$ are the usual raising
and lowering operators. The quantization procedure yields the
well-known result
\begin{equation*}
    \hat{H} = \hbar\omega\!\left(\hat{a}^{\dagger}\hat{a}+\tfrac{1}{2}\right),
    \quad [\hat{a},\hat{a}^{\dagger}] = 1,
\end{equation*}
with the corresponding energy eigenvalues
\begin{equation*}
    \hat{H}|n\rangle = \hbar\omega\!\left(n+\tfrac{1}{2}\right)|n\rangle.
\end{equation*}
We will use these results to discuss the quantization of the JJ. This
circuit consists of an insulating barrier that separates two
superconducting elements. It was observed by
Josephson~\cite{JOSEPHSON1962251} that when this circuit is operating
at low temperatures, the superconducting elements and the barrier
present a quantum phenomenon: the tunneling of Cooper pairs. The
circuit's dynamics is governed by the Josephson equations:
\begin{equation}
\label{eq:Josephson_equations}
    I = I_c \sin\!\left(\frac{\phi}{\phi_{0}}\right),\qquad
    \frac{d\phi}{dt} = \frac{1}{\phi_{0}}V,
\end{equation}
where $I_c$ is the critical current (the maximum current the junction
can sustain while remaining superconducting), $\phi$ is the
gauge-invariant phase difference across the junction, and
$\phi_0 = \hbar/2e$ is the reduced magnetic flux quantum.
Equations~\eqref{eq:Josephson_equations} describe a non-linear
inductor
\begin{equation}
   L_{J} = \frac{V}{\dot{I}} = \frac{\phi_0}{I_c \cos(\phi/\phi_0)},
\end{equation}
such that the spacing between the superconducting elements acts like
a parallel plate capacitor with effective capacitance $C_{J}$. These
characteristics are described by the Shunt Capacitive
model~\cite{PhysRevB.36.3548}, where the capacitor is connected in
parallel with the inductor as shown in Fig.~\ref{fig:JJ_circuit_diagram}.

We obtain the JJ circuit's Lagrangian by considering the potential
energy
\begin{equation}
\label{eq:JJ_potential_energy}
    U_{J} = \int V\,I\,dt
          = \int \phi_0\frac{d\phi}{dt}\,I_c\sin\!\left(\frac{\phi}{\phi_0}\right)dt
          = -\phi_0 I_c\cos\!\left(\frac{\phi}{\phi_0}\right)
          = -E_J\cos\!\left(\frac{\phi}{\phi_0}\right),
\end{equation}
which follows from Eq.~\eqref{eq:Josephson_equations}. Here
$E_J = \phi_0 I_c$. The kinetic energy is
\begin{equation}
\label{eq:JJ_kinetic_energy}
    K=\frac{C_{J}\dot{\phi}^{2}}{2}.
\end{equation}
Combining Eqs.~\eqref{eq:JJ_potential_energy}
and~\eqref{eq:JJ_kinetic_energy} we arrive at the JJ's Lagrangian
\begin{equation}
\label{eq:JJ_Lagrangian}
    \mathcal{L}(\phi,\dot{\phi})=\frac{C_{J}}{2}\dot{\phi}^{2}+E_{J}\cos\!\left(\frac{\phi}{\phi_{0}}\right),
\end{equation}
and its Hamiltonian
\begin{equation}
\label{eq:JJ_Hamiltonian}
    H=\frac{Q^{2}}{2C_{J}}-E_{J}\cos\!\left(\frac{\phi}{\phi_{0}}\right).
\end{equation}

The dynamics dictated by the Hamiltonian~\eqref{eq:JJ_Hamiltonian} is
controlled by a single dimensionless parameter: the ratio $E_J/E_C$,
where $E_C \equiv e^2/(2C_J)$ is the single-electron charging energy.
This ratio determines the depth of the cosine well relative to the
quantum of charging energy, and consequently the character of the
low-lying spectrum:
\begin{itemize}
\item For $E_J/E_C \ll 1$ (charge regime), the charge $Q$ is a
  nearly good quantum number and the cosine is a weak perturbation.
  The spectrum is dominated by electrostatic effects and is highly
  sensitive to offset charges.
\item For $E_J/E_C \gg 1$ (phase regime, or transmon
  regime~\cite{Koch2007}), the phase $\phi$ is well-localized near
  the bottom of the cosine well, and the low-lying states resemble
  those of a weakly anharmonic oscillator. This is the regime of
  primary interest in modern circuit QED, as the exponential
  suppression of charge dispersion makes the system robust against
  charge noise~\cite{Koch2007}.
\item In the intermediate regime $E_J/E_C \sim 10$--$50$, the cosine
  nonlinearity is most strongly felt: the fourth-order (Kerr)
  expansion of Eq.~\eqref{eq:JJ_perturbative} begins to fail, and
  non-perturbative methods such as the Gaussian closure developed in
  Sec.~\ref{section:effective_formalism} become essential.
\end{itemize}
Throughout this work, we treat $E_J/E_C$ as a free parameter and
study the effective dynamics across all three regimes, with
particular attention to the crossover where our all-orders Gaussian
closure offers a qualitative improvement over the Kerr truncation.

Expanding $\cos(\phi/\phi_0)$ in a Taylor series about $\phi=0$, and
dropping the constant term (which does not affect the dynamics), we
obtain the quartic (Kerr) approximation to the JJ Hamiltonian:
\begin{equation}
\label{eq:JJ_perturbative}
    H\approx\frac{Q^{2}}{2C_{J}}+\frac{1}{2!}\frac{E_{J}}{\phi_{0}^{2}}\phi^{2}
            -\frac{1}{4!}\frac{E_{J}}{\phi_{0}^{4}}\phi^{4},
\end{equation}
where we observe a harmonic oscillator modified by the fourth-order
perturbation term. Substituting the creation and annihilation
operators (Eq.~\eqref{eq:creation_annihilation_operators}) with
zero-point fluctuation amplitude
$\phizpf = (2E_C/E_J)^{1/4}\phi_0$ at the transmon operating point,
we obtain
\begin{equation}
\label{eq:Usual_Quantum_sol_JJ}
    \hat{H}_{\rm Kerr}
    = \hbar\omega\!\left(\hat{a}^{\dagger}\hat{a}+\tfrac{1}{2}\right)
      -\frac{E_{C}}{12}\!\left(\hat{a}+\hat{a}^{\dagger}\right)^{4}
    \approx \hbar\Omega\,\hat{a}^{\dagger}\hat{a}
            -\frac{E_{C}}{2}\,\hat{a}^{\dagger2}\hat{a}^{2},
\end{equation}
where $\Omega = \omega - E_C/\hbar$, and
$\tfrac{E_C}{2}\hat{a}^{\dagger2}\hat{a}^{2}$ is the nonlinear
contribution described as the Kerr effect~\cite{walls2008quantum}.

It is worth emphasizing that the Kerr
Hamiltonian~\eqref{eq:Usual_Quantum_sol_JJ} arises from two
independent approximations applied in succession: (i)~a Taylor
truncation of the cosine potential at fourth order
(Eq.~\eqref{eq:JJ_perturbative}), and (ii)~a normal-ordering
approximation that discards fast rotating terms. Within the effective
method framework of Sec.~\ref{section:effective_formalism},
approximation~(i) corresponds to truncating the quantum Hamiltonian
Eq.~\eqref{eq:H_q} at second-order moments, while the Gaussian
closure Eq.~\eqref{eq:Hierarchy} provides a systematic way to retain
all orders of the cosine without the Taylor truncation. The Kerr
description is thus a special case of the more general effective
treatment, recovered in the combined limits of small phase
fluctuations ($G^{2,0}\ll\phi_0^2$) and low truncation order. One type of this
hierarchy of approximations is among the central points of the
present work.

\subsection{Quantization of the cavity--Josephson Junction}
\label{subsection: cavity-JJ}
\begin{figure}[h]
    \centering
    \ctikzset{bipoles/length=1cm}
    \begin{circuitikz}
    \draw 
     (4,3) -- (5,3) 
     (5,3) to[capacitor=$C_{0}$](7,3)
     (7,3) -- (8,3)
     (4,3) -- (4,2)
     (3,2) -- (5,2)
     (3,2) -- (3,1.5)
     (3,1.5) to[capacitor=$C_{1}$](3,1)
     (3,1) -- (3,0.5)
     (5,2) -- (5,1.5)
     (5,1.5) to[inductor=$L_{1}$](5,1)
     (5,1) -- (5,0.5)
     (8,3) -- (8,2)
     (7,2) -- (9,2)
     (7,2) -- (7,1.5)
     (7,1.5) to[capacitor=$C_{2}$](7,1)
     (7,1) -- (7,0.5)
     (9,2) -- (9,1.5)
     (9,1.5) to[barrier=$L_{J}$](9,1)
     (9,1) -- (9,0.5)
     (3,0.5) -- (5,0.5)
     (7,0.5) -- (9,0.5)
     ;
     \filldraw[black] (4,3) circle (1pt) node[anchor=east]{$\phi_{1}$};
     \filldraw[black] (8,3) circle (1pt) node[anchor=west]{$\phi_{2}$};
    \end{circuitikz}
    \caption{Circuit diagram of the capacitively coupled
    cavity--Josephson Junction system. The LC oscillator (left, with
    capacitance $C_1$ and inductance $L_1$) is coupled to the
    Josephson junction (right, with Josephson energy $E_J$ and
    junction capacitance $C_2$) through the coupling capacitance
    $C_0$.}
    \label{fig:cav-JJ_circuit_diagram}
\end{figure}
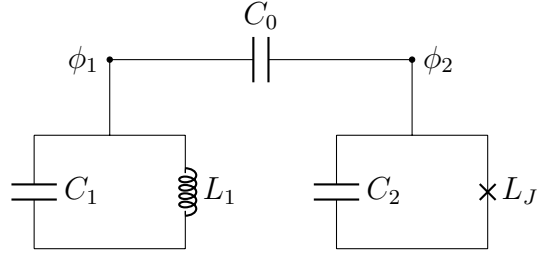

The next circuit configuration of interest is the coupling of a
resonant cavity ($LC$ oscillator) to a Josephson junction, shown in
Fig.~\ref{fig:cav-JJ_circuit_diagram}. Due to the anharmonicity
introduced by the Josephson junction, the energy levels are not
equally spaced. By restricting the dynamics to the two lowest levels,
the JJ effectively behaves as a two-level system (qubit), so its
dynamics is analogous to the light-matter interaction, as we will see
shortly.

First, we define the system's Lagrangian following the steps shown
in~\cite{Quantum_Machines}:
\begin{equation}
    \mathcal{L}_{cav\text{-}JJ}
    = \frac{1}{2}C_{1}\dot{\phi}_{1}^{2}+\frac{1}{2}C_{2}\dot{\phi}_{2}^{2}
      +\frac{C_{0}}{2}\!\left(\dot{\phi}_{1}-\dot{\phi}_{2}\right)^{2}
      -\frac{1}{2L_{1}}\phi_{1}^{2}+E_J \cos\!\left(\frac{\phi_{2}}{\phi_{0}}\right),
\end{equation}
and its corresponding Hamiltonian
\begin{equation}
    H_{cav\text{-}JJ}
    = \frac{1}{2C_{J_1}}Q_{1}^{2}+\frac{1}{2C_{J_2}}Q_{2}^{2}
      +\frac{C_{0}}{C_{T}}Q_{1}Q_{2}+\frac{1}{2L_{1}}\phi_{1}^{2}
      -E_{J}\cos\!\left(\frac{\phi_{2}}{\phi_{0}}\right),
\end{equation}
with
\begin{equation*}
    \frac{1}{C_{J_1}} = \frac{C_{0}+C_{2}}{C_{T}},\qquad
    \frac{1}{C_{J_2}} = \frac{C_{0}+C_{1}}{C_{T}},\qquad
    C_{T} = C_{0}C_{1}+C_{1}C_{2}+C_{0}C_{2}.
\end{equation*}
Using the canonical relations~\eqref{eq:canonical_relations} for each
variable $(\phi_1, \phi_2)$ we obtain the quantum Hamiltonian for the
cavity--JJ interaction up to fourth order:
\begin{equation}
    \hat{H}_{cav\text{-}JJ}
    = \hbar\omega\hat{a}^{\dagger}\hat{a}+\hbar\Omega\hat{b}^{\dagger}\hat{b}
      -\frac{E_{C}}{2}\hat{b}^{\dagger^{2}}\hat{b}^2
      -\frac{C_{0}}{C_{T}}Q_{{\rm zpf},1}\,Q_{{\rm zpf},2}
       \!\left(\hat{a}-\hat{a}^{\dagger}\right)\!\left(\hat{b}-\hat{b}^{\dagger}\right),
\end{equation}
where $(\hat{a}, \hat{a}^{\dagger})$ are the ladder operators for
variable $\phi_1$ and $(\hat{b}, \hat{b}^{\dagger})$ for variable
$\phi_2$.

In the superconducting circuits community one is usually interested
in exploring the dynamics of the first two energy levels, as they
define a computational basis. Restricting to these levels we can
disregard the nonlinear term coming from the JJ:
\begin{equation}
\label{eq:cav_JJ_second_fut_Linblad}
    \hat{H}_{cav\text{-}JJ}
    \approx \hbar\omega\hat{a}^{\dagger}\hat{a}+\hbar\Omega\hat{b}^{\dagger}\hat{b}
      -\frac{C_{0}}{C_{T}}Q_{{\rm zpf},1}\,Q_{{\rm zpf},2}
       \!\left(\hat{a}-\hat{a}^{\dagger}\right)\!\left(\hat{b}-\hat{b}^{\dagger}\right).
\end{equation}
We can further approximate this expression by performing the rotating
wave approximation (RWA)~\cite{Blais2004}:
\begin{equation}
    \hat{H}_{cav\text{-}JJ}
    = \hbar\omega\hat{a}^{\dagger}\hat{a}+\hbar\Omega\hat{b}^{\dagger}\hat{b}
      +\hbar g\!\left(\hat{a}^{\dagger}\hat{b}+\hat{a}\hat{b}^{\dagger}\right),
    \qquad
    g = \frac{C_0}{2 C_T}\sqrt{C_{J_1}C_{J_2}\,\omega\Omega}.
\label{eq:cav-jj}
\end{equation}
Provided that the JJ anharmonicity is large enough to make leakage to
higher levels negligible, we may replace the bosonic operators by
two-level (spin-$\tfrac{1}{2}$) operators, $\hat{b}\to\hat{\sigma}_{-}$
and $\hat{b}^{\dagger}\to\hat{\sigma}_{+}$, recovering the
Jaynes--Cummings Hamiltonian~\cite{Jaynes1963}:
\begin{equation}
        \hat{H}_{cav\text{-}JJ}
        = \hbar\omega\hat{a}^{\dagger}\hat{a}
          +\frac{\hbar\Omega}{2}\,\hat{\sigma}_{z}
          +\hbar g\!\left(\hat{a}^{\dagger}\hat{\sigma}_{-}+\hat{a}\hat{\sigma}_{+}\right),
\end{equation}
which admits an exact analytical solution.
\section{Effective Formalism}
\label{section:effective_formalism}

This section introduces the MQM formalism and derives the all-orders
effective Hamiltonian under the Gaussian closure,
Eq.~\eqref{eq:Ham_all_orders}, which forms the theoretical core of
the present work.

\subsection{Momentous Quantum Mechanics}
\label{section:formalism_momentous}
Exact analytical solutions in physics are rare; one therefore
typically resorts to perturbation theory or effective methods to
obtain approximate solutions. In this sense, the Momentous Quantum
Mechanics framework~\cite{doi:10.1142/S0129055X06002772,Quantum_Cosmology,doi:10.1142/S0219887807001941}
allows us to extract physical information from which we can study
semiclassically complex quantum phenomena through effective
classical-like equations that approximate the system's behaviour.

This formalism is endowed with a Poisson bracket structure, which
describes the algebra of all observables by using linearity and the
Leibniz rule~\cite{doi:10.1142/S0129055X06002772,PhysRevD.31.1341,griffiths2018introduction}:
\begin{equation}
\label{eq:quantum_phase_space}
\{\langle\hat{A}\rangle,\langle\hat{B}\rangle\}
  =\frac{1}{i\hbar}\langle[\hat{A},\hat{B}]\rangle.
\end{equation}
From this we readily identify that the above expression satisfies the
usual commutation relation for classical variables, e.g.\
$x=\langle\hat{x}\rangle$, $p=\langle\hat{p}\rangle$. However, in
general we have that $\langle\hat{A}^{n}\rangle\neq\langle\hat{A}\rangle^{n}$,
meaning that $\langle\hat{A}^{n}\rangle$ contains more information
that is not accessible by simply taking $\langle\hat{A}\rangle^{n}$,
as one does in classical mechanics. Moreover, we have that for products of operators
 $\langle\hat{A}^{a}\hat{B}^{b}\cdots\rangle \neq
\langle\hat{A}\rangle^{a}\langle\hat{B}\rangle^{b}\cdots$, indicating
that expectation values of products carry information not accessible
from products of expectation values alone\footnote{In cumulant
expansion methods, one approximates the expectation value of these
operators as products of lower-order
ones~\cite{Kubo1962,scholl2016control}. For instance
\begin{equation*}
\langle\hat{A}\hat{B}\hat{C}\rangle \approx
  \langle\hat{A}\hat{B}\rangle\langle\hat{C}\rangle
  +\langle\hat{A}\hat{C}\rangle\langle\hat{B}\rangle
  +\langle\hat{B}\hat{C}\rangle\langle\hat{A}\rangle
  - 2\langle\hat{A}\rangle\langle\hat{B}\rangle\langle\hat{C}\rangle.
\end{equation*}}.
It is therefore important to define observables that describe purely
quantum degrees of freedom systematically. In the Momentous
formalism, this is achieved by requiring the quantum variables to be
symplectically orthogonal to the classical ones
(Eq.~\eqref{eq:difference_quantum_classical}):
\begin{equation}
\label{Generalized effective dynamical variables}
    G^{a_{1},b_{1},\dots,a_{k},b_{k}}:=
    \big\langle (\hat{x}_{1} -\langle\hat{x}_{1}\rangle)^{a_{1}}
               (\hat{p}_{1} -\langle\hat{p}_{1}\rangle)^{b_{1}}\cdots
               (\hat{x}_{k} -\langle\hat{x}_{k}\rangle)^{a_{k}}
               (\hat{p}_{k} -\langle\hat{p}_{k}\rangle)^{b_{k}}
    \big\rangle_{\textrm{Weyl}},
\end{equation}
where $k$ is the number of the system's degrees of freedom, Weyl (fully
symmetric) ordering is employed, and the above variables satisfy
\begin{equation}
\label{eq:difference_quantum_classical}
    \{x_{k},G^{a_{1},b_{1},...,a_{k},b_{k}}\}
    =\{p_{k},G^{a_{1},b_{1},...,a_{k},b_{k}}\}=0,
\end{equation}
as required.

This definition is convenient because it yields generalised
uncertainty relations that serve as lower bounds for certain moments.
For $k=1$, Heisenberg's uncertainty relation takes the form
\begin{equation}
       G^{2,0}G^{0,2} - (G^{1,1})^{2} \geq \frac{\hbar^2}{4},
\end{equation}
where $G^{2,0}=(\Delta x)^{2}$, $G^{0,2}=(\Delta p)^{2}$, and
$G^{1,1}=\tfrac{1}{2}\langle\hat{x}\hat{p}+\hat{p}\hat{x}\rangle
-\langle\hat{x}\rangle\langle\hat{p}\rangle$ is the symmetrised
covariance.

To obtain the system's dynamics one has to deal with a Hamiltonian.
In the Momentous formalism the classical-like structure is recovered
via the Poisson bracket of Eq.~\eqref{eq:quantum_phase_space}, in
direct analogy with the Heisenberg equations of motion. Given that we
have defined classical and quantum variables, it is
convenient to rewrite the expectation value of the Hamiltonian in
terms of these variables,
$H_{Q}(x_{1},p_{1},\ldots,x_{k},p_{k};\,G^{a_{1},b_{1},...,a_{k},b_{k}})
= \langle\hat{H}\rangle$.
The system's dynamical evolution is then given by the effective
Hamiltonian function $H_{Q}$:
\begin{align}
\label{eq:H_q}
    H_{Q}=\langle\hat{H}\rangle
    &=\sum_{a_{1},b_{1},...,a_{k},b_{k}}^{\infty}
      \frac{1}{a_{1}!b_{1}!\cdots a_{k}!b_{k}!}
      \frac{\partial^{a_{1}+b_{1}+\cdots+a_{k}+b_{k}}H_{\rm class}}
           {\partial x_{1}^{a_{1}}\partial p_{1}^{b_{1}}\cdots
            \partial x_{k}^{a_{k}}\partial p_{k}^{b_{k}}}
      G^{a_{1},b_{1},\dots,a_{k},b_{k}}\nonumber\\
    &= H_{\rm class}
      +\sum_{a_{1}+b_{1}+\cdots+a_{k}+b_{k}\geq 2}^{\infty}
      \frac{1}{a_{1}!b_{1}!\cdots a_{k}!b_{k}!}
      \frac{\partial^{a_{1}+b_{1}+\cdots+a_{k}+b_{k}}H_{\rm class}}
           {\partial x_{1}^{a_{1}}\partial p_{1}^{b_{1}}\cdots
            \partial x_{k}^{a_{k}}\partial p_{k}^{b_{k}}}
      G^{a_{1},b_{1},\dots,a_{k},b_{k}},
\end{align}
where the classical Hamiltonian
$H_{\rm class}=H(x_{1},p_{1},\ldots,x_{k},p_{k})$ is recovered in
the first term of the sum\footnote{In this semiclassical description
the correspondence principle is straightforwardly satisfied: the
classical dynamics is recovered by imposing
$\hbar\to 0$, $G^{a_{1},b_{1},...,a_{k},b_{k}}\to 0$.}.
We have thus defined a phase space for classical and quantum
observables, as well as the Hamiltonian function, from which we
obtain equations of motion in the effective phase space:
\begin{equation}
\label{eq: Hamilton_formalism}
    \frac{d\langle\hat{f}\rangle}{dt}=\{\langle\hat{f}\rangle,H_{Q}\}.
\end{equation}
Classical-like variables follow a canonical algebra, in contrast with
the algebra of quantum variables, which cannot be considered as
canonical coordinates. For instance,
\begin{equation}
    \{G^{2,0},G^{1,1}\}=2G^{2,0},\quad
    \{G^{2,0},G^{0,2}\}=4G^{1,1}, \quad
    \{G^{1,1},G^{0,2}\}=2G^{0,2}.
\end{equation}

From this prescription we can understand that the Hamiltonian
Eq.~\eqref{eq:H_q} provides a classical dynamics modified by quantum
corrections, which in turn have back-reaction contributions on the
classical ones. As a result, the arising dynamics given by the
equations of motion is equivalent to that provided by the
Schr\"odinger equation~\cite{ashtekar_geometrical_1997,doi:10.1142/S0129055X06002772,Quantum_Cosmology,doi:10.1142/S0219887807001941}.

There are certain scenarios under which we can stress even further
the similarities this method has with the classical Hamiltonian
formalism. For example,  it is
possible to construct canonical variables in the following
way~\cite{Prezhdo2000,prezhdo2006quantized,PhysRevA.99.042114,PhysRevA.98.063417}:
\begin{equation}
\label{Canonical variables s and p_{s}}
    s = \sqrt{G^{2,0}}, \quad
    p_{s} = \frac{G^{1,1}}{\sqrt{G^{2,0}}}, \quad
    U = G^{2,0}G^{0,2} -(G^{1,1})^{2}\geq \frac{\hbar^{2}}{4},
\end{equation}
satisfying\footnote{Linearity and the Leibniz rule allow us to compute
the Poisson brackets between them.}
\begin{equation}
    \{s,p_{s}\} = 1,\quad \{s,U\} = \{p_{s},U\} = 0.
\end{equation}

With these new variables we can rewrite, for one degree
of freedom, the second-order truncation of the Hamiltonian
Eq.~\eqref{eq:H_q} as
\begin{equation}
\label{eq:second_order_canonical}
\begin{split}
    H_{Q}
    &= \frac{p^{2}+p_{s}^{2}}{2m}+\frac{U}{2ms^{2}}+V(x)
       +\frac{1}{2!}\frac{d^{2}V(x)}{dx^{2}}s^{2}\\
    &=\frac{p^{2}+p_{s}^{2}}{2m}+V_{\rm eff}(x,s),
\end{split}
\end{equation}
where $V_{\rm eff}(x,s)$ is an effective potential.

Even with the advantages that come from its similarities with the
classical Hamiltonian formalism, we are still facing the problem of
the infinite set of equations that arise from Eq.~\eqref{eq:H_q} and
the algebra of moments. An approach to deal with this is to implement
a closure on the hierarchy of moments: instead of truncating the
Hamiltonian, one can assume that higher-order moments can be expressed
in terms of lower-order ones. We are interested in expressing
higher-order moments as
follows~\cite{bojowald2014canonical,PhysRevA.99.042114,scholl2016control,bellac1992quantum}:
\begin{equation}
\label{eq:Hierarchy}
    \langle (\hat{x}-\langle\hat{x}\rangle)^{n}\rangle=G^{n,0}\simeq
    \begin{cases} 
      s^{n} & \text{for $n$ even,} \\
      0 & \text{for $n$ odd.} 
   \end{cases}
\end{equation}
The ansatz Eq.~\eqref{eq:Hierarchy} is equivalent to approximating
the quantum state by a Gaussian wave packet at all times: in a
Gaussian state all cumulants of order $n\geq 3$ vanish, so
$G^{n,0}\propto s^n$ for even $n$ and zero for odd $n$, exactly as
prescribed, and this choice preserves the periodicity and boundedness
of the Josephson cosine in closed form.
We can now rewrite compactly the full expansion of the quantum
Hamiltonian Eq.~\eqref{eq:second_order_canonical}:
\begin{equation}
\label{eq:Ham_all_orders}
\begin{split}
    H_{Q}
    &= \frac{p^{2}+p_{s}^{2}}{2m}+\frac{U}{2ms^{2}}+V(x)
       +\sum_{n=1}^{\infty}\frac{1}{(2n)!}\frac{d^{2n}V(x)}{dx^{2n}}s^{2n}\\
    &= \frac{p^{2}+p_{s}^{2}}{2m}+\frac{U}{2ms^{2}}
       +\frac{1}{2}\left[V(x+s)+V(x-s)\right].
\end{split}
\end{equation}

Equation~\eqref{eq:Ham_all_orders} is the all-orders quantum
Hamiltonian under the Gaussian closure. It has a remarkable
structure: the effect of quantum fluctuations on an arbitrary
potential $V(x)$ is captured by replacing the classical potential
with its average over two symmetrically displaced copies,
$\tfrac{1}{2}[V(x+s)+V(x-s)]$, where the displacement
$s=\sqrt{G^{2,0}}$ is the width of the quantum state. This can be
understood as a non-perturbative resummation of the infinite Taylor
series in Eq.~\eqref{eq:H_q}: rather than truncating at a fixed
order (second order gives a harmonic potential, fourth order gives the
Kerr/Duffing Hamiltonian), the Gaussian closure sums the entire
series into a closed-form expression. For periodic potentials such as
the Josephson cosine, this resummation is particularly valuable, as
it preserves the periodicity and boundedness of the potential,
properties that are lost in any finite Taylor truncation.

An analogous resummation strategy appears in other areas of
theoretical physics: in quantum field theory, effective potentials
are obtained by integrating out fluctuations to all loop
orders~\cite{coleman1973radiative}; in quantum cosmology, the
Gaussian closure has been used to study anharmonic corrections to the
Wheeler--DeWitt equation~\cite{doi:10.1142/S0129055X06002772,PhysRevD.110.043506,bojowald2014canonical}.
The application to circuit QED developed in the present work extends
this family of ideas to a new physical domain.
\section{Effective Description of Superconducting Circuits}
\label{Sec:EffectiveJJ}
Previously we presented a short review of the effective formulation of quantum mechanics, and how it illuminates the study of complex quantum systems through a clear separation of quantum and classical-like observables, and how a Gaussian closure truncates the otherwise infinite hierarchy of equations of motion to a closed finite set.

In this section we will use this effective method to study the behavior of the superconducting quantum circuits introduced in section \ref{section: SQC}, offering a different set of tools helpful for the analysis of even more complex setups.

\subsection{The Josephson Junction}
\label{subsection: Effective_Josephson}
In Sec.~\ref{subsection: QJJ}, we reviewed the canonical quantization of the JJ in terms of creation and annihilation operators, following the harmonic-oscillator analogy. We now apply the effective formalism to this system.

First, we retake the Hamiltonian for a single JJ Eq.~(\ref{eq:JJ_Hamiltonian}), and we apply the following $n = Q/2e$, $\theta = \phi/\phi_{0}$ and $E_{C} = e^{2}/2C_{J}$ \cite{Stirpe_2024}, resulting in
\begin{equation}
    H = 4E_{c}n^{2}-E_{J}\text{Cos}(\theta),\quad\text{with}\quad \{\theta,n\} = \frac{1}{\hbar} 
\end{equation}
and after promoting the classical variables to quantum ones, we obtain the JJ quantum Hamiltonian as described by Eq.~(\ref{eq:H_q})
\begin{equation*}
    \langle H\rangle = H(\theta,n)+\sum_{a+b\geq 2}^{\infty}\frac{1}{a!b!}\frac{\partial^{a+b}H(\theta,n)}{\partial\theta ^{a}\partial n^{b}}G^{a,b}
\end{equation*}
where $H(\theta,n) = H(\langle\hat{\theta}\rangle,\langle\hat{n}\rangle)$ is equivalent and corresponds to the classical Hamiltonian Eq.~(\ref{eq:JJ_Hamiltonian}). Explicitly
\begin{equation}
\label{eq: H_Q JJ}
\begin{split}
    H_{Q} &= H(\theta,n) + 4E_{c}G^{0,2}-E_{J}\cos(\theta)\sum_{n=1}^{\infty}\frac{(-1)^{n}}{(2n)!}G^{2n,0}+ E_{J}\sin(\theta)\sum_{m=1}^{\infty}\frac{(-1)^{m}}{(2m+1)!}G^{2m+1,0}
    \end{split}
\end{equation}
which provides a description of the JJ quantum dynamics in terms of moments of the position operator. Details on the expectation values of trigonometric operators are collected in Appendix~\ref{sec:Expect_val_Trig}.

Now, this is where the implementation of the effective method becomes handy, as we can study not only its dynamics by truncating the series but also approximate the all-orders behavior by performing the moment closure shown before. Let us examine the latter. If we assume that the condition given in Eq.~(\ref{eq:Hierarchy}) is applicable, then the odd terms in Eq.~(\ref{eq: H_Q JJ}) vanish, and by defining the following canonical variables
\begin{equation}
\label{Canonical variables s and p_{s}}
\begin{split}
    \theta_{s}=\sqrt{G^{2,0}}, \quad\quad n_{s}=\frac{G^{1,1}}{\sqrt{G^{2,0}}},\quad
    U&=\theta_{s}^{2}G^{0,2}-\theta_{s}^{2}n_{s}^{2} \geq 1/4,\quad \{\theta_{s},n_{s}\} = 1/\hbar
\end{split}
\end{equation}
we arrive at the all-orders effective Josephson Junction Hamiltonian,
\begin{equation}
\label{eq:All_orders_effective}
\begin{split}
    \langle H\rangle &= 4E_{c}(n^{2}+n_{s}^{2})+\frac{4E_{c}U}{\theta_{s}^{2}}-E_{J}\text{Cos}(\theta)\text{Cos}(\theta_{s})
\end{split}
\end{equation}
From this expression, we can obtain the system's equations of motion (EOM)
\begin{equation}
    \begin{split}
        \dot{\theta}&=\frac{8E_{c}n}{\hbar},\quad \dot{\theta}_{s}=\frac{8E_{c}n_{s}}{\hbar},\quad \dot{n}=-\frac{E_{J}}{\hbar}\text{Sin}(\theta)\text{Cos}(\theta_{s}),\quad \dot{n}_{s}=-\frac{E_{J}}{\hbar}\text{Cos}(\theta)\text{Sin}(\theta_{s})+\frac{8E_{c}U}{\hbar\theta_{s}^{3}}\\
    \end{split}
 \label{eq:EOM-JJ}
\end{equation}
Two features are worth emphasising, as they recur in the coupled and dissipative circuits below. The classical force on $n$ is the symmetric average $\sin\!\big(\theta+\theta_s\big)+\sin\!\big(\theta-\theta_s\big)=2\sin(\theta)\cos(\theta_s)$, while the force on the width $n_s$ is the antisymmetric combination $\sin\!\big(\theta+\theta_s\big)-\sin\!\big(\theta-\theta_s\big)=2\cos(\theta)\sin(\theta_s)$. Both reduce to the bare-junction force as $\theta_s\to0$.

In this last equation, we can perform the following transformation to simplify the system
\begin{equation}
    y = \theta+\theta_{s}\rightarrow\frac{d^{2} y}{dt^{2}}=-\frac{8E_{c}E_{J}}{\hbar^{2}}\sin(y)+\frac{64E_{c}^{2}U}{\hbar^{2}\theta_{s}^{3}}
\end{equation}
so we can rewrite it in a more suitable way by using the lower bound of $U$: $U = 1/4$, and the Josephson plasma frequency $\Omega_{p}=\sqrt{8E_{c}E_{J}}/\hbar$ 
\begin{equation}
    \frac{d^2 y}{dt^2}=-\Omega_p^{2}\sin(y)+\frac{16E_{c}^{2}}{\hbar^{2}\theta_{s}^{3}}
\label{eq:double-pendulum}
\end{equation}

We can observe how the quantum corrections modify the classical dynamics of the pendulum's equation. The structure of Eq.~\eqref{eq:double-pendulum} reveals how $E_J/E_C$ controls the relative weight of the classical and quantum contributions. The classical term scales as $\Omega_p^2\propto E_JE_C$, while the quantum correction scales as $64UE_{c}^{2}/(\hbar^{2}\theta_s^3)$. Evaluated at the ground-state width $\theta_s=\theta_{\rm zpf}$, the ratio of the quantum to the classical term is $(2E_C/E_J)^{1/4}$, so it is parametrically small in the transmon regime and grows toward the intermediate regime. In the deep phase regime ($E_J/E_C \gg 1$) the classical pendulum dynamics dominates and the quantum correction is a small perturbation, so the Gaussian closure and the Kerr approximation are both close to the harmonic limit. In the intermediate regime ($E_J/E_C \sim 10$--$50$) the quantum correction becomes comparable to the anharmonic part of the classical term, and the non-perturbative structure of the closure---which preserves the periodicity of the sine in the equations of motion---is fully active. The dynamical consequences of Eq.~\eqref{eq:double-pendulum} are examined in Sec.~\ref{sec:comparison}.

It is instructive to examine two limiting cases of the all-orders Hamiltonian Eq.~\eqref{eq:All_orders_effective}. In the limit $\theta_s \to 0$ (vanishing quantum fluctuations), the dressing factor $\cos(\theta_s) \to 1$ and we recover the classical JJ Hamiltonian Eq.~\eqref{eq:JJ_Hamiltonian}, as expected from the correspondence principle. In the opposite limit $\theta_s \gg 1$ (large quantum fluctuations), the cosine factor oscillates rapidly and its effect is suppressed, washing out the Josephson potential---a behaviour reminiscent of the Debye--Waller factor in solid-state physics, where fluctuations suppress Bragg peaks. Between these extremes the closure interpolates smoothly, respecting the periodicity and boundedness of the cosine at all values of $\theta_s$.

\subsubsection{Exact spectrum via Mathieu functions}
\label{sec:benchmark}
To contrast the effective dynamics we use the exact spectrum of the JJ Hamiltonian Eq.~\eqref{eq:JJ_Hamiltonian}, whose time-independent form reduces to a Mathieu equation~\cite{Koch2007,braumuller2018quantum}
\begin{equation}
\label{eq:Appendix_Mathieu's_equation}
\begin{split}
    \left[4E_{c}\left(-i \frac{\partial}{\partial \theta}\right)^{2}-E_{J}\cos\left({\theta}\right)\right]\Psi = E\Psi\rightarrow \frac{\partial^{2}g(x)}{\partial x^{2}}+ \left[\frac{E}{E_{C}}+\frac{E_{J}}{E_{C}}\cos(2x)\right]g(x) = 0,
\end{split}
\end{equation}
the latter expression obtained with $x = \theta/2$. Its eigenvalues are expressed in terms of Mathieu's characteristic values
\begin{equation}
\label{eq:Appendix_Mathieus_eigenvalues}
    E_{n} = E_{C}\begin{cases}
      M_{A}\left(k+1,-\frac{E_{J}}{2 E_{C}}\right) & \text{for } k = 2n  \\
    M_{B}\left(k, \frac{E_{J}}{2 E_{C}}\right) & \text{for } k = 2n+1
   \end{cases}
   , \quad n = 0,1,2,3,\dots
\end{equation}
Only bound states with $E<E_{J}$ represent superconducting states; levels above the barrier correspond to currents exceeding the critical current $I_{c} = E_{J}/\phi_{0}$ and lie outside the model description~\cite{Quantum_Machines, dassonneville:tel-02274266}. The spectrum $\{E_n\}$ provides the exact frequency $(E_1-E_0)/\hbar$ used in Secs.~\ref{sec:Veff_freq}, and an energy basis $\hat H|n\rangle=E_n|n\rangle$ from which the time-dependent solution of Sec.~\ref{sec:comparison} is reconstructed.

\subsubsection{Effective potential, quantum dressing, and dressed frequency}
\label{sec:Veff_freq}
The two limiting behaviours just described---recovery of the classical cosine as $\theta_s \to 0$ and suppression of the Josephson potential at $\theta_s \gg 1$---are directly visible in the effective potential felt by the classical degree of freedom. Setting $\theta_s$ to the ground-state value $\theta_{zpf} = (2E_C/E_J)^{1/4}$ and $U$ to its minimum-uncertainty value $1/4$ in Eq.~\eqref{eq:All_orders_effective}, we obtain
\begin{equation}
    V_{\rm eff}(\phi) = -E_J\text{Cos}(\theta)\text{Cos}(\theta_{zpf})+ \frac{E_C}{\theta_\text{zpf}^{2}},
\label{eq:Veff_explicit}
\end{equation}
where the dressing factor $\cos(\theta_{\rm zpf}) < 1$ reduces the Josephson well depth relative to the bare cosine. The effective potential preserves the periodicity and boundedness of the cosine at all phase values. This is the central structural advantage of the closure over the Kerr (fourth-order) truncation: the Kerr quartic $V_{\rm K}(\theta)=E_J\big(\theta^2/2-\theta^4/24\big)$ turns over at $|\theta|=\sqrt{6}$ and is unbounded below beyond it, so it cannot describe phase excursions approaching the top of the well.

\begin{figure}[H]
\centering
\includegraphics[width=0.9\textwidth]{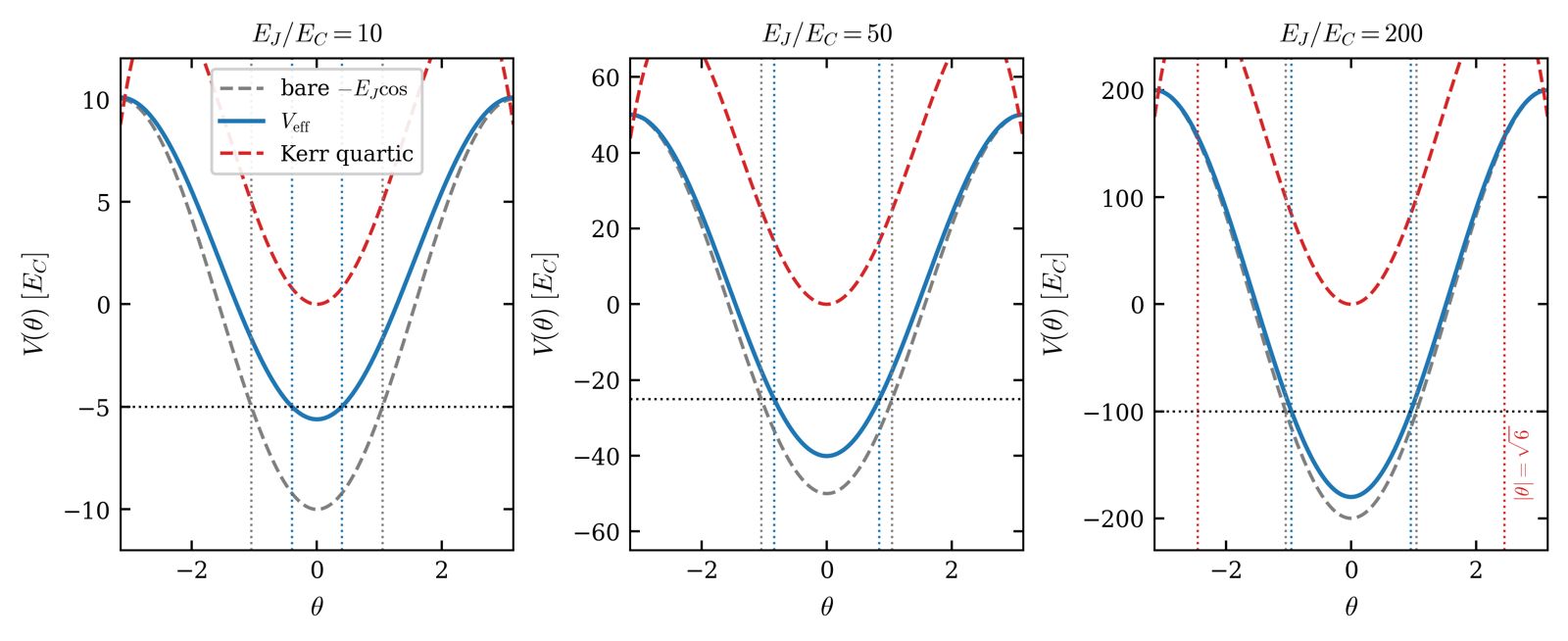}
\caption{Effective potential $V_{\rm eff}(\theta)$ (Eq.~\eqref{eq:Veff_explicit}, solid blue) compared with the bare cosine (gray dashed) and the fourth-order Kerr quartic (red dashed), for $E_J/E_C = 10,\,50,\,200$. The all-orders potential preserves the periodicity and boundedness of the cosine at all phase values; the Kerr quartic is unbounded below for $|\theta| > \sqrt{6}$. Horizontal dotted line: initial energy $E_{\rm initial}$; vertical markers: classical turning points for each potential.}
\label{fig:Veff}
\end{figure}

Figure~\ref{fig:Veff} compares $V_{\rm eff}$ with the bare cosine and the Kerr quartic. The classical turning points $\phi_{\rm tp}$, defined by $V(\phi_{\rm tp}) = E_{\rm initial}$, differ between the descriptions, giving a direct estimate of the amplitude error incurred by the Kerr truncation at moderate phase excursions.

The same dressing factor controls the small-oscillation frequency. Linearising the $\dot{n}$ equation of Eq.~\eqref{eq:EOM-JJ} about $\theta=n=0$, $\theta_s=\theta_{zpf}$,
\begin{equation}
    \frac{d^{2}\theta}{dt^{2}} = -\frac{8E_{c}E_{J}}{\hbar^{2}}\text{Sin}(\theta)\text{Cos}(\theta_{s})\;\approx\; -\Omega_{p}^{2} \text{Cos}(\theta_{s})\theta,
\end{equation}
so the dressed small-oscillation frequency is
\begin{equation}
\omega_{\rm eff} = \Omega_p\sqrt{\cos\!\left(\theta_{s}\right)},
\label{eq:omega_dressed}
\end{equation}
dressed by the same $\cos(\theta_{s})$ that appears in Eq.~\eqref{eq:Veff_explicit}. In the Kerr approximation the corresponding result is $\omega_{\mathrm{Kerr}}=\Omega_p-E_C/\hbar$, recovered from Eq.~\eqref{eq:omega_dressed} by setting $\phi_s=\phi_{\mathrm{zpf}}$, expanding the cosine to second order, and using $\sqrt{1-X}\approx 1-X/2$. The two analytic predictions coincide at leading order (the term $\propto\sqrt{E_C/E_J}$) and first differ at $\mathcal{O}(E_C/E_J)$. The linear frequency is consequently a regime in which the Kerr truncation already performs well: for $E_J/E_C=50$ we find $\omega_{\mathrm{eff}}/\Omega_p\approx 0.9496$ and $\omega_{\mathrm{Kerr}}/\Omega_p\approx 0.9500$, against the exact value $0.9471$; for $E_J/E_C=10$, $\omega_{\mathrm{eff}}/\Omega_p\approx 0.886$ and $\omega_{\mathrm{Kerr}}/\Omega_p\approx 0.888$, against $0.883$. Both reproduce the exact frequency to better than $0.6\%$ across this range and both reduce to $\Omega_p$ as $E_J/E_C\to\infty$, confirming that the all-orders Hamiltonian recovers the correct harmonic and weakly anharmonic limits.

\begin{figure}[H]
\centering
\includegraphics[width=0.9\textwidth]{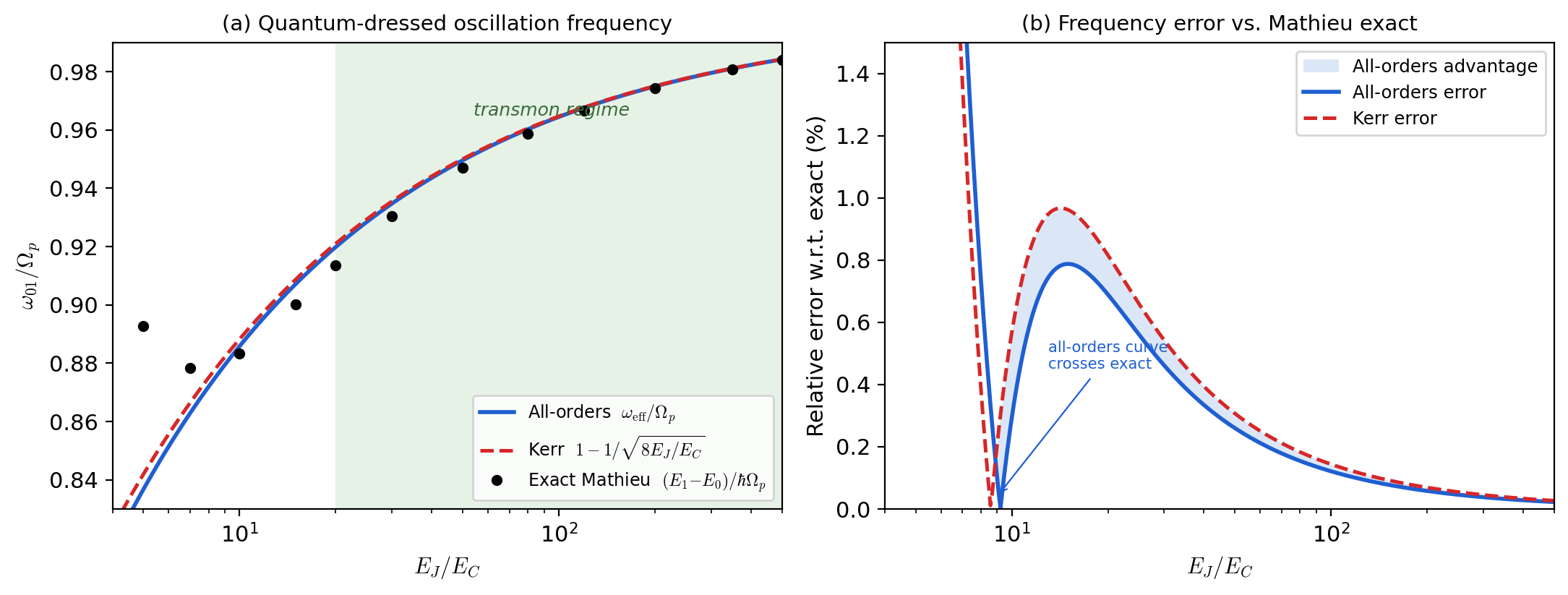}
\caption{Small-oscillation frequency $\omega_{01}/\Omega_p$ vs.\ $E_J/E_C$: all-orders (blue, Eq.~\eqref{eq:omega_dressed}), Kerr ($1-1/\sqrt{8E_J/E_C}$, red dashed), and exact diagonalization of Eq.~\eqref{eq:JJ_Hamiltonian} (black dots). The two analytic curves nearly coincide; panel (b) shows the relative error w.r.t.\ the exact result. Both analytic estimates reproduce the exact frequency to better than $0.6\%$ across the range, confirming that the linear spectroscopy is not where the closure and the Kerr truncation differ.}
\label{fig:omega_eff}
\end{figure}

This near-coincidence is itself informative: it shows that the value of the non-perturbative treatment lies not in the linear spectroscopy---where the harmonic frequency dominates and Kerr already suffices---but in the closed-form, bounded effective potential (Fig.~\ref{fig:Veff}) and in the nonlinear, large-excursion dynamics examined below. Figure~\ref{fig:omega_eff} plots $\omega_{\mathrm{eff}}/\Omega_p$ and $\omega_{\mathrm{Kerr}}/\Omega_p$ together with the exact frequency $\omega_{\mathrm{exact}}=(E_1-E_0)/\hbar$ from Eq.~\eqref{eq:Appendix_Mathieus_eigenvalues}; on the scale of panel~(a) the analytic curves are indistinguishable, and panel~(b) shows that both track the exact result within a fraction of a percent over the whole range.

\subsubsection{Dynamics of the classical phase}
\label{sec:comparison}
\begin{figure}[H]
\includegraphics[width=0.5\textwidth]{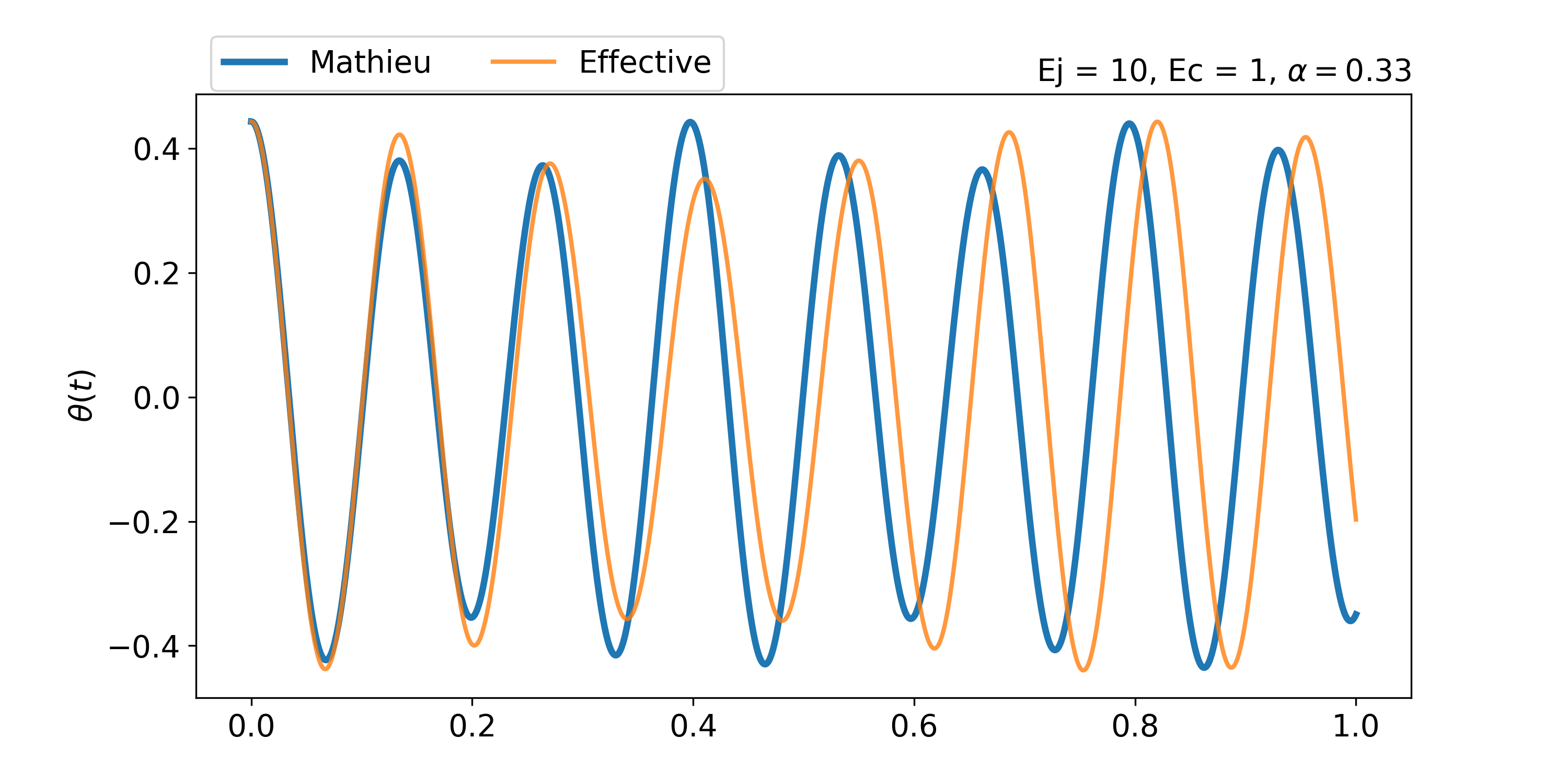}
\includegraphics[width=0.5\textwidth]{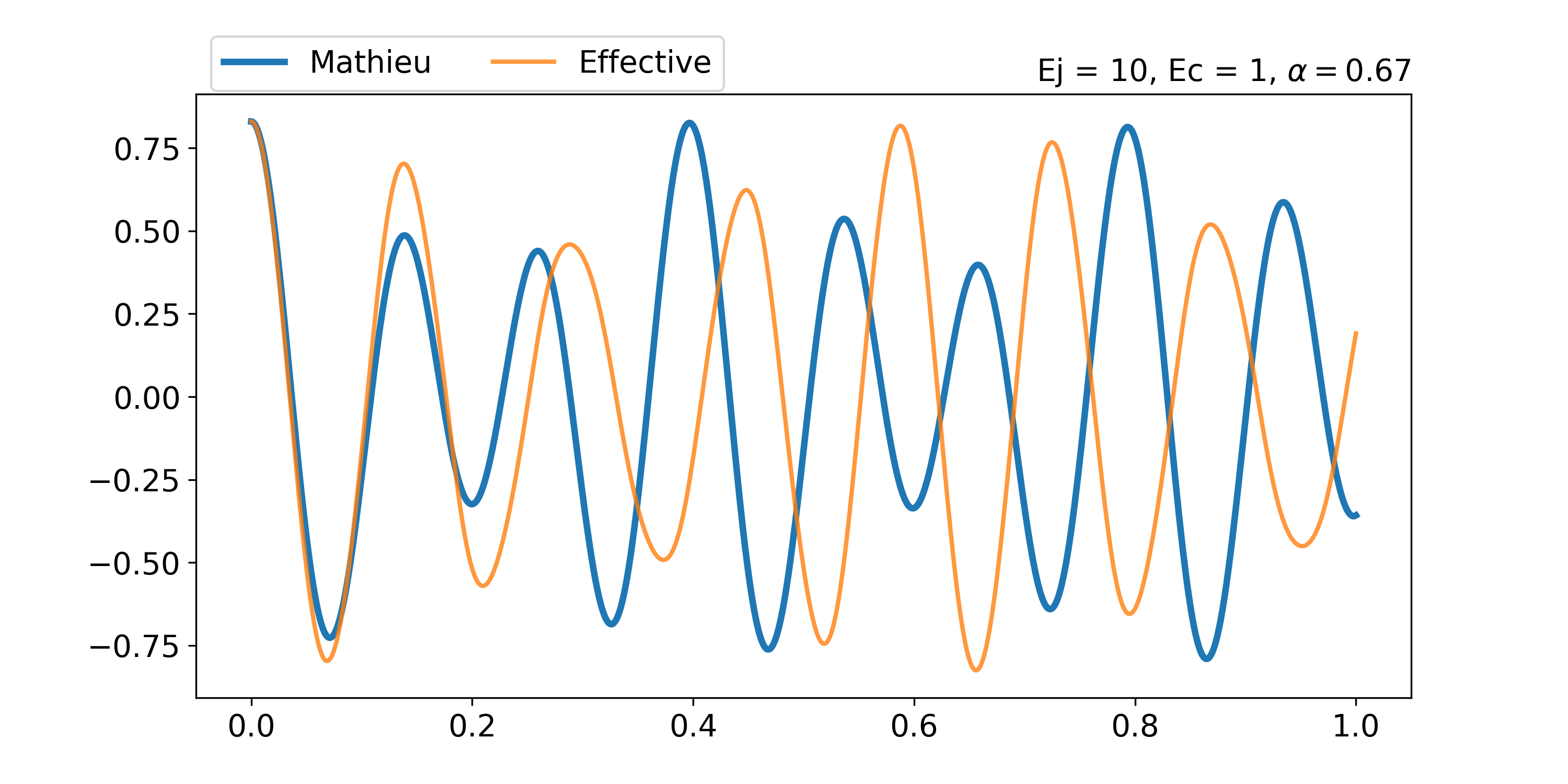}
\includegraphics[width=0.5\textwidth]{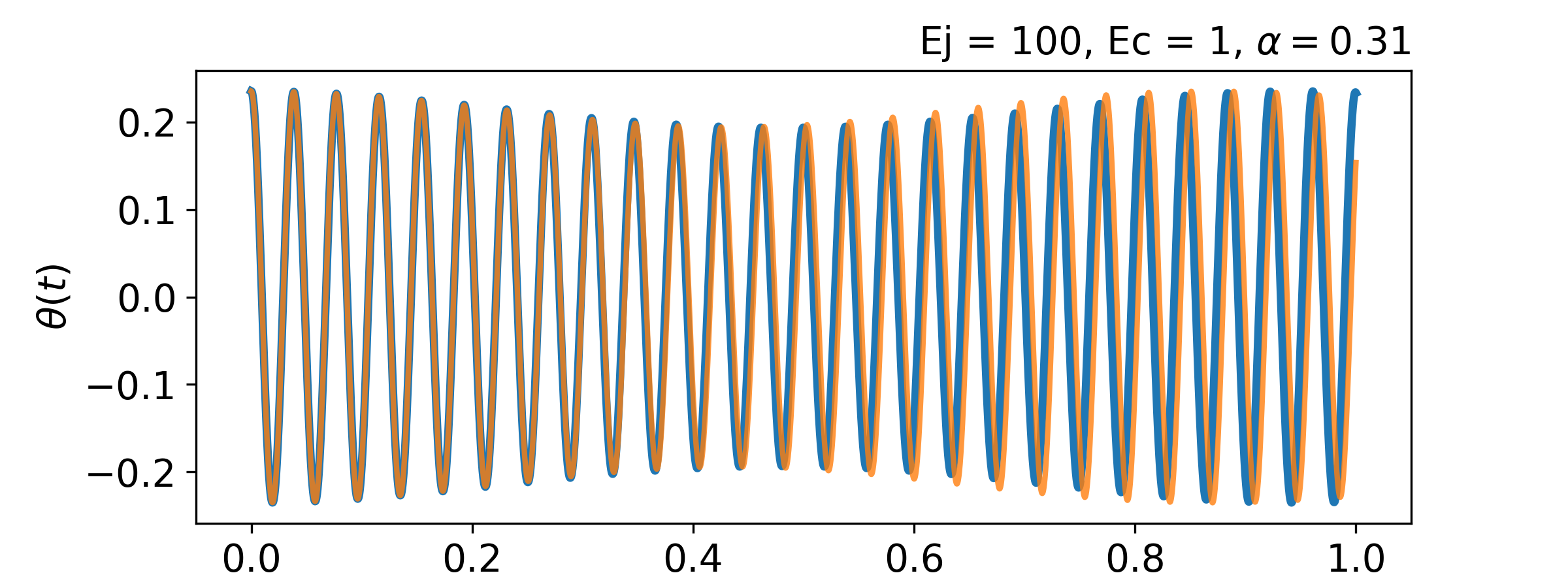}
\includegraphics[width=0.5\textwidth]{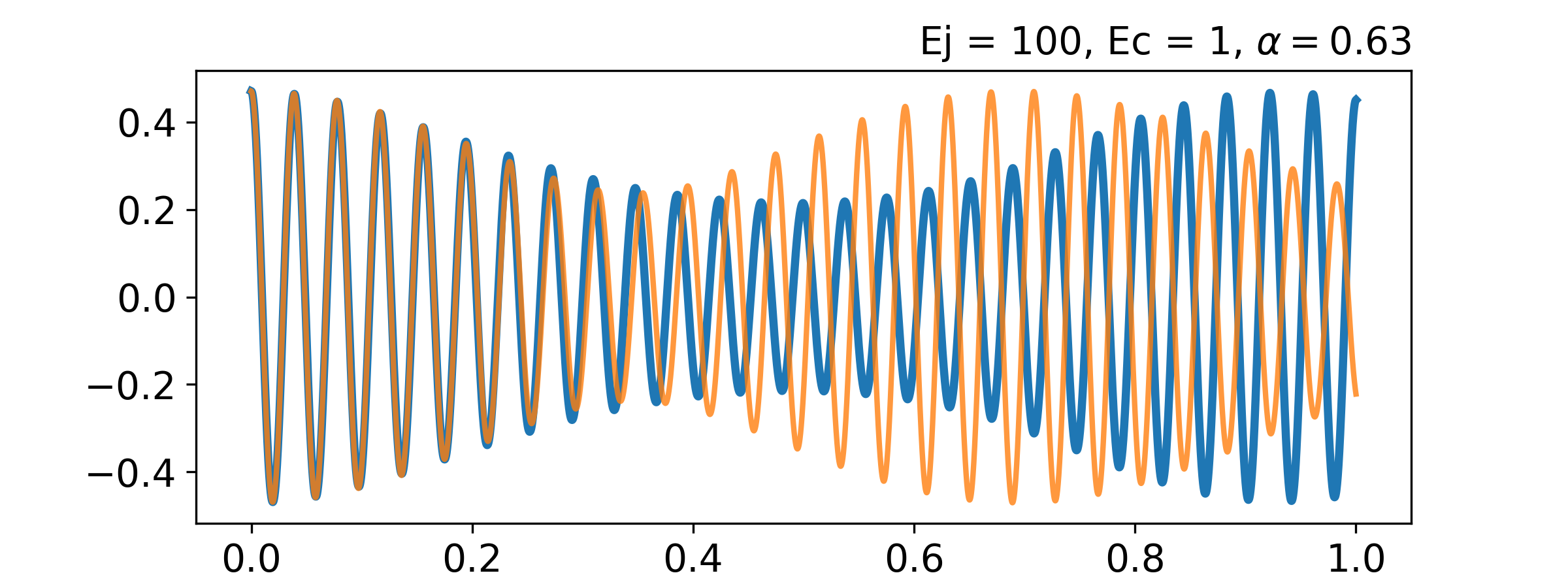}
\includegraphics[width=0.5\textwidth]{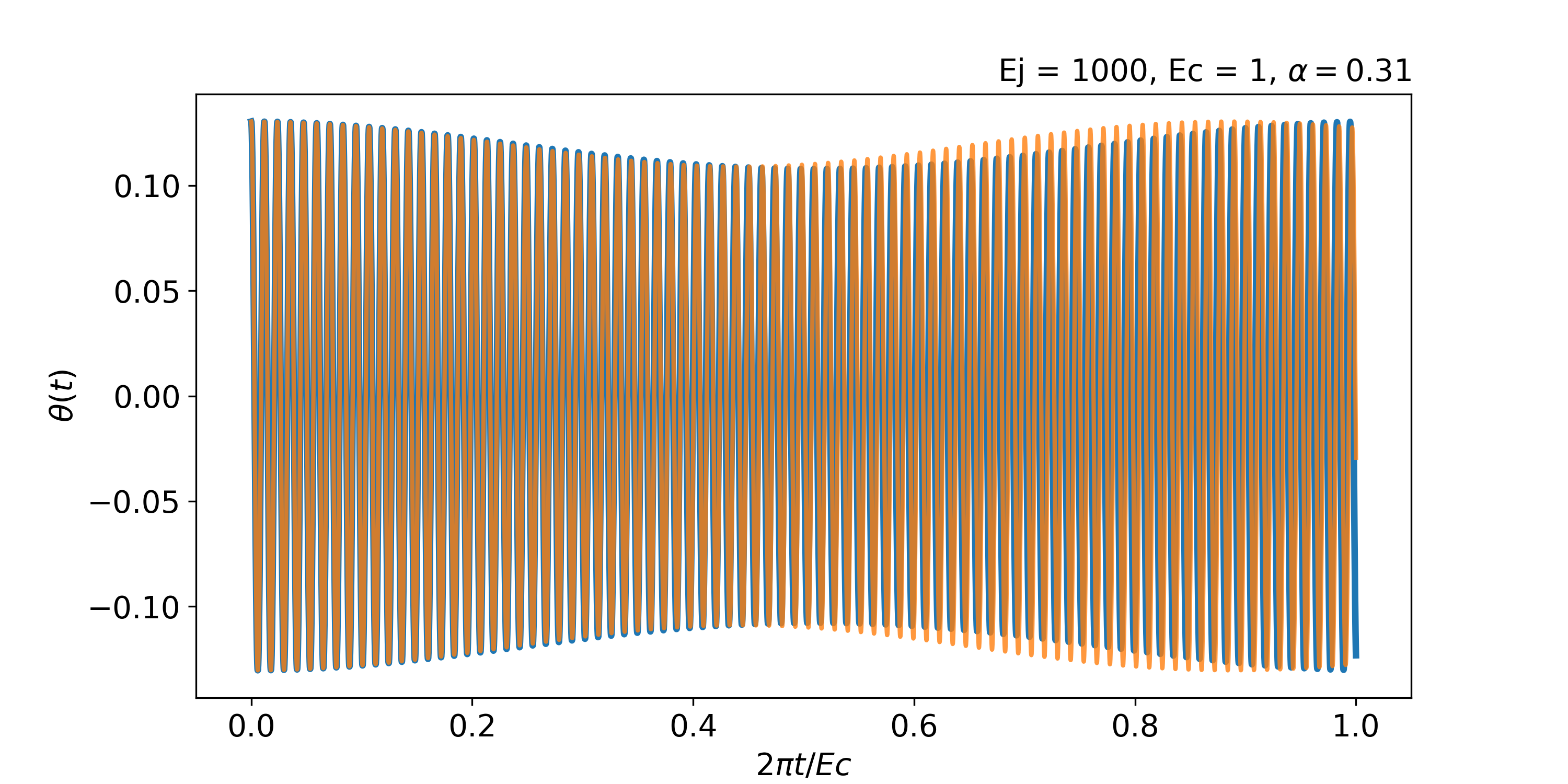}
\includegraphics[width=0.5\textwidth]{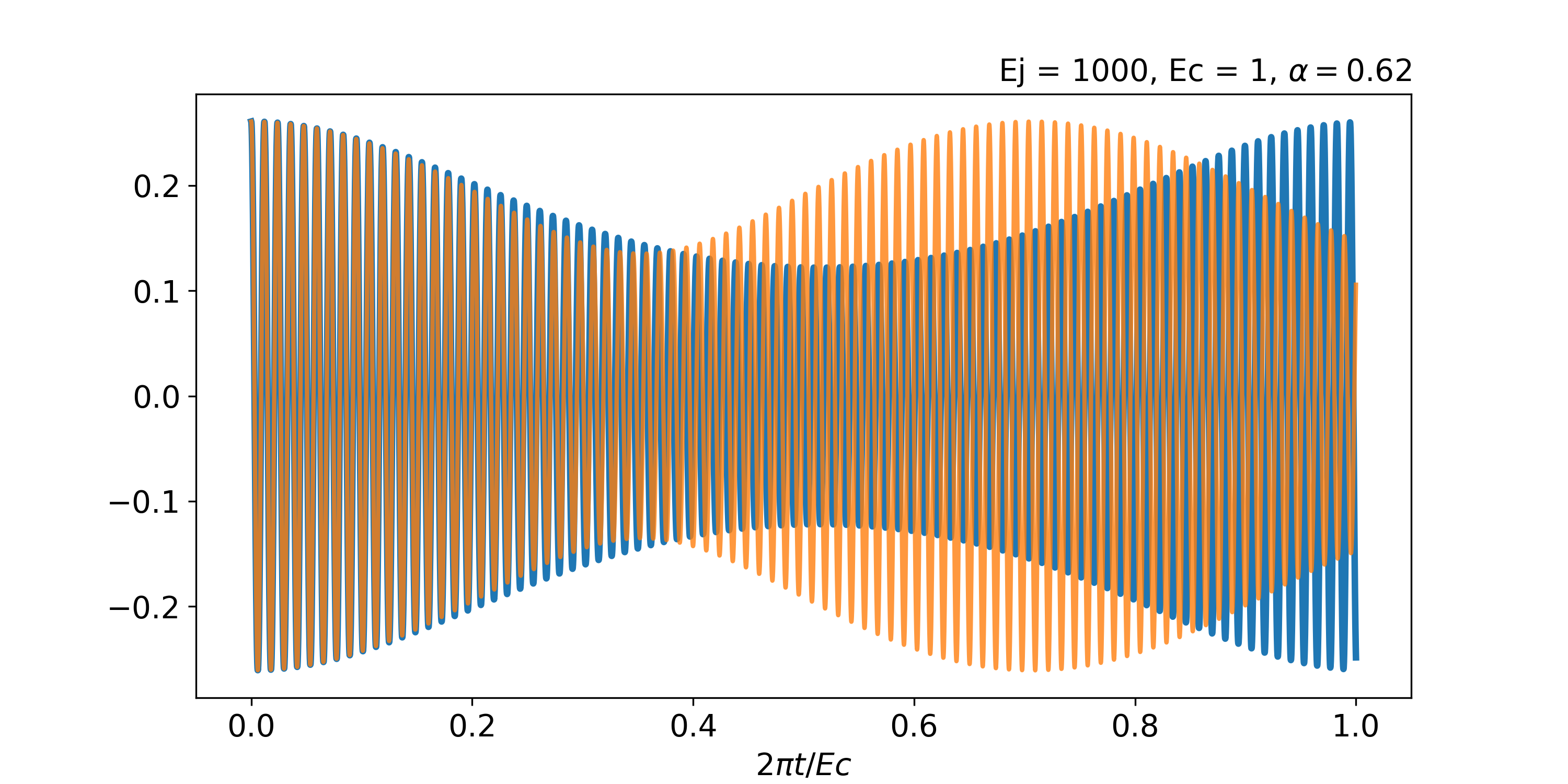}
\caption{Flux expectation value $\langle\hat\phi(t)\rangle$ of the all-orders
effective dynamics (Eq.~\eqref{eq:All_orders_effective}, orange) against the exact
finite-dimensional coherent-state evolution (Eq.~\eqref{eq:Appendix_Beta}, blue)
for the bare Josephson junction, as a function of dimensionless time over one
revival window $t\lesssim t_{\rm rev}=2\pi\hbar/E_C$. Panels: $E_J/E_C = 10$,
$100$, $1000$, with $\alpha = 0.3$ (left) and $\alpha = 0.6$ (right). On the left hand side, as $E_J/E_C$ increases the effective trajectory converges to the
exact result, whereas on the right hand side, for $E_J/E_C = 10$ the departure is already visible at
$t/t_{\rm rev}\approx 0.1$--$0.2$ due to the larger anharmonicity in
the intermediate regime; for $E_J/E_C = 100$ and $1000$ the agreement extends
much further and the residual departure near $t_{\rm rev}$ reflects the onset of
non-Gaussian (Yurke--Stoler cat-like) features (Sec.~\ref{sec:G20}).}
\label{fig:JJ_dynamics}
\end{figure}

We now test the effective dynamics in the time domain against the exact Schr\"odinger evolution. Using the energy basis of Sec.~\ref{sec:benchmark}, we construct a finite-dimensional coherent state adapted to the bounded JJ spectrum
\begin{equation}
    |\beta\rangle = e^{-\frac{|\alpha|^{2}}{2}}\sum_{m = 1}^{n} \frac{\alpha^{m}}{\sqrt{m!}}|m\rangle + e^{-\frac{|\alpha|^{2}}{2}} A|0\rangle,\quad A^{2} = e^{|\alpha|^{2}}-\sum_{m = 0}^{n} \frac{(|\alpha|^{2})^{m}}{m!}+1 ,
    \label{eq:finite-coherent}
\end{equation}
where the normalization constant $A$ is fixed by $\langle\beta|\beta\rangle = 1$ and satisfies $A\to1$ for $n\gg|\alpha|^2$, so that $|\beta\rangle$ reduces to the standard Glauber coherent state as $n\to\infty$. Its real and imaginary parts are fixed by the initial conditions of Eq.~\eqref{eq:EOM-JJ}, $\mathrm{Re}(\alpha) = \phi_{0}\theta(0)/(2\phizpf)$ and $\mathrm{Im}(\alpha) = 2e n(0)/(2Q_{\mathrm{zpf}})$, with $\theta_{zpf} = (2 E_C/E_J)^{1/4}$ and $n_{\mathrm{zpf}} = (1/2) (E_J/2E_C)^{1/4}$ the zero-point spreads of flux and charge. This construction follows~\cite{Perelomov1986} for finite-dimensional Hilbert spaces, adapted to the bounded JJ spectrum. The time-evolved state and the flux expectation value are
\begin{equation}
\label{eq:Appendix_Beta}
    |\beta(t)\rangle = e^{-\frac{|\alpha|^{2}}{2}}\sum_{m = 1}^{n} \frac{\alpha^{m}}{\sqrt{m!}}e^{-iE_{m}t/\hbar}|m\rangle + e^{-\frac{|\alpha|^{2}}{2}} Ae^{-iE_{0}t/\hbar}|0\rangle,
\end{equation}
\begin{equation}
\label{eq:Appendix_position_beta}
\begin{split}
 \frac{\langle\beta(t)|\hat{\theta}|\beta(t)\rangle}{\phizpf} &= e^{-|\alpha|^{2}}\sum_{m = 1}^{n-1} \frac{(|\alpha|^{2})^{m}}{m!}\left[\alpha e^{i(E_{m+1}-E_{m})t/\hbar} + \alpha^{*}e^{-i(E_{m+1}-E_{m})t/\hbar} \right]\\
 &+ e^{-|\alpha|^{2}} A\left[\alpha e^{i(E_{1}-E_{0})t/\hbar} + \alpha^{*}e^{-i(E_{1}-E_{0})t/\hbar} \right].
\end{split}
\end{equation}
Figure~\ref{fig:JJ_dynamics} compares $\langle\hat\theta(t)\rangle$ from the
all-orders effective dynamics, Eq.~\eqref{eq:EOM-JJ}, with the exact
evolution, Eq.~\eqref{eq:Appendix_Beta}, at representative values of $E_J/E_C$.
For low values of $\alpha$, the effective trajectory reproduces the exact carrier oscillation and its slow
amplitude modulation, and the agreement improves systematically as $E_J/E_C$
increases. However, as $\alpha$ gets bigger, the departure between the two trajectories becomes
visible already at $t/t_{\rm rev}\approx 0.1$--$0.2$ for $E_J/E_C = 10$, reflecting the larger
anharmonicity of the intermediate regime; for $E_J/E_C = 100$ and $1000$ the
agreement extends progressively further and the residual departure shifts toward
$t_{\rm rev}$, where the exact state develops non-Gaussian (Yurke--Stoler
cat-like) features that the second-order closure cannot represent.
We therefore restrict the
comparison window to $t\lesssim t_{\rm rev}$; the validity boundary is quantified
more sharply at the level of the quantum width in Sec.~\ref{sec:G20}.

\subsubsection{Dynamics of quantum fluctuations}
\label{sec:G20}
The comparison of Sec.~\ref{sec:comparison} tests the method at the level of the classical expectation value $\langle\hat\theta\rangle(t)$. A complementary and more sensitive diagnostic is the quantum width $\theta_s(t) = \sqrt{G^{2,0}(t)}$, whose equation of motion (second line of Eq.~\eqref{eq:EOM-JJ}) carries the nonlinear restoring force $\cos(\theta)\sin(\theta_{s})$, coupling the width to the classical phase. The closure therefore predicts a width that oscillates nonlinearly and exchanges energy with the classical sector---a Gaussian-level squeezing dynamics that a static second-order description does not contain.

The exact variance is computed from the finite-dimensional coherent-state evolution Eq.~\eqref{eq:Appendix_Beta} as
\begin{equation}
G^{2,0}_{\rm exact}(t) = \langle\beta(t)|\hat\theta^2|\beta(t)\rangle - \langle\beta(t)|\hat\theta|\beta(t)\rangle^2,
\label{eq:G20_exact}
\end{equation}
using the same Mathieu-basis matrix elements that enter Eq.~\eqref{eq:Appendix_position_beta}; the all-orders variance is $G^{2,0}(t) = \theta_s(t)^2$ from the ODE integration of Eq.~\eqref{eq:EOM-JJ}.

\begin{figure}[H]
\includegraphics[width=0.5\textwidth]{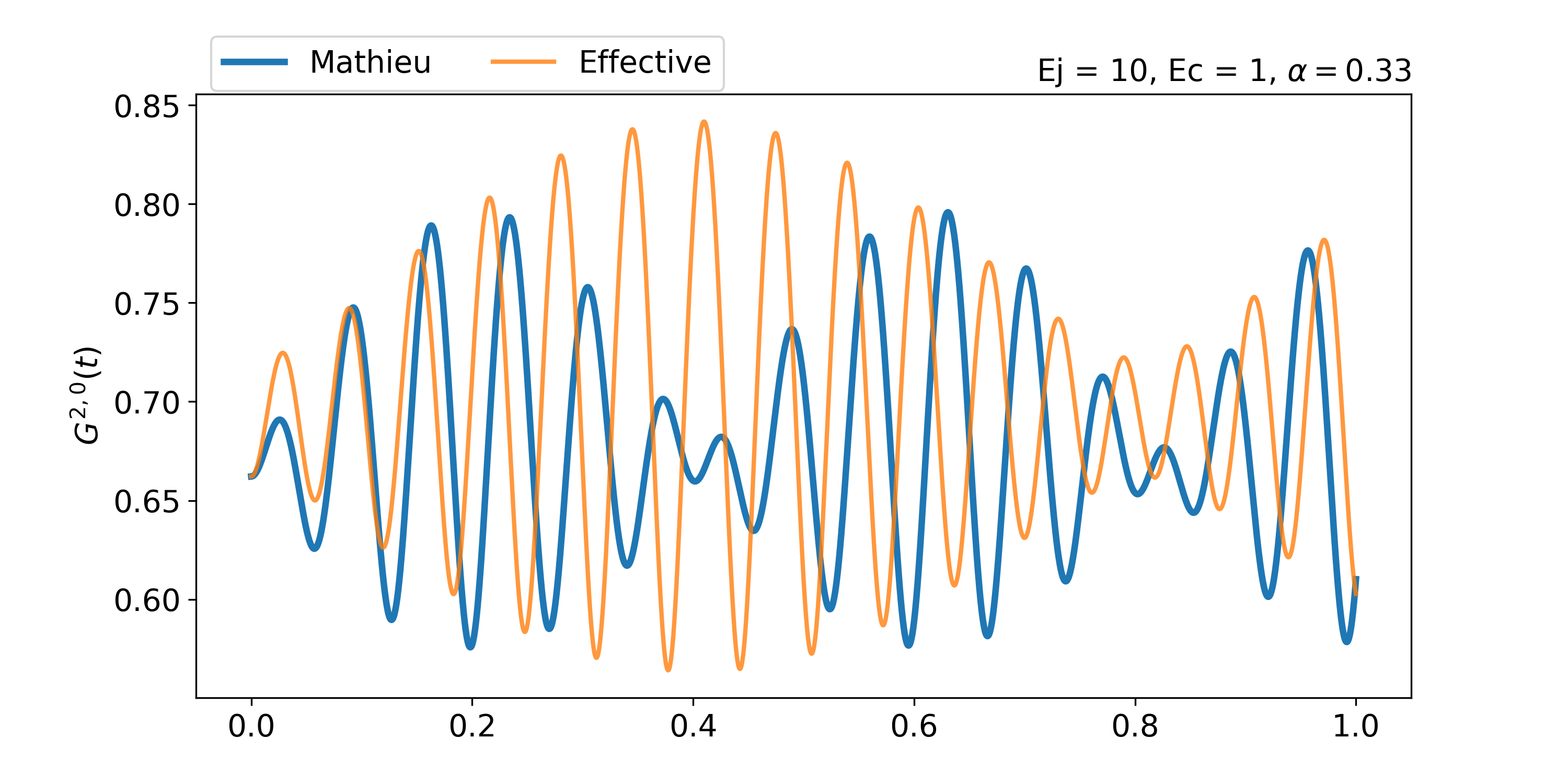}
\includegraphics[width=0.5\textwidth]{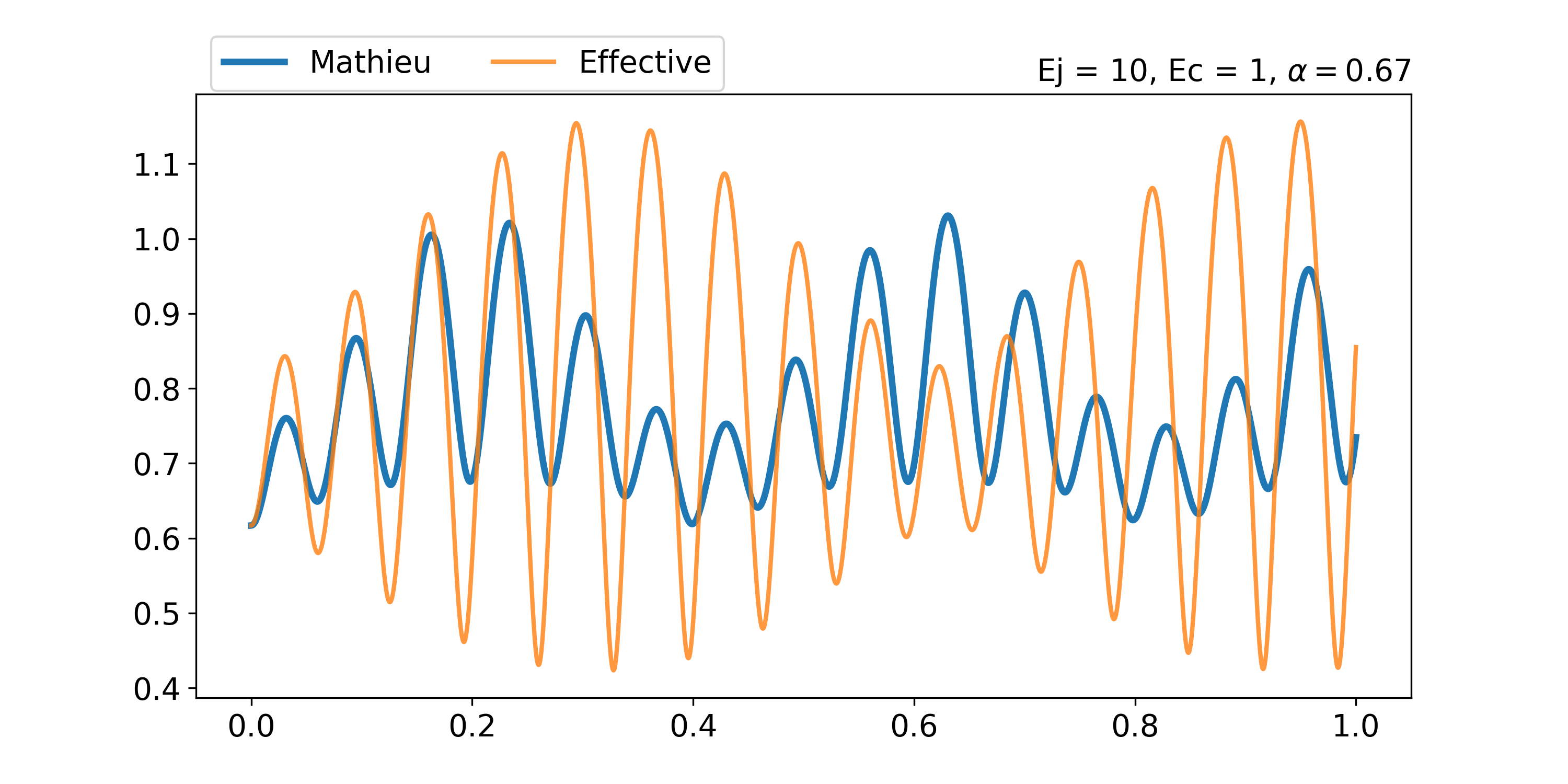}
\includegraphics[width=0.5\textwidth]{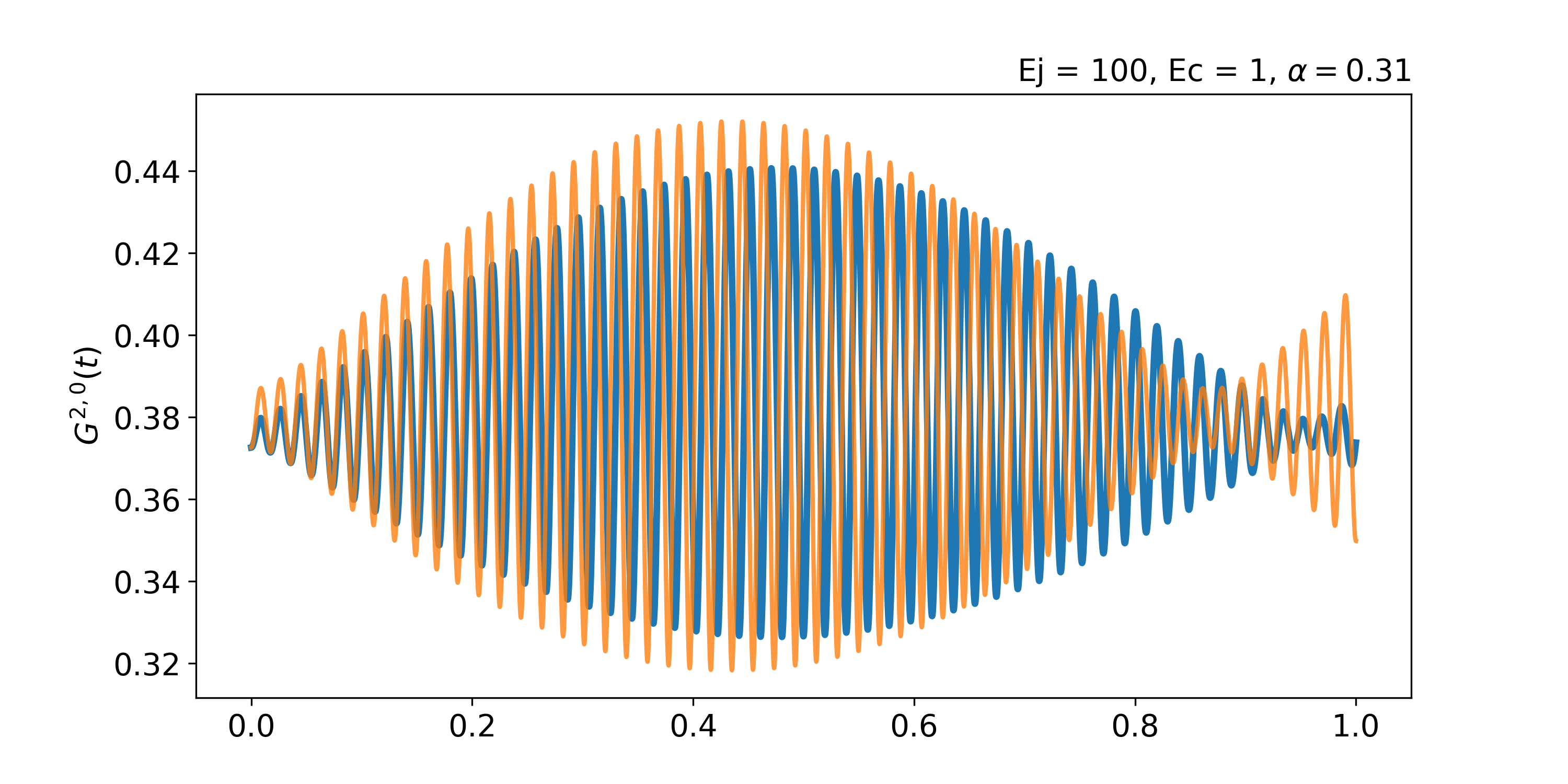}
\includegraphics[width=0.5\textwidth]{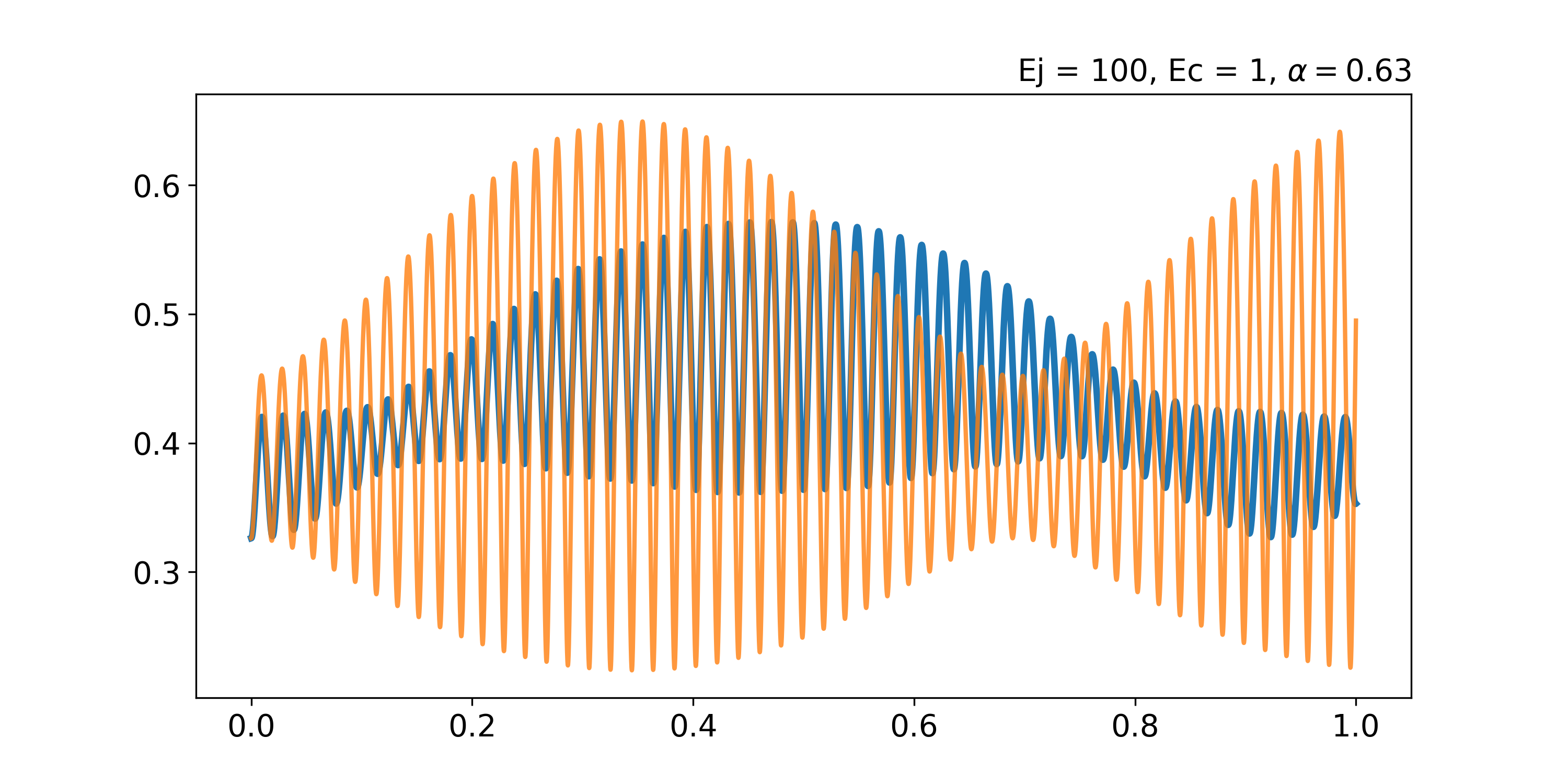}
\includegraphics[width=0.5\textwidth]{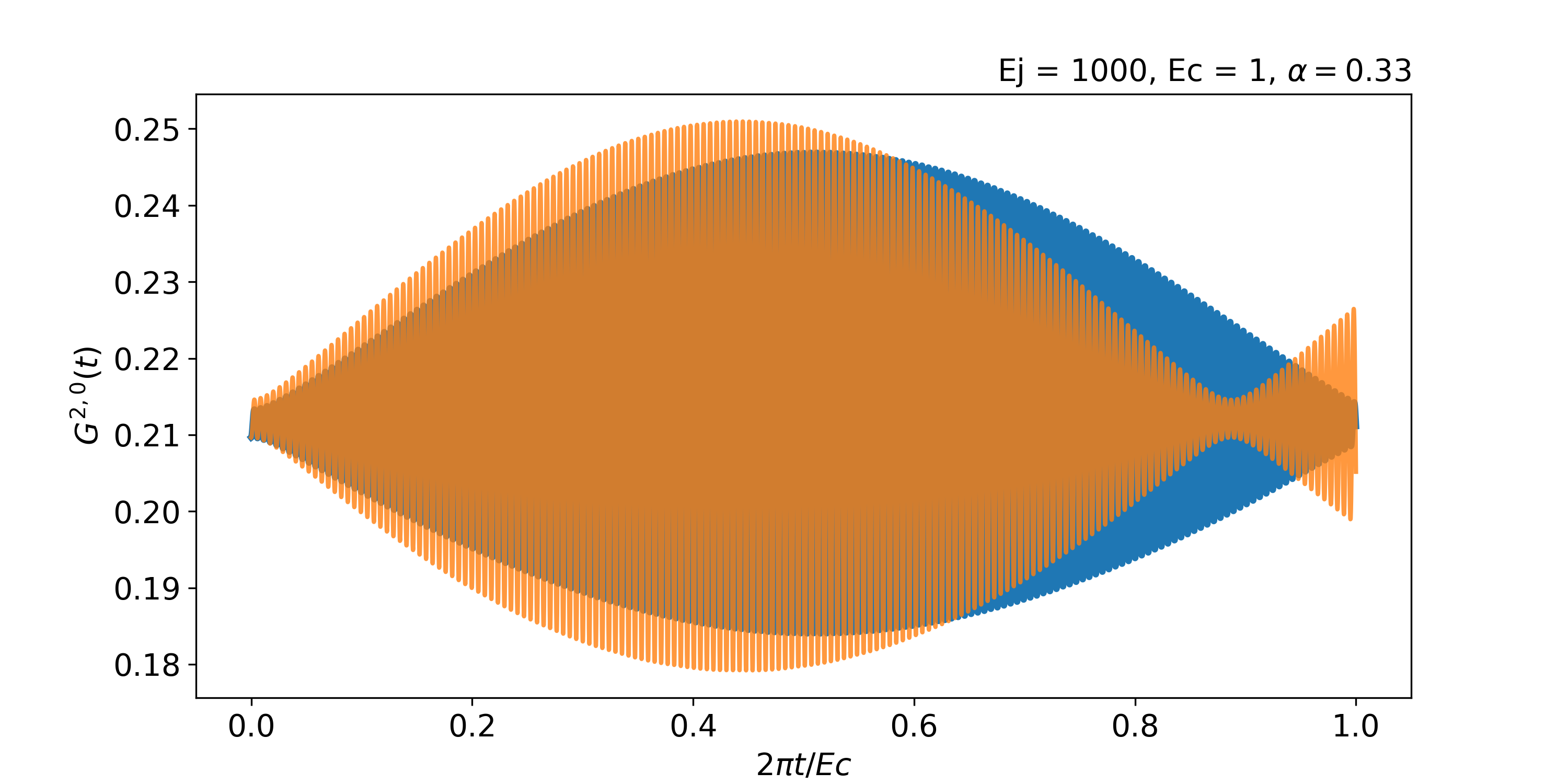}
\includegraphics[width=0.5\textwidth]{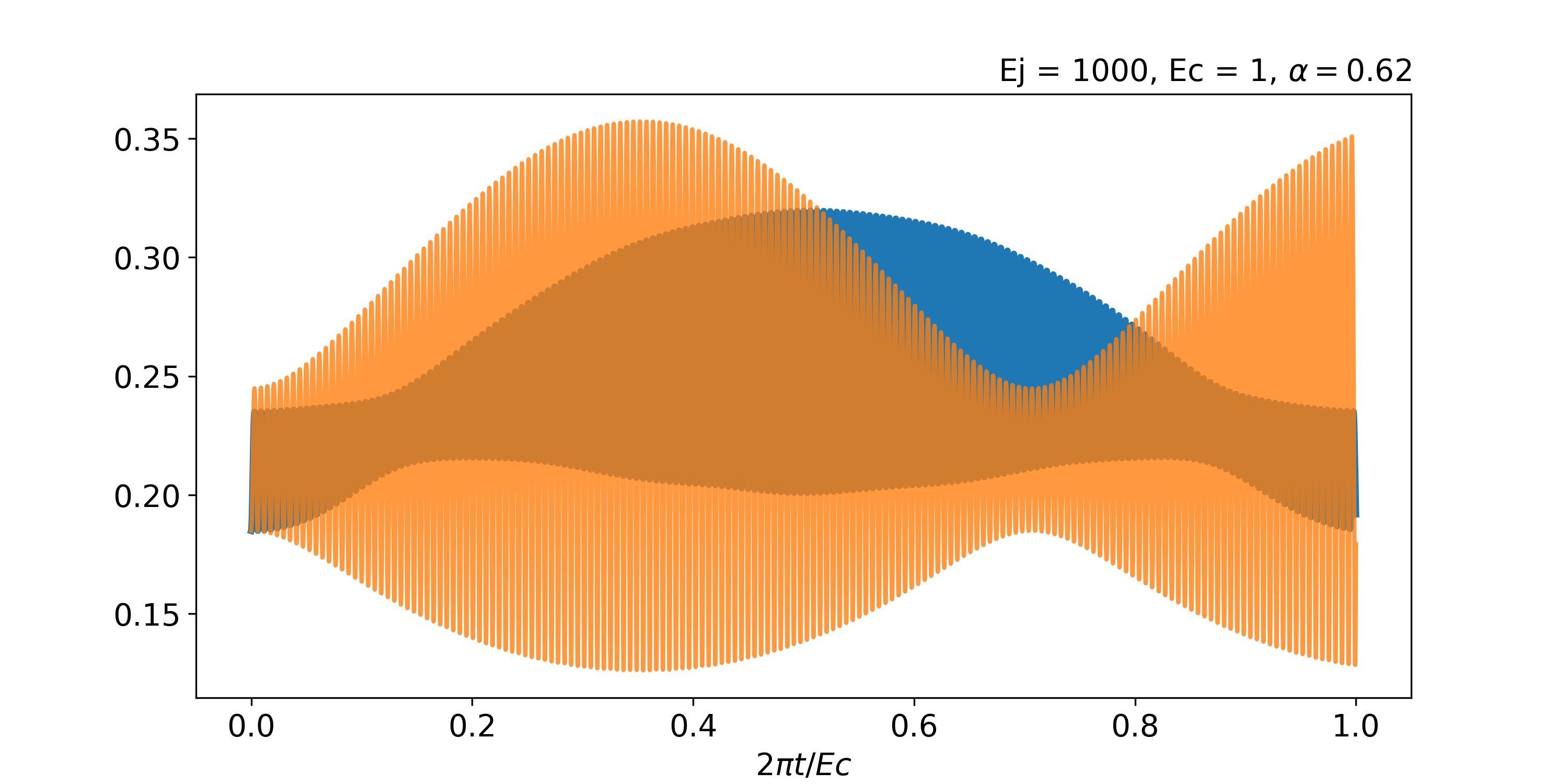}
\caption{Variance $G^{2,0}(t)$ for the bare Josephson junction
at $E_J/E_C = \{10,100,1000\}$, initial coherent state $|\alpha| = \{0.3,0.6\}$: exact (blue,
Eq.~\eqref{eq:G20_exact}) and all-orders (orange, $\theta_s(t)^2$ from
Eq.~\eqref{eq:EOM-JJ}). 
The closure correctly sets the initial variance and captures the order of magnitude
of the fluctuations, but predicts oscillations at the wrong timescale, making
$G^{2,0}(t)$ the most sensitive diagnostic of the Gaussian-closure limitation.}
\label{fig:G20}
\end{figure}

Figure~\ref{fig:G20} compares the two at $E_J/E_C = \{10,100,1000\}$, with  $|\alpha| = \{0.3,0.6\}$.
Both curves share the same initial value $G^{2,0}(0)=\phi_{\rm zpf}^2$
(inset of Fig.~\ref{fig:G20}), however, depending on $\alpha$'s value, their profile will either be similar ($\alpha = 0.3$), or will diverge almost immediately thereafter ($\alpha = 0.6$). The origin of this discrepancy resides in two things: the value of $\alpha$, and how well the closure employed approximates the energy eigenvalues of the cosine potential. To better understand this let us review Eq. (\ref{eq:Appendix_position_beta}), specifically this part
\begin{equation}
    e^{-|\alpha|^{2}}\sum_{m = 1}^{n-1} \frac{(|\alpha|^{2})^{m}}{m!} e^{i(E_{m+1}-E_{m})t/\hbar}
\end{equation}
from which we understand $e^{-|\alpha|^{2}}\frac{(|\alpha|^{2})^{m}}{m!}$ as a weight for the energy difference $E_{m+1}-E_{m} = \Delta E_{m}$. Thus, depending on $\alpha$, the contribution of the m-th value will be greater or lesser, in accordance to a Poissonian distribution. Now, as the the effective Hamiltonian is derived as a Taylor expansion, we can understand that higher energy values $E^{Eff}_{m}$ will be different from those exact ones, thus we should expect
\begin{equation}
    E^{Eff}_{m+1}-E^{Eff}_{m} = \Delta E_{m}+\epsilon_{m},\quad\text{with } \epsilon_{m+1}\geq \epsilon_{m}
\end{equation}
where $\epsilon_{m}$ is the error in the approximation of m-th term. It is from this that we observe that even if the ratio $E_{J}/E_{C}$ is big, if we get an $\alpha$ sufficiently big, it will have the contribution of higher energy terms, that ended up modifying the dynamics. Also, the opposite happens, even if we work with a highly anharmonic potential (e.g. $E_{J}/E_{C}\leq 10$), for small $\alpha$ values we see that the weights decay fast enough and we do not observe modifications on the dynamics.  
%
In the same manner, the Gaussian closure correctly sets the initial width and captures the general
magnitude of the fluctuations, but this starts to deviate from the exact one, a limitation rooted in the same truncation of higher cumulants that
causes the anharmonicity error. However, this difference is more pronounced than the one for $\langle\hat{\theta}\rangle$, as the weights act on the following energy difference
\begin{equation}
    e^{-|\alpha|^{2}}\sum_{m = 2}^{n-2} \frac{(|\alpha|^{2})^{m}}{m!} e^{i(E_{m+2}-E_{m})t/\hbar}
\end{equation}

This frequency mismatch makes $G^{2,0}(t)$ a sharper and earlier diagnostic of
the closure's limitations than the mean-field trajectory $\langle\hat\theta\rangle(t)$
of Fig.~\ref{fig:JJ_dynamics}: the error is exposed from $t=0$, before any
non-Gaussian features develop near $t_{\rm rev}$.

\subsection{Cavity-Josephson junction}
\label{subsection: Effective_Cavity-Josephson}
Having obtained the JJ effective dynamics, we proceed to the next circuit introduced in Sec.~\ref{section: SQC}. The motivation is that, to read out or drive the JJ, one couples it to a system that can be addressed externally---for example a resonant cavity (Fig.~\ref{fig:cav-JJ_circuit_diagram}), with Hamiltonian
\begin{equation}
\label{eq:H_cav-JJ_classical}
    H_{cav-JJ} = \frac{1}{2C_{J1}}Q_{1}^{2}+\frac{1}{2C_{J2}}Q_{2}^{2}+\frac{C_{0}}{C_{T}}Q_{1}Q_{2}+\frac{1}{2L_{1}}\phi_{1}^{2}-E_{J}\cos\left(\frac{\phi_{2}}{\phi_{0}}\right).
\end{equation}
This dynamics is more involved than the single JJ. In Appendix~\ref{appendix: II} we review the second-order truncation at the classical and quantum level. It is also possible to access higher-order terms by performing the closure of Eq.~(\ref{eq:Hierarchy}) on the moment hierarchy. Instead of the canonical pairs $(s,p_{s}, U)$, here it is convenient to keep the moments $G^{a,b,c,d}$ explicit and apply only the closure, giving
\begin{equation}
\label{eq:cavJJ_Hamiltonian}
    \begin{split}
         H_{Q} &= \frac{1}{2C_{J1}}Q_{1}^{2}+\frac{1}{2C_{J2}}Q_{2}^{2}+\frac{C_{0}}{C_{T}}Q_{1}Q_{2}+\frac{1}{2L_{1}}\phi_{1}^{2}\\
         &+ \frac{1}{C_{J1}}G^{0,2,0,0}+\frac{1}{C_{J2}}G^{0,0,0,2}+\frac{C_{0}}{C_{T}}G^{0,1,0,1}+\frac{1}{L_{1}}G^{2,0,0,0}+\sum_{n=0}^{\infty}\frac{1}{(2n)!}\frac{d^{2n}V(\phi_{2})}{d\phi_{2}^{2n}}\left(\sqrt{G^{0,0,2,0}}\right)^{2n}\\
         &= \frac{1}{2C_{J1}}Q_{1}^{2}+\frac{1}{2C_{J2}}Q_{2}^{2}+\frac{C_{0}}{C_{T}}Q_{1}Q_{2}+\frac{1}{2L_{1}}\phi_{1}^{2}+ \frac{1}{C_{J1}}G^{0,2,0,0}+\frac{1}{L_{1}}G^{2,0,0,0}+\frac{1}{C_{J2}}G^{0,0,0,2}\\
         &+\frac{C_{0}}{C_{T}}G^{0,1,0,1}+\frac{1}{2}\left[V\left(\phi_{2}+s\right)+V(\phi_{2}-s)\right],
    \end{split}
\end{equation}
where $s = \sqrt{G^{0,0,2,0}}$ and $V(\phi_{2}) = -E_{J}\cos(\phi_{2}/\phi_{0})$. To obtain the all-orders dynamics we evaluate the moment Poisson brackets with the Leibniz rule
\begin{equation*}
    \{G^{0,0,0,2},\sqrt{G^{0,0,2,0}}\} = -2\frac{G^{0,0,1,1}}{\sqrt{G^{0,0,2,0}}}\;\longrightarrow\; \{G^{0,0,0,2},\left(\sqrt{G^{0,0,2,0}}\right)^{2n}\} = -2\frac{G^{0,0,1,1}}{\sqrt{G^{0,0,2,0}}}\, 2n\left(\sqrt{G^{0,0,2,0}}\right)^{2n-1}.
\end{equation*}
The bracket of the charge variance with the effective potential then reads
\begin{equation*}
    \left\{G^{0,0,0,2}, \frac{V\left(\phi_{2}+s\right)+V\left(\phi_{2}-s\right)}{2}\right\}= -2\frac{G^{0,0,1,1}}{\sqrt{G^{0,0,2,0}}}\sum_{n=1}^{\infty}\frac{1}{(2n-1)!}\frac{d^{2n}V(\phi_{2})}{d\phi_{2}^{2n}}\left(\sqrt{G^{0,0,2,0}}\right)^{2n-1}.
\end{equation*}
With $k=2n-1$ the sum runs over odd $k$ only, and the odd part of the Taylor series of $V'(\phi_2+s)$ resums to the antisymmetric average $\tfrac12\big[V'(\phi_2+s)-V'(\phi_2-s)\big]$:
\begin{equation*}
    \left\{G^{0,0,0,2}, \frac{V\left(\phi_{2}+s\right)+V\left(\phi_{2}-s\right)}{2}\right\}= -\frac{G^{0,0,1,1}}{\sqrt{G^{0,0,2,0}}}\left[\frac{dV(\phi_{2}+s)}{d\phi_{2}}-\frac{dV(\phi_{2}-s)}{d\phi_{2}}\right],
\end{equation*}
with, for the Josephson cosine,
\begin{equation*}
    \frac{dV(\phi_{2}+s)}{d\phi_{2}}-\frac{dV(\phi_{2}-s)}{d\phi_{2}} = \frac{E_{J}}{\phi_{0}}\left[\sin\left(\frac{\phi_{2}+s}{\phi_{0}}\right)-\sin\left(\frac{\phi_{2}-s}{\phi_{0}}\right)\right]=\frac{2E_{J}}{\phi_{0}}\cos\!\left(\frac{\phi_2}{\phi_0}\right)\sin\!\left(\frac{s}{\phi_0}\right).
\end{equation*}
This is the coupled-circuit analogue of the width force in Eq.~\eqref{eq:EOM-JJ}: as in the bare junction, the classical force on $Q_2$ is the symmetric average $\tfrac12[V'(\phi_2+s)+V'(\phi_2-s)]$, while the moment force is the antisymmetric combination above; both reduce to the bare-JJ result when the cavity is decoupled ($C_0\to0$). The resulting EOM for the classical part are
\begin{equation}
\label{eq:classical_all_orders_cav-JJ}
    \begin{split}
        \dot{\phi}_{1} &= \frac{1}{C_{J_1}}Q_{1}+\frac{C_{0}}{C_{T}}Q_{2}\,,\qquad
        \dot{Q}_{1} = -\frac{1}{L_{1}}\phi_{1}\\
        \dot{\phi}_{2} &= \frac{1}{C_{J_2}}Q_{2}+\frac{C_{0}}{C_{T}}Q_{1}\,,\qquad
        \dot{Q}_{2} = -\frac{E_{J}}{2\phi_{0}}\left[\sin\left(\frac{\phi_{2}+s}{\phi_{0}}\right)+\sin\left(\frac{\phi_{2}-s}{\phi_{0}}\right)\right]=-\frac{E_J}{\phi_0}\sin\!\left(\frac{\phi_2}{\phi_0}\right)\cos\!\left(\frac{s}{\phi_0}\right)
    \end{split}
\end{equation}
and for the quantum part
\begin{equation}
    \begin{split}
        \dot{G}^{2,0,0,0} &= \frac{2}{C_{J_1}}G^{1,1,0,0}+\frac{2C_{0}}{C_{T}}G^{1,0,0,1},\qquad
        \dot{G}^{0,2,0,0} = -\frac{2}{L_{1}}G^{1,1,0,0}\\
        \dot{G}^{0,0,2,0} &= \frac{2}{C_{J_2}}G^{0,0,1,1}+\frac{2C_{0}}{C_{T}}G^{0,1,1,0},\qquad
        \dot{G}^{0,0,0,2} =\frac{-2G^{0,0,1,1}}{\sqrt{G^{0,0,2,0}}}\frac{E_{J}}{\phi_{0}}\left[\sin\left(\frac{\phi_{2}+s}{\phi_{0}}\right)-\sin\left(\frac{\phi_{2}-s}{\phi_{0}}\right)\right]\\
        \dot{G}^{1,0,0,1} &=\frac{1}{C_{J_1}}G^{0,1,0,1}+\frac{C_{0}}{C_{T}}G^{0,0,0,2}-\frac{G^{1,0,1,0}}{\sqrt{G^{0,0,2,0}}}\frac{E_{J}}{\phi_{0}}\left[\sin\left(\frac{\phi_{2}+s}{\phi_{0}}\right)-\sin\left(\frac{\phi_{2}-s}{\phi_{0}}\right)\right]\\
        \dot{G}^{1,0,1,0} &= \frac{1}{C_{J_1}}G^{0,1,1,0}+\frac{1}{C_{J_2}}G^{1,0,0,1}+\frac{C_{0}}{C_{T}}\left(G^{1,1,0,0}+G^{0,0,1,1}\right)\\
        \dot{G}^{0,1,0,1} &=-\frac{1}{L_{1}}G^{1,0,0,1}-\frac{G^{0,1,1,0}}{\sqrt{G^{0,0,2,0}}}\frac{E_{J}}{\phi_{0}}\left[\sin\left(\frac{\phi_{2}+s}{\phi_{0}}\right)-\sin\left(\frac{\phi_{2}-s}{\phi_{0}}\right)\right]\\
        \dot{G}^{0,1,1,0} &= -\frac{1}{L_{1}}G^{1,0,1,0}+\frac{1}{C_{J_2}}G^{0,1,0,1}+\frac{C_{0}}{C_{T}}G^{0,2,0,0}\\
        \dot{G}^{1,1,0,0} &= \frac{1}{C_{J_1}}G^{0,2,0,0}-\frac{1}{L_{1}}G^{2,0,0,0}+\frac{C_{0}}{C_{T}}G^{0,1,0,1}\\
        \dot{G}^{0,0,1,1} &= \frac{1}{C_{J_2}}G^{0,0,0,2}+\frac{C_{0}}{C_{T}}G^{0,1,0,1}-\sqrt{G^{0,0,2,0}}\,\frac{E_{J}}{\phi_{0}}\left[\sin\left(\frac{\phi_{2}+s}{\phi_{0}}\right)-\sin\left(\frac{\phi_{2}-s}{\phi_{0}}\right)\right]
    \end{split}
\label{eq:Gs-cav}
\end{equation}
\begin{figure}[H]
\centering
\subfloat[Cavity flux $\phi_1(t)$]{\includegraphics[width=0.45\linewidth]{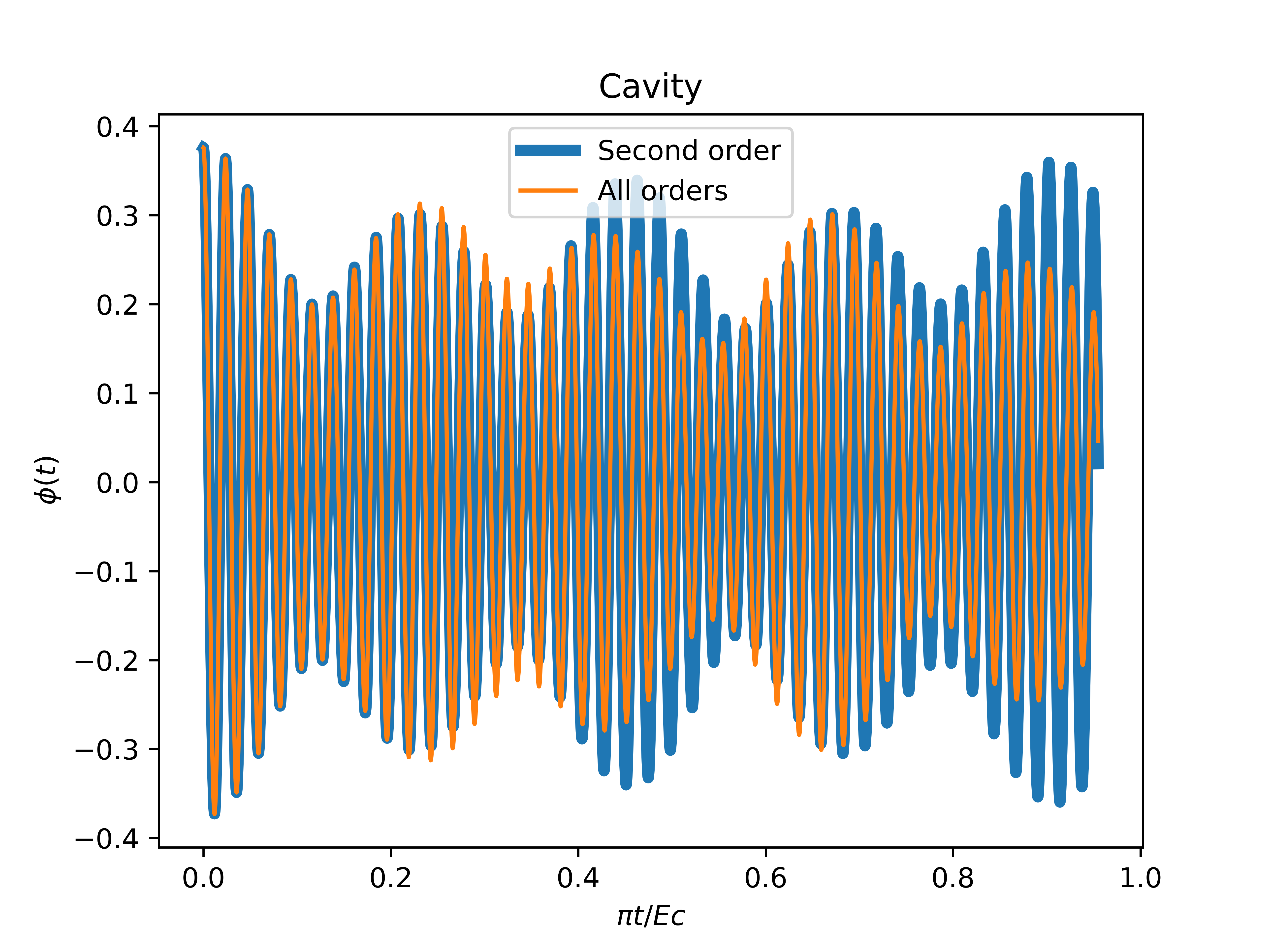}\label{subfigure:cav_dynamics_all_order}}
\hfill
\subfloat[Energy evolution]{\includegraphics[width=0.45\linewidth]{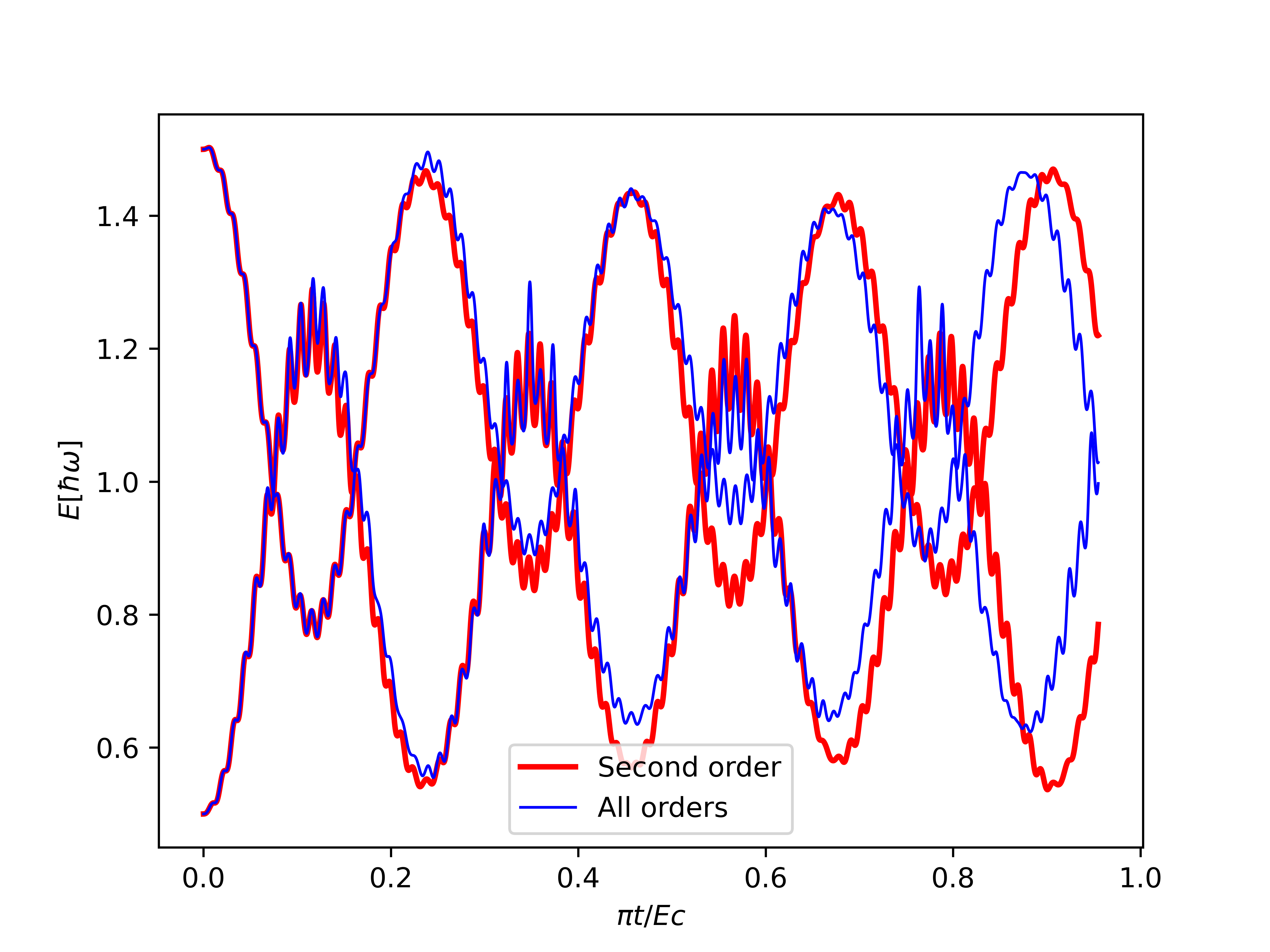}\label{subfigure:energy_dynamics_all_order}}
\caption{Comparison between the all-orders effective dynamics (Eqs.~\eqref{eq:classical_all_orders_cav-JJ}--\eqref{eq:Gs-cav}) and the second-order (cavity--cavity) approximation (Appendix~\ref{appendix: II}) for the cavity--Josephson junction. (a)~Cavity flux $\phi_1(t)$. (b)~Energy evolution, showing the quantum state transfer between the cavity and the JJ. At short times both approximations agree; at later times the all-orders dynamics departs in amplitude as the full cosine nonlinearity becomes important.}
\label{fig:all_order_cavity_cavity_interaction}
\end{figure}
Figure~\ref{fig:all_order_cavity_cavity_interaction} compares the all-orders system, Eqs.~\eqref{eq:classical_all_orders_cav-JJ}--\eqref{eq:Gs-cav}, with the second-order approximation of Appendix~\ref{appendix: II}. Panel~\ref{subfigure:cav_dynamics_all_order} shows that the cavity flux $\phi_1(t)$ agrees at short times and that the amplitudes begin to differ as time increases while the carrier frequency stays nearly the same; panel~\ref{subfigure:energy_dynamics_all_order} shows the energy exchange between the cavity and the JJ, including the fast oscillating terms generated by the coupling. The section thus demonstrates that the effective framework extends naturally to a coupled, multi-mode circuit without invoking a rotating-wave or dispersive approximation.

\subsection{Dissipative dynamics}
\label{sec:dissipative}
In physical realizations, superconducting circuits are inevitably coupled to their
environment, leading to energy dissipation and decoherence. A complete description
must therefore incorporate dissipative effects. In what follows we study three
approaches within the effective formalism: the Caldirola--Kanai (CK)
time-dependent Hamiltonian~\cite{CK_Caldirola,CK_Kanai}, the Bateman
dual-oscillator Hamiltonian~\cite{DEKKER19811,JavierValdez_2025}, and the
Lindblad master equation~\cite{Breuer2002,RevModPhys.93.025005}. The CK model is the
simplest and serves as a baseline; the Bateman approach preserves the canonical
structure at the cost of an auxiliary degree of freedom; and the Lindblad
framework is the gold standard for completely positive trace-preserving (CPTP)
dynamics.

A useful organizing principle, which we exploit throughout this section, is that
once the dynamics is projected onto the physical second-order moments
$\mathbf{G}$ of the cavity--cavity system, all three models can be cast in the
common linear form
\begin{equation}
  \dot{\mathbf{G}} = \mathbf{M}\,\mathbf{G} + \mathbf{d},
  \label{eq:moment_master}
\end{equation}
with one and the same drift matrix $\mathbf{M}$. The physics of dissipation of the three models however, is
encoded entirely in the inhomogeneous term $\mathbf{d}$: it is effectively
time dependent for CK, a constant vacuum
floor $\propto\gamma\hbar$ for Lindblad, and a set of bounded, oscillatory
auxiliary-oscillator moments for Bateman. A quantitative comparison and a
discussion of the respective regimes of validity are given in
Appendix~\ref{sec:quantitative_comparison}.

\subsubsection{Caldirola--Kanai dissipation}
\label{sec:CK}

The simplest Hamiltonian model of dissipation for the cavity oscillator is the
Caldirola--Kanai (CK) Hamiltonian~\cite{CK_Caldirola,CK_Kanai,caldirola1983quantum,DEKKER19811}:
\begin{equation}
  H_{\rm CK} = \frac{Q_1^2}{2\CJ}\,e^{-\lambda t}
              + \frac{\CJ\,\omega_1^2}{2}\,\phi_1^2\,e^{\lambda t},
  \label{eq:CK_ham}
\end{equation}
where $Q_1$ is the canonical momentum conjugate to $\phi_1$ and $\lambda$
is the damping rate, which plays the same role as the parameter $\lambda$ of
the Bateman Hamiltonian, Eq.~\eqref{eq:bateman_full}. The kinetic (mechanical)
momentum is
\begin{equation}
  p_{\rm mech} = \CJ\,\dot\phi_1 = Q_1\,e^{-\lambda t},
  \label{eq:CK_pmech}
\end{equation}
and the classical equation of motion is that of a damped oscillator,
\begin{equation}
  \ddot\phi_1 + \lambda\,\dot\phi_1 + \omega_1^2\,\phi_1 = 0.
  \label{eq:CK_eom_class}
\end{equation}

Since $H_{\rm CK}$ is quadratic in the canonical variables, the moment expansion
of Eq.~\eqref{eq:H_q} terminates exactly at second order and no Gaussian
closure is needed. The effective Hamiltonian is
\begin{equation}
  H_{\rm CK}^Q = H_{\rm CK}(\phi_1, Q_1)
               + \frac{\GT{0}{2}}{2\CJ}\,e^{-\lambda t}
               + \frac{\CJ\,\omega_1^2}{2}\,\GT{2}{0}\,e^{\lambda t}.
  \label{eq:CK_eff}
\end{equation}

Applying $\dot{f} = \{f, H_{\rm CK}^Q\}$ to the second-order moments
$(\GT{2}{0},\,\GT{0}{2},\,\GT{1}{1})$ yields
\begin{align}
  \dot{G}^{2,0} &= \frac{2}{\CJ}\,\GT{1}{1}\,e^{-\lambda t}, &
  \dot{G}^{0,2} &= -2\CJ\omega_1^2\,\GT{1}{1}\,e^{\lambda t}, &
  \dot{G}^{1,1} &= \frac{\GT{0}{2}}{\CJ}\,e^{-\lambda t}
                    -\CJ\omega_1^2\,\GT{2}{0}\,e^{\lambda t}.
  \label{eq:CK_Gs}
\end{align}
A direct computation using Eqs.~\eqref{eq:CK_Gs} shows that the canonical
uncertainty product
\begin{equation}
  \Ucan(t) \equiv \GT{2}{0}\,\GT{0}{2} - \bigl(\GT{1}{1}\bigr)^2
  \label{eq:CK_Ucan}
\end{equation}
satisfies $\dot{U}_{\text{can}} = 0$; it is conserved at its initial value
$\Ucan(0) = \hbar^2/4$ for a minimum-uncertainty (coherent) initial state
(see Appendix~\ref{app:CK}).

This formal conservation is, however, deceptive. The physically meaningful
uncertainty product is built from the kinetic momentum $p_{\rm mech}$ of
Eq.~\eqref{eq:CK_pmech}. Using $\Delta^2 p_{\rm mech}=e^{-2\lambda t}\,\GT{0}{2}$
and $\Delta(\phi_1 p_{\rm mech})=e^{-\lambda t}\,\GT{1}{1}$, one finds
\begin{equation}
  \Uphys(t) \equiv \GT{2}{0}\,\Delta^2 p_{\rm mech}
             - \bigl(\Delta(\phi_1\, p_{\rm mech})\bigr)^2
           = e^{-2\lambda t}\,\Ucan(t)
           = \frac{\hbar^2}{4}\,e^{-2\lambda t}
             \xrightarrow{t\to\infty} 0 .
  \label{eq:CK_violation}
\end{equation}
Thus $\Uphys(t) < \hbar^2/4$ for all $t>0$, in violation of the Heisenberg
uncertainty principle in the physical variables~\cite{DEKKER19811,UM200263}.
As already noted by Caldirola and Kanai~\cite{CK_Kanai}, this fundamental
shortcoming motivates the Bateman approach developed in the next subsection.

\emph{CK coupled to a second oscillator.}

A first inquiry into dissipation in the cavity--JJ problem can be carried out on
the cavity--cavity system. Starting from the second-order Hamiltonian
Eq.~\eqref{eq:H_cav-cav_classical}, which can be read as two harmonic
oscillators plus their interaction, we replace the conservative dynamics of one
oscillator by the CK Hamiltonian Eq.~\eqref{eq:CK_ham}. Identifying the
interaction term from the limit $\lambda\to0$, the full Hamiltonian is
\begin{align}
  H_T &= H_{\rm CK}(\phi_1,Q_1)
        + \frac{Q_2^2}{2\CJtwo} + \frac{\phi_2^2}{2L_2}
        + \frac{C_0}{C_T}\,e^{-\lambda t}\,Q_1\,Q_2,
  \label{eq:CK_coupled}
\end{align}
where the factor $e^{-\lambda t}$ multiplying the coupling follows from using
$\CJ\dot\phi_1 = Q_1\,e^{-\lambda t}$ to express the interaction in terms of the
canonical charge; it is a direct consequence of the CK time-dependent
rescaling rather than an independent modelling assumption. The resulting
classical EOM and moment equations are collected in
Appendix~\ref{app:CK_coupled}. As anticipated, this straightforward
quantization of the CK model within the momentous formalism inherits the
Heisenberg-violation problem identified above: the physical uncertainty product
of the dissipating mode still decays to zero. This is precisely the deficiency
that the Bateman construction is designed to cure.

\subsubsection{Bateman dissipation}
\label{sec:bateman}

To overcome the Heisenberg-violation problem of the CK model while retaining a
canonical Hamiltonian structure, one introduces an auxiliary degree of freedom
that absorbs the dissipated energy. This is the strategy of the Bateman
dual-oscillator Hamiltonian~\cite{DEKKER19811}, studied within the effective
formalism in~\cite{JavierValdez_2025}. Proceeding as for the CK case, we now
replace one of the two oscillators by the Bateman Hamiltonian,
\begin{equation}
\begin{split}
    H_{\text{Bat}} &= \frac{p_{x}p_{y}}{C_{J_1}}+\lambda\left(yp_{y}-xp_{x}\right)+C_{J_1}\Omega^{2}xy\\
    &= \left(\frac{p_{1}^{2}}{2C_{J_1}}+\frac{1}{2}C_{J_1}\Omega^{2}x_{1}^{2}\right)-\left(\frac{p_{2}^{2}}{2C_{J_1}}+\frac{1}{2}C_{J_1}\Omega^{2}x_{2}^{2}\right)-\lambda\left(x_{1}p_{2}+x_{2}p_{1}\right),
\end{split}
\label{eq:bateman_full}
\end{equation}
where the normal coordinates are defined by
\begin{align}
\label{eq:Bateman_transformation}
        x&=\tfrac{1}{\sqrt{2}}(x_{1}+x_{2}), & y&=\tfrac{1}{\sqrt{2}}(x_{1}-x_{2}), & p_{x}&=\tfrac{1}{\sqrt{2}}(p_{1}+p_{2}), & p_{y}&=\tfrac{1}{\sqrt{2}}(p_{1}-p_{2}),
\end{align}
and $\Omega^{2}=\dfrac{1}{L_{1}C_{J_1}}-\lambda^{2}=\omega_1^2-\lambda^2$. Here
$x$ is the physical (damped) coordinate and $y$ the time-reversed mirror
coordinate that stores the dissipated energy. A characteristic feature of this
Hamiltonian is the mismatch between the canonical and the usual mechanical
momentum,
\begin{equation}
    p_{y}=C_{J_1}(\dot{x}+\lambda x),\qquad p_{x}=C_{J_1}(\dot{y}-\lambda y),
\label{eq:Original_algebra}
\end{equation}
which yields a more involved interaction Hamiltonian. Before imposing the
quantum structure described below, the mechanical momentum of the physical mode
is $p_y$, so we write
\begin{equation*}
    H_{I} = \frac{C_{0}}{C_{T}}p_{y}Q_{2}
          = \frac{C_{0}}{\sqrt{2}\,C_{T}}(p_{1}-p_{2})Q_{2},
\end{equation*}
because in the limit $\lambda\to0$ this reduces to
\begin{equation}
\label{eq:expected_int_Hamiltonian}
    H_{I}= \frac{C_{0}}{C_{T}}\,(C_{J_1}\dot{x})\,Q_{2},
\end{equation}
in agreement with the interaction term of Eq.~\eqref{eq:H_cav-cav_classical}. 

A consistent quantum description requires more care. As shown
in~\cite{JavierValdez_2025}, in order to obtain a quantum dynamics that
preserves the Heisenberg uncertainty\footnote{In the sense described in
Sec.~\ref{section:formalism_momentous}.} the quantization of the Bateman
Hamiltonian must implement the switched commutation relations
\begin{equation}
\begin{split}
    \{x_{1},p_{1}\}=\{x_{2},p_{2}\}=1 &\;\rightarrow\; [\hat{x}_{1},\hat{p}_{1}]=[\hat{p}_{2},\hat{x}_{2}]=i\hbar,\\
    \{x_{2},p_{1}\}=\{x_{1},p_{2}\}=0 &\;\rightarrow\; [\hat{x}_{2},\hat{p}_{1}]=[\hat{x}_{1},\hat{p}_{2}]=0,
\end{split}
 \label{eq:switch_commutation_Bateman}
\end{equation}
under which the momentum relations become
\begin{equation*}
    p_{x}=C_{J_1}(\dot{x}-\lambda y),\qquad p_{y}=C_{J_1}(\dot{y}+\lambda x),
\end{equation*}
so that in the limit $\lambda\to0$ the canonical and mechanical momenta
coincide. With this prescription, and using
Eq.~\eqref{eq:expected_int_Hamiltonian}, the physical quantum mechanical momentum of
mode~1 is now $p_x=(p_1+p_2)/\sqrt2$, and the quantum interacting dissipative
Hamiltonian reads
\begin{equation}
\begin{split}
    \hat{H}_{\text{Bate}_1}+\hat{H}_{\text{Osc}_2}+\hat{H}_{I}
    &= \left(\frac{\hat{p}_{1}^{2}}{2C_{J_1}}+\frac{1}{2}C_{J_1}\Omega^{2}\hat{x}_{1}^{2}\right)-\left(\frac{\hat{p}_{2}^{2}}{2C_{J_1}}+\frac{1}{2}C_{J_1}\Omega^{2}\hat{x}_{2}^{2}\right)\\
    &\quad-\lambda\left(\hat{x}_{1}\hat{p}_{2}+\hat{x}_{2}\hat{p}_{1}\right)
    +\frac{1}{2C_{J_2}}\hat{Q}_{2}^{2}+\frac{1}{2L_{2}}\hat{\phi}_{2}^{2}
    +\frac{C_{0}}{\sqrt{2}\,C_{T}}(\hat{p}_{1}+\hat{p}_{2})\hat{Q}_{2}.
\end{split}
\label{eq:Bateman_full}
\end{equation}
The use of the $(\hat{p}_1+\hat{p}_2)$ combination here (rather than $\hat{p}_1-\hat{p}_2$) is dictated
by the switched algebra Eq.~\eqref{eq:switch_commutation_Bateman}: it is
$p_x=(p_1+p_2)/\sqrt2$ that becomes the mechanical momentum of the physical mode,
and it is this same combination that enters the quantum Hamiltonian below and
the drift matrix Eq.~\eqref{eq:Bateman_M_matrix}.

The classical dynamics generated by Eqs.(\ref{eq:bateman_full},\ref{eq:expected_int_Hamiltonian}) is
\begin{align}
\label{eq:class_damp_cav_cav}
    \dot{x}_{1}&= \frac{1}{C_{J_1}}p_{1}-\lambda x_{2}+k Q_{2}, &
    \dot{x}_{2}&= -\frac{1}{C_{J_1}}p_{2}-\lambda x_{1}+k Q_{2}, &
    \dot{\phi}_{2}&= \frac{1}{C_{J_2}}Q_{2}+k(p_{1}-p_{2}), \nonumber\\
    \dot{p}_{1}&= -C_{J_1}\Omega^{2}x_{1}+\lambda p_{2}, &
    \dot{p}_{2}&= C_{J_1}\Omega^{2}x_{2}+\lambda p_{1}, &
    \dot{Q}_{2}&= -\frac{1}{L_{2}}\phi_{2},
\end{align}
with $k=C_{0}/(\sqrt{2}\,C_{T})$.The classical evolution of the coupled
system is shown in Fig.~\ref{subfig:classical_cav_cav_dis}: the magnetic-flux
amplitude $\phi(t)$ of the physical mode experiences an overall dissipation,
while energy is transferred coherently to the second cavity.

The total quantum dynamics is richer. Following the method of
Sec.~\ref{section:formalism_momentous}, the effective Hamiltonian,
Eq.~\eqref{eq:H_q}, generated by the quantum (switched-algebra), and the classical (Eqs.(\ref{eq:bateman_full},\ref{eq:expected_int_Hamiltonian})) Bateman
Hamiltonian reads
\begin{equation}
\label{eq:SBTH}
\begin{split}
    \langle\hat{H}\rangle &= H_{\textit{Bate}_1}+H_{osc_2}+k(p_{1}-p_{2})Q_{2}+\frac{1}{2C_{J_1}}G^{0,2,0,0,0,0}_{1}+\frac{C_{J_1}\Omega^{2}}{2}G^{2,0,0,0,0,0}_{1}\\
    &\quad-\frac{1}{2C_{J_1}}G^{0,0,2,0,0,0}_{1}+\frac{C_{J_1}\Omega^{2}}{2}G^{0,0,0,2,0,0}_{1}-\lambda\left(G^{1,0,1,0,0,0}_{1}+G^{0,1,0,1,0,0}_{1}\right)\\
    &\quad+\frac{1}{2C_{J2}}G^{0,0,0,0,0,2}_{1}
    +\frac{1}{2L_{2}}G^{0,0,0,0,2,0}_{1}+k\left(G^{0,1,0,0,0,1}_{1}+G^{0,0,1,0,0,1}_{1}\right),
\end{split}
\end{equation}
where the quantum variables of the doubled phase space are
\begin{equation*}
G_{1}^{a,b,c,d,e,f}=\big\langle(\hat{x}_{1}-\langle\hat{x}_{1}\rangle)^{a}(\hat{p}_{1}-\langle\hat{p}_{1}\rangle)^{b}(\hat{p}_{2}-\langle\hat{p}_{2}\rangle)^{c}(\hat{x}_{2}-\langle\hat{x}_{2}\rangle)^{d}(\hat{\phi}_{2}-\langle\hat{\phi}_{2}\rangle)^{e}(\hat{Q}_{2}-\langle\hat{Q}_{2}\rangle)^{f}\big\rangle_{\text{Weyl}}.
\label{eq:quant_damp_cav_cav}
\end{equation*}
This generates equations of motion for the $21$ second-order moments of the
doubled space. We are ultimately interested in the untransformed
(physical) moments, which carry the physical interpretation and allow a direct
comparison with the CK and Lindblad descriptions. The physical second-order
moments of the first oscillator are recovered as
\begin{equation}
\begin{split}
    G^{2,0,0,0} &= \tfrac{1}{2}\left[G^{2,0,0,0,0,0}_{1}+G^{0,0,0,2,0,0}_{1}+2G_{1}^{1,0,0,1,0,0}\right],\quad
    G^{0,2,0,0} = \tfrac{1}{2}\left[G^{0,2,0,0,0,0}_{1}+G^{0,0,2,0,0,0}_{1}+2G^{0,1,1,0,0,0}_{1}\right],\\
    G^{1,1,0,0} &= \tfrac{1}{2}\left[G^{1,1,0,0,0,0}_{1}+G^{0,0,1,1,0,0}_{1}+G^{0,1,0,1,0,0}_{1}+G^{1,0,1,0,0,0}_{1}\right],
\end{split}
\label{eq:Bateman_phys_reduction}
\end{equation}
which are nothing but the variances of $\phi_1=(x_1+x_2)/\sqrt2$ and
$Q_1=(p_1+p_2)/\sqrt2$ and their covariance. The complete system of physical
EOM consists of $10$ coupled equations that can be written in the master form
Eq.~\eqref{eq:moment_master}, $\dot{\mathbf{G}}=\mathbf{M}_B\mathbf{G}+\mathbf{d}_B$,
with
\begin{equation}
    \mathbf{G}^{T} = (G^{2,0,0,0},G^{0,2,0,0},G^{0,0,2,0}, G^{0,0,0,2}, G^{1,1,0,0}, G^{1,0,1,0}, G^{1,0,0,1}, G^{0,1,0,1}, G^{0,1,1,0},G^{0,0,1,1}).
\end{equation}
In the Bateman approach the inhomogeneous term $\mathbf{d}_{B}$ contains the
auxiliary-oscillator (mirror) moments $G^{a,b,c,d,e,f}_{1}$,
\begin{align}
    d_{1} &= 2\lambda[G_{1}^{2,0,0,0,0,0}+G_{1}^{1,0,0,1,0,0}] & d_{6} &= \frac{2\lambda}{\sqrt{2}}G^{1,0,0,0,1,0}_{1} & d_{9} &=\frac{2\lambda}{\sqrt{2}}G^{0,0,1,0,1,0}_{1}\nonumber\\
    d_{2} &=2\lambda[G_{1}^{0,2,0,0,0,0}+G_{1}^{0,1,1,0,0,0}] & d_{7} &=\frac{2\lambda}{\sqrt{2}}G^{1,0,0,0,0,1}_{1} & d_{8} &=\frac{2\lambda}{\sqrt{2}}G^{0,0,1,0,0,1}_{1}\nonumber\\
    d_{5} &= \lambda[2G_{1}^{1,0,1,0,0,0}+G_{1}^{1,1,0,0,0,0}+G_{1}^{0,0,1,1,0,0}]
\end{align}
all remaining components being zero, and the drift matrix $\mathbf{M}_B$ is
\begin{equation}
\label{eq:Bateman_M_matrix}
\mathbf{M}_B=
\begin{pmatrix}
-2\lambda & 0 & 0 & 0 & 2/C_{J_1} & 0 & 2C_{0}/C_{T} & 0 & 0 & 0\\
0 & -2\lambda & 0 & 0 & -2C_{J_1}\Omega^{2} & 0 & 0 & 0 & 0 & 0\\
0 & 0 & 0 & 0 & 0 & 0 & 0 & 0 & 2C_{0}/C_{T} & 2/C_{J_2}\\
0 & 0 & 0 & 0 & 0 & 0 & 0 & 0 & 0 & -2/L_{2}\\
-C_{J_1}\Omega^{2} & 1/C_{J_1} & 0 & 0 & -2\lambda & 0 & 0 & C_{0}/C_{T} & 0 & 0\\
0 & 0 & 0 & 0 & C_{0}/C_{T} & -\lambda & 1/C_{J_2} & 0 & 1/C_{J_1} & C_{0}/C_{T}\\
0 & 0 & 0 & C_{0}/C_{T} & 0 & -1/L_{2} & -\lambda & 1/C_{J_1} & 0 & 0\\
0 & 0 & 0 & 0 & 0 & -C_{J_1}\Omega^{2} & -\lambda & -1/L_{2} & 0 & 0\\
0 & C_{0}/C_{T} & 0 & 0 & 0 & -C_{J_1}\Omega^{2} & 0 & 1/C_{J_2} & -\lambda & 0\\
0 & 0 & -1/L_{2} & 1/C_{J_2} & 0 & 0 & 0 & 0 & C_{0}/C_{T} & 0
\end{pmatrix}.
\end{equation}
%

A crucial consequence of the switched algebra
Eq.~\eqref{eq:switch_commutation_Bateman} is that the auxiliary drive partly
cancels the damping in the physical moments. For instance, the $\phi_1$
variance obeys
$\dot{G}^{2,0,0,0}=-2\lambda G^{2,0,0,0}+(2/C_{J_1})G^{1,1,0,0}
+(2C_{0}/C_{T})G^{1,0,0,1}+d_{1}$, and using
$d_{1}=2\lambda[\,\mathrm{Var}(x_1)+\mathrm{Cov}(x_1,x_2)\,]$ together with
Eq.~\eqref{eq:Bateman_phys_reduction} one finds that the damping term and the
drive combine into $\lambda\,[\mathrm{Var}(x_1)-\mathrm{Var}(x_2)]$. The net
moment dynamics is therefore undamped and oscillatory: the doubled
quantum system has purely imaginary characteristic frequencies, so the physical
uncertainty product remains bounded and oscillates around the vacuum value
rather than decaying. This is the precise mechanism by which the Bateman model
repairs the Heisenberg violation of the CK model. A more detailed explanation is shown in section \ref{sec:discussion_dissipation}. The corresponding moment
evolution is shown in Fig.~\ref{subfig:moments_cav_cav_dis}.

It should be stressed that this faithful behavior holds for the
quantum moments; the classical mean trajectories obey
Eq.~\eqref{eq:class_damp_cav_cav}, in which the mirror coordinate $y(t)$ grows
as the physical coordinate $x(t)$ decays (so as to keep the doubled Hamiltonian
conservative). This sets a finite validity window for the numerical integration
of the mean dynamics, which for the parameters employed we found it to be of order $t\lesssim 700/\gamma$, beyond which the growth
of $y(t)$ contaminates the solution. The quantitative analysis of this window
and of the agreement with Lindblad is given in
Appendix~\ref{sec:quantitative_comparison}.

\subsubsection{Lindblad master equation}
\label{sec:lindblad}

The third and most fundamental approach to dissipation considered here is the
Lindblad (GKSL) master equation~\cite{Breuer2002,Lindblad}, which provides the
most general completely positive and trace-preserving (CPTP) Markovian evolution
of a quantum state. Unlike the Caldirola--Kanai and Bateman approaches, in the
Lindblad framework the effect of the environment is encoded in a set of jump
operators acting directly on the system's density matrix, with no need for an
explicit bath Hamiltonian or auxiliary degree of freedom.

\emph{Physical model.}

We take the system Hamiltonian of Eq.~\eqref{eq:cav-jj},
\begin{equation}
  \hat{H}_{\rm sys} = \hbar\omega'_{1}\,\hat{a}^{\dagger}\hat{a}
    + \hbar\omega'_{2}\,\hat{b}^{\dagger}\hat{b}
    - \frac{C_0}{C_T}Q_{{\rm zpf},1}Q_{{\rm zpf},2}
      \bigl(\hat{a}-\hat{a}^{\dagger}\bigr)
      \bigl(\hat{b}-\hat{b}^{\dagger}\bigr),
  \label{eq:H_sys_Lindblad}
\end{equation}
in which oscillator~1 (the cavity, mode $\hat a$) is embedded in a
zero-temperature Markovian reservoir\footnote{In cavity QED this dissipation
arises from the coupling of the resonator to a thermal reservoir~\cite{carmichael2008statistical}; in superconducting circuits it originates from the coupling to a transmission line~\cite{RevModPhys.93.025005}.}.
The zero-temperature approximation is standard for superconducting microwave
circuits, since at operating temperatures $T\sim10$--$50\,\mathrm{mK}$ the mean
thermal photon number $\bar{n}_{\rm th}=(e^{\hbar\omega/k_BT}-1)^{-1}\lesssim
10^{-3}$ is negligible~\cite{RevModPhys.93.025005}. Under the Born--Markov
approximation~\cite{Breuer2002}, the reduced density matrix evolves according to
the GKSL equation
\begin{equation}
\label{eq:Lindblad_cav-cav}
  \dot{\rho}
    = -\frac{i}{\hbar}\bigl[\hat{H}_{\rm sys},\,\rho\bigr]
      + \frac{\gamma}{2}
        \bigl(2\hat{a}\,\rho\,\hat{a}^{\dagger}
              - \hat{a}^{\dagger}\hat{a}\,\rho
              - \rho\,\hat{a}^{\dagger}\hat{a}\bigr).
\end{equation}
The first term is the von Neumann (coherent) part; the second is the Lindblad
dissipator $\mathcal{D}[\hat{a}]\rho$, with jump operator $\hat{a}$ and photon
loss rate $\gamma$, describing the incoherent dynamics induced by the bath. The
Josephson junction (oscillator~2, mode $\hat b$) experiences only an indirect
damping in this model: no jump operator acts on $\hat b$, so its coherent
dynamics is affected solely through the coupling with oscillator~1.

Expanding the commutator $[\hat{H}_{\rm sys},\rho]$ using
Eq.~\eqref{eq:H_sys_Lindblad} gives
\begin{equation}
\label{eq:Lindblad_cav-cav_expanded}
  \dot{\rho}
    = -i\omega'_{1}[\hat{a}^{\dagger}\hat{a},\rho]
      - i\omega'_{2}[\hat{b}^{\dagger}\hat{b},\rho]
      - ig\bigl[\hat{a}^{\dagger}\hat{b}
                +\hat{a}\hat{b}^{\dagger}
                -\hat{a}\hat{b}
                -\hat{a}^{\dagger}\hat{b}^{\dagger},\,\rho\bigr]
      + \frac{\gamma}{2}
        \bigl(2\hat{a}\rho\hat{a}^{\dagger}
              -\hat{a}^{\dagger}\hat{a}\rho
              -\rho\hat{a}^{\dagger}\hat{a}\bigr),
\end{equation}
where $g = C_0 Q_{{\rm zpf},1}Q_{{\rm zpf},2}/(C_T\hbar)$ is the coupling
constant and we set $\hbar=1$ for compactness. The interaction term expands as
$(\hat{a}-\hat{a}^{\dagger})(\hat{b}-\hat{b}^{\dagger}) =
-(\hat{a}^{\dagger}\hat{b}+\hat{a}\hat{b}^{\dagger}
  -\hat{a}\hat{b}-\hat{a}^{\dagger}\hat{b}^{\dagger})$,
retaining all four terms (no rotating-wave approximation is made).

\emph{First-order expectation values.}

The mean dynamics follows from $d\langle\hat{O}\rangle/dt =
\mathrm{Tr}[\hat{O}\,\dot{\rho}]$, equivalently from the adjoint
form~\cite{Mathematical_methods_optics}
\begin{equation}
    \frac{d\langle\hat{O}\rangle}{dt}
    = -\frac{i}{\hbar}\langle[\hat{O},\hat{H}_{\rm sys}]\rangle
      +\frac{\gamma}{2}\big\langle\hat{a}^{\dagger}[\hat{O},\hat{a}]
       +[\hat{a}^{\dagger},\hat{O}]\,\hat{a}\,\big\rangle ,
  \label{eq:adjoint_lindblad}
\end{equation}
where the explicit $\gamma/2$ prefactor on the dissipator contribution is
retained. Applied to Eq.~\eqref{eq:Lindblad_cav-cav_expanded} this yields
\begin{align}
\label{eq:eom_lindblad}
  \frac{d\langle\hat{\phi}_{1}\rangle}{dt}
    &= -\frac{\gamma}{2}\langle\hat{\phi}_{1}\rangle
       +\frac{1}{C_{J_{1}}}\langle\hat{Q}_{1}\rangle
       +\frac{C_{0}}{C_{T}}\langle\hat{Q}_{2}\rangle,
  &
  \frac{d\langle\hat{\phi}_{2}\rangle}{dt}
    &= \frac{1}{C_{J_{2}}}\langle\hat{Q}_{2}\rangle
       +\frac{C_{0}}{C_{T}}\langle\hat{Q}_{1}\rangle,
  \nonumber\\
  \frac{d\langle\hat{Q}_{1}\rangle}{dt}
    &= -\frac{\gamma}{2}\langle\hat{Q}_{1}\rangle
       -C_{J_{1}}\omega'^{2}_{ 1}\langle\hat{\phi}_{1}\rangle,
  &
  \frac{d\langle\hat{Q}_{2}\rangle}{dt}
    &= -C_{J_{2}}\omega'^{2}_{2}\langle\hat{\phi}_{2}\rangle.
\end{align}
The damping term $-(\gamma/2)\langle\hat{O}\rangle$ appears only for the
observables of mode $\hat a$ ($\hat\phi_1,\hat Q_1$), confirming that the
reservoir couples exclusively to the cavity; it is absent from the equations
for $\langle\hat\phi_2\rangle$ and $\langle\hat Q_2\rangle$. Dissipation
nonetheless reaches oscillator~2 indirectly, through the coupling terms
$\tfrac{C_0}{C_T}\langle\hat Q_2\rangle$ and
$\tfrac{C_0}{C_T}\langle\hat Q_1\rangle$. Remarkably, this set of first-order
EOM is identical to the Bateman classical equations
Eq.~\eqref{eq:class_damp_cav_cav} under the identification $\gamma=2\lambda$, $\Omega = \omega'_{1}$ and $1/L_{2} = C_{J_2}\omega'^{2}_{2}$,
so the two frameworks agree at the level of mean-field dynamics.

\emph{Second-order moments.}

Given the mean-field agreement, it is natural to ask whether the two approaches
also share the same quantum (moment) dynamics. The information beyond the mean
trajectories is carried by the moments
$G_L^{a,b,c,d} = \langle(\hat\phi_1-\langle\hat\phi_1\rangle)^a
(\hat Q_1-\langle\hat Q_1\rangle)^b
(\hat\phi_2-\langle\hat\phi_2\rangle)^c
(\hat Q_2-\langle\hat Q_2\rangle)^d\rangle_{\rm Weyl}$.
As a worked example we derive the EOM for
$G_L^{2,0,0,0}\equiv\langle\hat\phi_1^2\rangle-\langle\hat\phi_1\rangle^2$:
\begin{equation}
\begin{split}
\label{eq:Lindblad_moment_deriv}
  \dot{G}_L^{2,0,0,0}
      &= \overbrace{-\gamma \langle \hat{\phi}_{1}^{2}\rangle +\frac{1}{C_{J_1}}\langle\hat{\phi}_{1}\hat{Q}_{1}+\hat{Q}_{1}\hat{\phi}_{1}\rangle+\frac{2 C_{0}}{C_{T}}\langle\hat{\phi}_{1}\hat{Q}_{2}\rangle+\frac{\gamma\hbar}{2 C_{J_1}\omega'_{1}}}^{d\langle\hat{\phi}_{1}^{2}\rangle/dt}\quad-2\langle\hat{\phi}_{1}\rangle\overbrace{\left[-\frac{\gamma}{2}\langle\hat{\phi}_{1}\rangle+\frac{1}{C_{J_1}}\langle\hat{Q}_{1}\rangle+\frac{C_{0}}{C_{T}}\langle\hat{Q}_{2}\rangle\right]}^{d\langle\hat{\phi}_{1}\rangle/dt}\\
      &= -\gamma\bigl(\langle\hat{\phi}_{1}^{2}\rangle-\langle\hat{\phi}_{1}\rangle^{2}\bigr)
        +\frac{2}{C_{J_1}}\left(\frac{\langle\hat{\phi}_{1}\hat{Q}_{1}+\hat{Q}_{1}\hat{\phi}_{1}\rangle}{2}-\langle\hat{\phi}_{1}\rangle\langle\hat{Q}_{1}\rangle\right)
        +\frac{2C_{0}}{C_{T}}\bigl(\langle\hat{\phi}_{1}\hat{Q}_{2}\rangle-\langle\hat{\phi}_{1}\rangle\langle\hat{Q}_{2}\rangle\bigr)+\frac{\gamma\hbar}{2 C_{J_1}\omega'_{1}}\\
      &=-\gamma\, G^{2,0,0,0}_{L}+\frac{2}{C_{J_1}}G^{1,1,0,0}_{L}+\frac{2C_{0}}{C_{T}}G^{1,0,0,1}_{L}+\frac{\gamma\hbar}{2C_{J_1}\omega'_{1}},
\end{split}
\end{equation}
where we identified the vacuum-noise coefficient $\gamma\,\phi_{{\rm zpf},1}^2 =
\gamma\hbar/(2C_{J_1}\omega'_1)$.  The remaining moment equations follow
analogously.

In matrix form, Eq.~\eqref{eq:moment_master} becomes
$\dot{\mathbf{G}}=\mathbf{M}_{L}\mathbf{G}+\mathbf{d}_{L}$. The drift matrix
$\mathbf{M}_{L}$ coincides with the Bateman matrix
Eq.~\eqref{eq:Bateman_M_matrix}, if we make  $\gamma=2\lambda$, $\Omega = \omega'_{1}$ and $1/L_{2} = C_{J_2}\omega'^{2}_{2}$, while the inhomogeneous term is now a
constant vacuum floor,
\begin{equation}
    \mathbf{d}_{L}^{T} = \left(\frac{\gamma\hbar}{2C_{J_1}\omega'_{1}},\; \frac{\gamma\hbar\, C_{J_1}\omega'_{1}}{2},\,0,0,0,0,0,0,0,0\right).
\end{equation}
The undamped-JJ moment equations ($G_L^{0,0,2,0}$, $G_L^{0,0,0,2}$,
$G_L^{0,0,1,1}$) and the cross-moment equations ($G_L^{1,0,0,1}$,
$G_L^{0,1,1,0}$, $G_L^{1,0,1,0}$, $G_L^{0,1,0,1}$) have the same structure as
the corresponding conservative equations in Appendix~\ref{appendix: II}, but
with an additional $-(\gamma/2)$ prefactor on every moment that involves a
mode-$\hat a$ index.
\\

\emph{Physical interpretation of the noise terms.}

The nonzero entries of $\mathbf{d}_{L}$ are the diffusion coefficients
\cite{dekker_fundamental_1984,SANDULESCU1987277}, representing the quantum
vacuum noise injected into the cavity by the zero-temperature reservoir.
Physically, even as the cavity decays at rate $\gamma$ (driving the mean
amplitude to zero), the zero-point fluctuations continuously repopulate the
variance; the balance between decay and repopulation fixes the steady state.
Setting $\dot{\mathbf{G}}=0$ in the uncoupled limit
($C_0=0$)\footnote{In this limit the two oscillators decouple and their moment
equations can be solved independently.} gives
\begin{equation}
  G_L^{2,0,0,0}\big|_\infty = \phi_{{\rm zpf},1}^2 = \frac{\hbar}{2C_{J_1}\omega'_1},
  \qquad
  G_L^{0,2,0,0}\big|_\infty = Q_{{\rm zpf},1}^2 = \frac{\hbar C_{J_1}\omega'_1}{2},
  \qquad
  G_L^{1,1,0,0}\big|_\infty = 0,
  \label{eq:Lindblad_ss_local}
\end{equation}
so that the steady-state uncertainty product saturates the Heisenberg bound,
$G_L^{2,0,0,0}|_\infty\,G_L^{0,2,0,0}|_\infty =
\phi_{{\rm zpf},1}^2\,Q_{{\rm zpf},1}^2 = \hbar^2/4$,
corresponding to the quantum vacuum. This is in stark contrast with the
Caldirola--Kanai model, whose physical uncertainty $\Uphys(t)\to0$
(Sec.~\ref{sec:CK}), and with the Bateman model, which has no genuine steady
state but instead oscillates around the Lindblad vacuum value.
\\

\emph{Comparison of the two moment descriptions.}

\begin{figure}[H]
  \centering
  \subfloat[Flux expectation values $\langle\hat\phi_i\rangle$ with their
    one-standard-deviation quantum uncertainty bands
    $\pm\sqrt{G^{2,0,0,0}}$ (cavity~1) and $\pm\sqrt{G^{0,0,2,0}}$
    (cavity~2).]{%
    \includegraphics[width=0.49\textwidth]{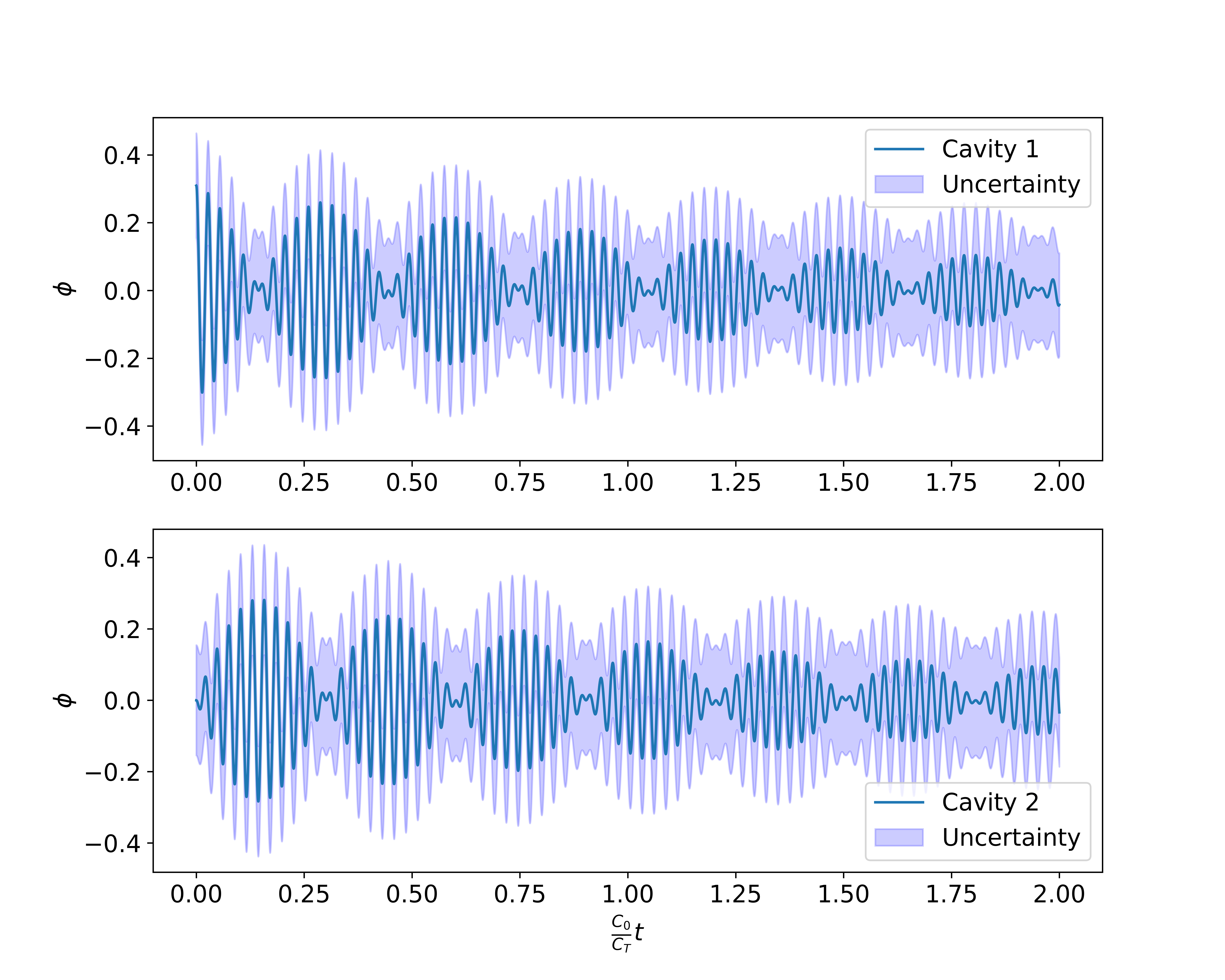}%
    \label{subfig:classical_cav_cav_dis}}
  \hfill
  \subfloat[Second-order moments (cavity~1): uncertainty product
    $G^{2,0,0,0}G^{0,2,0,0}$ (top) and covariance $G^{1,1,0,0}$ (bottom),
    for the Bateman and Lindblad models.]{%
    \includegraphics[width=0.49\textwidth]{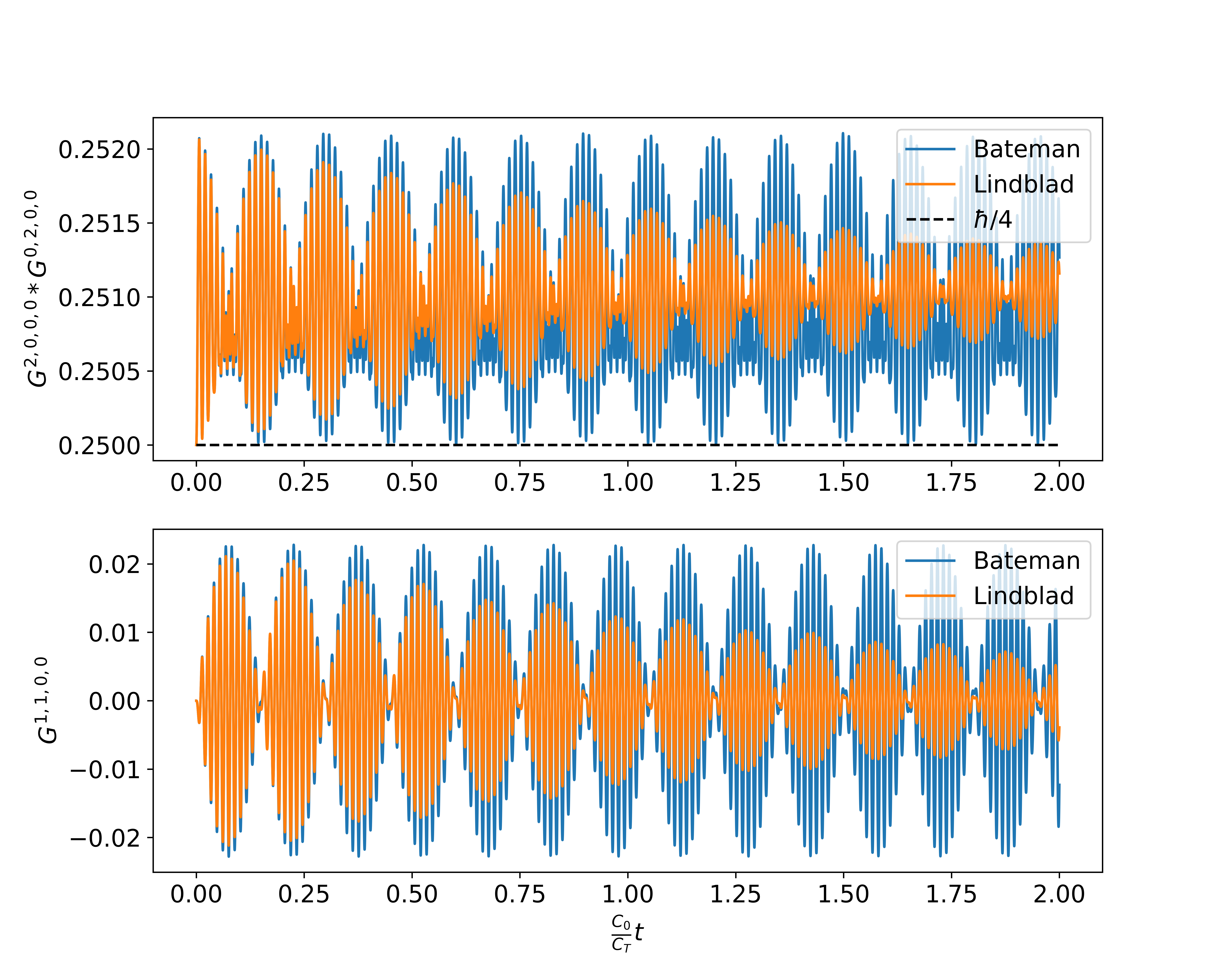}%
    \label{subfig:moments_cav_cav_dis}}
  \caption{Dissipative dynamics of the coupled cavity--cavity system, with
    dissipation acting on cavity~1 (rate $\gamma=2\lambda$) and cavity~2
    damped only indirectly through the capacitive coupling $C_0/C_T$.
    \textbf{(a)}~The flux expectation value of cavity~1 decays as energy is
    coherently transferred to cavity~2, while the quantum uncertainty bands
    remain of order the zero-point width, signaling that the state stays close
    to minimum uncertainty. \textbf{(b)}~The uncertainty product
    $G^{2,0,0,0}G^{0,2,0,0}$ stays at or above the Heisenberg floor
    $\hbar^{2}/4$ (dashed line) for both models: the Lindblad product relaxes
    slightly above the vacuum floor, whereas the Bateman product oscillates around it with
    a larger, sustained envelope; the covariance $G^{1,1,0,0}$ oscillates about
    zero with the same pattern. All quantities are in units $\hbar=1$; time is
    shown in cavity cycles $C_{0}t/C_{T}$. Parameters: $C_{J_1}=C_{J_2}\approx 1.09$,
    $\omega'_1 = \omega'_2\approx 19.15$, $C_0/C_T\approx 0.083$, $\lambda=0.1$.}
  \label{fig:Bateman_Lindblad_cav_cav_dis}
\end{figure}

The Lindblad and Bateman moment systems thus share the common structure
Eq.~\eqref{eq:moment_master} with the same drift matrix $\mathbf{M}$. The sole difference lies in the driving vector:
for Lindblad $\mathbf{d}_{L}$ is a constant noise floor $\propto\gamma\hbar$,
whereas $\mathbf{d}_{B}(t)$ is built from the bounded, oscillatory
auxiliary-oscillator moments. This distinction governs the long-time behaviour:
the Lindblad system relaxes to the vacuum ground state, while the Bateman
description oscillates around it. The agreement of the two frameworks at the
level of mean trajectories (Eq.~\eqref{eq:eom_lindblad}) and their shared
drift matrix $\mathbf{M}$ confirm that the Bateman approach is a valid
semiclassical model of the dissipative dynamics within its validity window.
The two descriptions are compared in
Fig.~\ref{fig:Bateman_Lindblad_cav_cav_dis}; a fully quantitative comparison,
including the error threshold and the validity window, is given in
Appendix~\ref{sec:quantitative_comparison}.

\section{Discussion}
\label{Discussion}

\subsection{Quantitative comparison of the three dissipation models}
\label{sec:discussion_dissipation}

\begin{figure}[t]
  \centering
  \includegraphics[width=\textwidth]{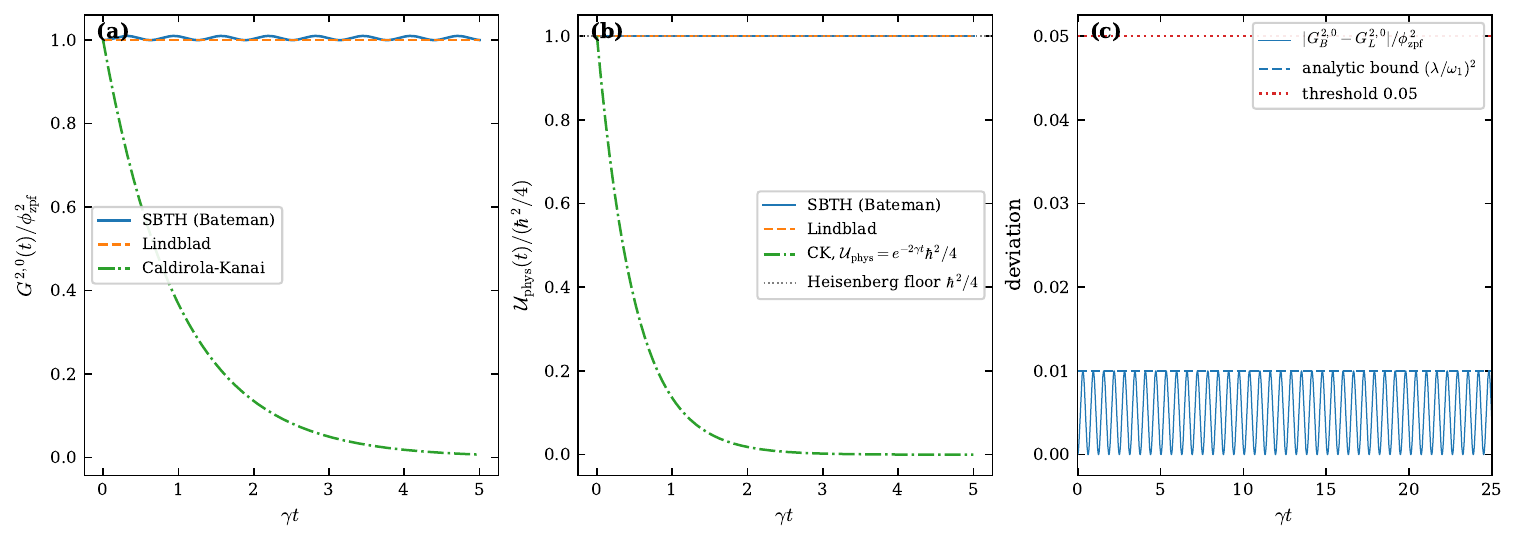}
  \caption{Quantitative comparison of the three dissipation models for
  the single cavity oscillator ($C_0=0$), with $\hbar=1$,
  $\CJ=0.5$, $\omega_1=1$, $\lambda/\omega_1=0.1$ and the
  identifications $\gamma=2\lambda$ (Lindblad) and
  $\lambda_{\rm CK}=\gamma$ (so that all three models share the same
  mean-field decay envelope $e^{-\gamma t/2}$). The initial state is a
  minimum-uncertainty state with vacuum widths.
  \textbf{(a)}~Flux variance $\GT{2}{0}(t)/\phizpf^2$: the Lindblad
  moment remains at the vacuum floor, the Bateman (SBTH) moment
  oscillates around it according to the exact solution
  Eq.~\eqref{eq:SBTH_vacuum_moments} (inset), and the Caldirola--Kanai
  variance decays to zero. \textbf{(b)}~Physical uncertainty product
  $\Uphys(t)/(\hbar^2/4)$: Lindblad and Bateman respect the Heisenberg
  floor (dotted) for all times, whereas the CK product decays as
  $e^{-2\gamma t}$. \textbf{(c)}~Deviation
  $|\GT{2}{0}_{B}-\GT{2}{0}_{L}|/\phizpf^2$ over a longer window: it
  saturates the analytic bound $(\lambda/\omega_1)^2$ (dashed,
  Eq.~\eqref{eq:SBTH_deviation_bound}) and stays well below the $5\%$
  threshold (dotted) used in
  Appendix~\ref{sec:quantitative_comparison}.}
  \label{fig:Bateman_Lindblad}
\end{figure}

In Sec.~\ref{sec:dissipative} we introduced three descriptions of
dissipation in the cavity oscillator: the Caldirola--Kanai (CK)
Hamiltonian Eq.~\eqref{eq:CK_ham}, the Bateman effective dynamics
generated by Eq.~\eqref{eq:SBTH}, and the Lindblad master equation
Eq.~\eqref{eq:Lindblad_cav-cav}. The CK model provides a useful
baseline but violates the Heisenberg bound in the kinetic variables,
as shown in Sec.~\ref{sec:CK}. Here we compare the two
Heisenberg-consistent approaches, Bateman and Lindblad,%
quantitatively, keeping the CK result as a reference; the three models
are confronted side by side in Fig.~\ref{fig:Bateman_Lindblad}, and
the full numerical analysis (error threshold and validity window) is
collected in Appendix~\ref{sec:quantitative_comparison}.\\

\emph{Classical equations of motion.\\}
At the level of the expectation values $\langle\hat\phi_1\rangle$ and
$\langle\hat Q_1\rangle$, the Bateman and Lindblad approaches yield
identical equations of motion (compare
Eqs.~\eqref{eq:class_damp_cav_cav} and~\eqref{eq:eom_lindblad}) under the identification $\gamma=2\lambda$, $\Omega = \omega'_{1}$ and $1/L_{2} = C_{J_2}\omega'^{2}_{2}$: the Lindblad photon-loss rate
equals twice the Bateman parameter. This agreement at the level of
mean values was already established in Sec.~\ref{sec:lindblad}; the
two frameworks differ at the level of second-order moments, as the
matrix analysis below shows.\\

\emph{Matrix form for second-order moments.\\}
For the single cavity oscillator (no coupling,
$C_0=0$)\footnote{Since the two oscillators decouple in this limit, the
full $10\times 10$ system splits into independent blocks and the block
of the dissipating mode can be solved on its own.}, both the Lindblad
and the Bateman moment equations reduce to the form
\begin{equation}
  \dot{\mathbf{G}} = \mathbf{M}\,\mathbf{G} + \mathbf{d},
  \qquad
  \mathbf{G} = \begin{pmatrix} \GT{2}{0} \\ \GT{0}{2} \\ \GT{1}{1} \end{pmatrix},
  \label{eq:matrix_form}
\end{equation}
with one and the same drift matrix
\begin{equation}
  \mathbf{M} = \begin{pmatrix}
    -\gamma & 0 & 2/\CJ \\
    0 & -\gamma & -2\CJ\omega'^{2}_{1} \\
    -\CJ\omega'^{2}_{1} & 1/\CJ & -\gamma
  \end{pmatrix},
  \label{eq:matrix_M}
\end{equation}
obtained from Eq.~\eqref{eq:Bateman_M_matrix} by setting $C_0=0$ and
identifying $\gamma=2\lambda$ and $\omega'_{1} = \Omega$. The two approaches differ only in the
inhomogeneous driving term:
\begin{align}
  \mathbf{d}_{\rm Lindblad}
    &= \begin{pmatrix} \gamma\hbar/(2\CJ\omega'_1) \\
                        \gamma\hbar\CJ\omega'_1/2  \\ 0 \end{pmatrix}
     = \begin{pmatrix} \gamma\,\phi_{\text{zpf},1}^2 \\
                        \gamma\,Q_{\text{zpf},1}^2   \\ 0 \end{pmatrix},
  \label{eq:d_lindblad} \\[4pt]
  \mathbf{d}_{\rm Bateman}(t)
    &= \begin{pmatrix}
        \gamma\bigl[G^{2,0,0,0,0,0}_1 + G^{1,0,0,1,0,0}_1\bigr] \\[2pt]
        \gamma\bigl[G^{0,2,0,0,0,0}_1 + G^{0,1,1,0,0,0}_1\bigr] \\[2pt]
        \tfrac{\gamma}{2}\bigl[2G^{1,0,1,0,0,0}_{1}+G^{1,1,0,0,0,0}_{1}+G^{0,0,1,1,0,0}_{1}\bigr]
       \end{pmatrix}.
  \label{eq:d_bateman}
\end{align}
The key distinction is structural: $\mathbf{d}_{\rm Lindblad}$ is a
constant quantum-noise floor, while $\mathbf{d}_{\rm Bateman}(t)$
is built from the auxiliary-oscillator moments and is therefore
time dependent (bounded and oscillatory, as shown below).\\

\emph{Lindblad steady state.}

Under Lindblad evolution, the stationary condition
$\dot{\mathbf{G}}=0$ gives
$\mathbf{G}_\infty = -\mathbf{M}^{-1}\mathbf{d}_{\rm Lindblad}$, from
which one finds immediately
\begin{equation}
  \GT{2}{0}_\infty = \frac{\hbar}{2\CJ\omega'_1} = \phizpf^2,
  \qquad
  \GT{0}{2}_\infty = \frac{\hbar\CJ\omega'_1}{2} = \Qzpf^2,
  \qquad
  \GT{1}{1}_\infty = 0.
  \label{eq:Lindblad_ss}
\end{equation}
This is readily verified by inserting Eq.~\eqref{eq:Lindblad_ss} into
Eqs.~\eqref{eq:matrix_form}--\eqref{eq:d_lindblad}: the first row
gives $-\gamma\phizpf^2 + (2/\CJ)\cdot 0 + \gamma\phizpf^2 = 0$, and
the third row gives
$-\CJ\omega'^{2}_{1}\,\phizpf^{2} + \Qzpf^{2}/\CJ
 = -\hbar\omega'_{1}/2 + \hbar\omega'_{1}/2 = 0$, with the second row
following analogously. The steady-state uncertainty product
$\GT{2}{0}_\infty\GT{0}{2}_\infty = \phizpf^2\Qzpf^2 = \hbar^2/4$
saturates the Heisenberg bound and corresponds to the quantum vacuum
(zero-temperature ground state): the Lindblad dynamics drives the
oscillator to the ground state at $T=0$.\\

\emph{Bateman moment dynamics and its geometric origin.\\}
For the Bateman approach it was already observed
in~\cite{JavierValdez_2025} that the switched Poisson algebra
Eq.~\eqref{eq:switch_commutation_Bateman} turns the moment sector into
that of two undamped harmonic oscillators. The geometric reason
was clarified in~\cite{BLASONE1996115}, where
 the meaning of the auxiliary system and of the doubled
variables $(x_1,x_2,p_1,p_2)$ was discussed. Consider the oscillator equation
\begin{equation}
    \ddot{r}+\Omega^{2} r=0,
    \qquad
    \Omega^{2} = \omega_1^{2}-\lambda^{2},
    \label{eq:Blasone_osc}
\end{equation}
written in the rate convention of Sec.~\ref{sec:bateman} (damping rate
$\lambda$, frequency $\omega_1$, mass $\CJ$). Depending on the metric
adopted in the doubled $(x_1,x_2)$ plane, Eq.~\eqref{eq:Blasone_osc}
describes two physically different dynamics (Table~\ref{tab:Metric}):
\begin{table}[h]
    \centering
    \begin{tabular}{|c|c|c|}
    \hline
         & Pseudo-Euclidean & Euclidean  \\
    \hline
    Metric & $r^{2}=x^{2}_1-x^{2}_2$ & $r^{2}=x^{2}_1+x^{2}_2$ \\
    \hline
    Equations of motion & $\ddot{x}+2\lambda\dot{x}+\omega_1^{2} x = 0$ & $\ddot{x}+\Omega^{2}x=0$ \\
        & $\ddot{y}-2\lambda\dot{y}+\omega_1^{2} y = 0$ & $\ddot{y}+\Omega^{2}y=0$ \\
    \hline
    \end{tabular}
    \caption{Dynamics of the doubled oscillator with respect to the
    metric of the $(x_1,x_2)$ plane~\cite{BLASONE1996115}, in the rate
    convention $\Omega^2=\omega_1^2-\lambda^2$ of
    Sec.~\ref{sec:bateman}.}
    \label{tab:Metric}
\end{table}
as can be seen by composing the canonical transformation to the
Bateman variables, Eq.~\eqref{eq:Bateman_transformation}, with
\begin{align}
    &\text{Pseudo-Euclidean} & &\text{Euclidean}\nonumber\\
    x_{1} &= r(t)\cosh u(t), & x_{1} &=r(t)\cos\eta,\nonumber\\
    x_{2} &= r(t)\sinh u(t), & x_{2} &=r(t)\sin\eta,
    \label{eq:Blasone_transformation}
\end{align}
where $u(t) = -\lambda t$ and $\eta$ is constant. The pseudo-Euclidean
choice produces the familiar damped/amplified pair, while the
Euclidean choice produces two conservative oscillators. The switched
algebra selects the latter: imposing
$\{x_{1},p_{1}\}=\{p_{2},x_{2}\}=1$ of
Eq.~\eqref{eq:switch_commutation_Bateman} on the Bateman Hamiltonian
yields the fluctuation equations of motion
\begin{equation}
    \begin{split}
        \dot{x}_{1}&=\frac{1}{\CJ}p_{1}-\lambda x_{2},\qquad \dot{p}_{1}=\lambda p_{2}-\CJ\Omega^{2}x_{1},\\
        \dot{x}_{2}&=\frac{1}{\CJ}p_{2}+\lambda x_{1},\qquad \dot{p}_{2}=-\lambda p_{1}-\CJ\Omega^{2}x_{2},
    \end{split}
    \label{eq:switched_EOM}
\end{equation}
whose kinetic equations collapse, using
$\Omega^2+\lambda^2=\omega_1^2$, to
\begin{equation}
    \ddot{x}_{1}+\omega_{1}^{2}x_{1}=0,\qquad
    \ddot{x}_{2}+\omega_{1}^{2}x_{2}=0
    \;\;\Longrightarrow\;\;
    \ddot{x}+\omega_{1}^{2} x = 0, \qquad \ddot{y}+\omega_{1}^{2} y = 0,
    \label{eq:switched_kinetic}
\end{equation}
i.e.\ precisely the Euclidean column of Table~\ref{tab:Metric}. This
makes transparent why the choice of algebra for the purely
quantum variables preserves the uncertainty relation
$\GT{2}{0}\GT{0}{2}-(\GT{1}{1})^{2}\geq\hbar^{2}/4$: the moment sector
inherits the conservative (Euclidean) dynamics, with purely imaginary
characteristic frequencies.\\

\emph{Closed-form moment solutions and deviation bound.\\}
Because Eq.~\eqref{eq:switched_EOM} is linear with spectrum
$\pm i\omega_1$ (doubly degenerate), the second-order moments evolve
as constants plus oscillations at $2\omega_1$---no exponential growth
or decay is present---with coefficients fixed by the initial
conditions (explicit general expressions are given
in~\cite{JavierValdez_2025}). For the physically relevant initial
condition of a minimum-uncertainty state with vacuum widths,
$\GT{2}{0}(0)=\phizpf^2$, $\GT{0}{2}(0)=\Qzpf^2$, $\GT{1}{1}(0)=0$,
the projection Eq.~\eqref{eq:Bateman_phys_reduction} onto the physical
moments admits the exact closed forms
\begin{equation}
\begin{split}
    \GT{2}{0}(t) &= \phizpf^{2}\left[1+\frac{\lambda^{2}}{\omega_{1}^{2}}\,\sin^{2}(\omega_{1}t)\right],\\
    \GT{0}{2}(t) &= \Qzpf^{2}\left[1-\frac{\lambda^{2}}{\omega_{1}^{2}}\left(1-\frac{\lambda^{2}}{\omega_{1}^{2}}\right)\sin^{2}(\omega_{1}t)\right],\\
    \GT{1}{1}(t) &= -\frac{\hbar\,\lambda^{2}}{4\,\omega_{1}^{2}}\,\sin(2\omega_{1}t),
\end{split}
\label{eq:SBTH_vacuum_moments}
\end{equation}
from which the physical uncertainty product follows exactly,
\begin{equation}
    \Uphys(t)
    = \GT{2}{0}\GT{0}{2}-\bigl(\GT{1}{1}\bigr)^{2}
    = \frac{\hbar^{2}}{4}\left[1+\frac{\lambda^{6}}{\omega_{1}^{6}}\,\sin^{4}(\omega_{1}t)\right]
    \;\geq\; \frac{\hbar^{2}}{4}
    \qquad \forall\, t,
\label{eq:SBTH_uncertainty}
\end{equation}
so the Heisenberg floor is respected at all times, with excursions
above it only at order $(\lambda/\omega_1)^6$. Since the Lindblad
moments in the limit $t\rightarrow \infty$ approach the vacuum values Eq.~\eqref{eq:Lindblad_ss}
for this initial state, Eq.~\eqref{eq:SBTH_vacuum_moments} also yields
an analytic bound on the Bateman--Lindblad discrepancy,
\begin{equation}
    \max_{t}\;\bigl|\GT{2}{0}_{B}(t)-\GT{2}{0}_{L}(t)\bigr|
    = \left(\frac{\lambda}{\omega_{1}}\right)^{2}\phizpf^{2}
    = \left(\frac{\gamma}{2\,\omega_{1}}\right)^{2}\phizpf^{2}.
\label{eq:SBTH_deviation_bound}
\end{equation}
For example, for the weak damping used in the example shown in Fig. \ref{fig:Bateman_Lindblad},
$\lambda/\omega_1 = 0.1$, the bound is a $1\%$ effect, comfortably
below the $5\%$ acceptance threshold adopted in
Appendix~\ref{sec:quantitative_comparison}, where the agreement is
verified numerically out to $t\approx 700/\gamma$ and extended to the
coupled cavity--cavity system.

Figure~\ref{fig:Bateman_Lindblad} summarises the comparison of the
three models. Panel~(a) shows the flux variance: the Lindblad moment limit $t\rightarrow \infty)$
sits at the vacuum floor, the Bateman moment oscillates around it with
the amplitude $(\lambda/\omega_1)^2\phizpf^2$ of
Eq.~\eqref{eq:SBTH_vacuum_moments}, and the CK variance decays to
zero---the time-domain signature of the missing vacuum noise floor.
Panel~(b) makes the same point at the level of the physical
uncertainty product: Lindblad preserves it exactly, Bateman keeps it
at or marginally above the bound,
cf.~Eq.~\eqref{eq:SBTH_uncertainty}, while the CK product decays as
$e^{-2\gamma t}$ in violation of the Heisenberg principle
(Sec.~\ref{sec:CK}). Panel~(c) confirms that the Bateman--Lindblad
deviation saturates the analytic bound
Eq.~\eqref{eq:SBTH_deviation_bound} and remains well below the
Appendix-\ref{sec:quantitative_comparison} threshold over the whole
window shown.
\\

\emph{Validity window.\\}
The faithful behaviour established above concerns the quantum
moments. The classical mean trajectories obey
Eq.~\eqref{eq:class_damp_cav_cav}, in which the mirror coordinate
$y(t)\sim e^{+\lambda t}$ grows as the physical coordinate
$x(t)\sim e^{-\lambda t}$ decays, so as to keep the doubled
Hamiltonian conservative. This exponential growth eventually
contaminates the numerical integration of the mean dynamics and sets
a finite validity window, which for the parameters employed was $t\lesssim 700/\gamma$ quantified in
Appendix~\ref{sec:quantitative_comparison}
(Fig.~\ref{fig:classical_constr}). Since the experimentally relevant coherence
window is typically $t\lesssim 5\,T_1 = 10/\gamma$, the Bateman
effective description is reliable throughout the regime of practical
interest.\\

\emph{Outlook of the comparison.\\}
Although the two frameworks differ in the fine aspects of the
quantum sector, the Bateman Hamiltonian thus provides a canonical,
Heisenberg-consistent approximation to the dissipative dynamics that
tracks the Lindblad benchmark within controlled error bounds. This is
valuable because the effective (moment) formulation extends
immediately to more involved systems: it supplies a framework in which
the Gaussian closure of Sec.~\ref{section:formalism_momentous} can be
performed on the full moment hierarchy, opening the way to a
non-perturbative study of dissipation in the cavity--Josephson
junction system of
Sec.~\ref{subsection: Effective_Cavity-Josephson}. By contrast,
including the higher orders of the cosine in $\hat{H}_{\rm sys}$
within the Lindblad description leads to highly complex operator
equations of motion for which a cumulant expansion would be required,
and a closure of the type presented in
Sec.~\ref{section:formalism_momentous} is not straightforward at the
level of the master equation.

\subsection{Relation to other semiclassical and effective methods}
\label{sec:comparison_methods}
The momentous quantum mechanics framework employed in this work is
one of several semiclassical approaches that augment classical
equations of motion with dynamical variables encoding quantum
fluctuations.  In this subsection we situate MQM within the landscape
of related methods most relevant to the cQED and many-body
communities, and identify the specific advantages and limitations of
each approach as they apply to the Josephson junction problem studied
here.\\

\emph{Cumulant expansion methods.\\}
Cumulant expansions---used extensively in quantum optics, nuclear
physics, and open quantum systems---close the moment hierarchy by
expressing connected correlators of order $n$ in terms of products
of correlators of order $n-1$ or lower, thereby truncating the
Bogoliubov--Born--Green--Kirkwood--Yvon (BBGKY)
chain~\cite{Kubo1962,scholl2016control,bellac1992quantum}.
At second order, for instance, one sets
$\langle\hat{A}\hat{B}\hat{C}\rangle \approx
\langle\hat{A}\hat{B}\rangle\langle\hat{C}\rangle
+\langle\hat{A}\hat{C}\rangle\langle\hat{B}\rangle
+\langle\hat{B}\hat{C}\rangle\langle\hat{A}\rangle
-2\langle\hat{A}\rangle\langle\hat{B}\rangle\langle\hat{C}\rangle$,
which is equivalent to the Gaussian-state approximation at the
level of three-point functions.
MQM and cumulant expansions share the same physical approximation at
the Gaussian level: both truncate at second-order fluctuations and
both reproduce the Kerr/Duffing limit when applied to the cosine
potential.  The key structural difference is in the
implementation of the closure.  In cumulant methods the
closure is imposed on products of operator expectation values, so
operator-ordering ambiguities arise already at second order and
become increasingly problematic at higher orders: there is no
canonical prescription for which ordering to choose when
factorising $\langle\hat{A}^2\hat{B}^2\rangle$ in terms of lower
cumulants, and different choices yield different equations of
motion.  In MQM, the variables $G^{a,b}$ are defined with Weyl
(fully symmetric) ordering by construction
(Eq.~\eqref{Generalized effective dynamical variables}), and the
Poisson bracket structure (Eq.~\eqref{eq:quantum_phase_space})
determines the algebra uniquely.  The closure
(Eq.~\eqref{eq:Hierarchy}) is then applied not to operator products
but to moments in the symplectic sense, which eliminates
ordering ambiguities entirely.  A further advantage specific to the
Josephson problem is the closed-form resummation
$\tfrac{1}{2}[V(\phi+s)+V(\phi-s)]$
(Eq.~\eqref{eq:Ham_all_orders}): because MQM operates at the level
of the Hamiltonian function in phase space rather than at the level
of operator products, the all-orders sum of the cosine series
collapses into a closed-form expression that preserves the
periodicity of the potential.  Cumulant methods, applied to the
same cosine, would still require a Taylor expansion at each
truncation order and cannot produce this resummation.\\

\emph{Prezhdo's Quantized Hamilton Dynamics (QHD).\\}
Quantized Hamilton dynamics, introduced by Prezhdo and
Pereverzev~\cite{Prezhdo2000} and reviewed
in~\cite{prezhdo2006quantized}, has been developed extensively for
molecular quantum dynamics; it extends classical Hamiltonian
mechanics by promoting the position--momentum pair $(x,p)$ to an
enlarged set that includes second-order fluctuation variables
$((\Delta x)^2, (\Delta p)^2, \Delta(xp))$.  At second order,
QHD-2 evolves this extended set under equations of motion derived
from a Taylor-expanded Hamiltonian, with third- and higher-order
moments set to zero (the Gaussian closure).  The resulting
equations are formally identical to the second-order MQM equations
of Sec.~\ref{section:formalism_momentous}: the variables
$(s, p_s, U)$ of Eq.~\eqref{Canonical variables s and p_{s}}
are in one-to-one correspondence with the QHD-2 variables
$((\Delta x)^2,\, \Delta(xp),\, (\Delta p)^2)$ through
$s = \sqrt{G^{2,0}}$, $p_s = G^{1,1}/s$,
$U = G^{2,0}G^{0,2}-(G^{1,1})^2$.
The two frameworks are therefore equivalent at second order
for a single degree of freedom, a fact already noted
in~\cite{PhysRevA.99.042114}.  The distinction becomes substantial
at the all-orders level.  QHD-2 truncates the Taylor expansion of
the potential at fourth order (or at whatever fixed order is
specified), so applied to the Josephson cosine it reproduces the
Kerr Hamiltonian and cannot access the non-perturbative regime.
Higher-order QHD (QHD-4, etc.) would require explicit computation
of higher-order closure relations, which grow rapidly in complexity
and do not in general produce a closed-form Hamiltonian.
MQM with the Gaussian closure, by contrast, directly yields
the all-orders effective Hamiltonian
$\tfrac{1}{2}[\cos(\phi+s)+\cos(\phi-s)]$ without
truncating the Taylor series at any finite order, and does so
within the same second-order framework.  This is the
essential technical advantage of MQM over QHD for the periodic
Josephson potential: the resummation is achieved not by going to
higher-order moments, but by retaining the full nonlinearity of
the potential in the Hamiltonian function before computing the
Poisson brackets.  Additionally, QHD has been developed primarily
for chemical-physics applications (molecular dynamics, vibrational
energy transfer, dephasing), while MQM has an explicit connection
to the geometric formulation of quantum mechanics and to quantum
gravity, which may prove useful for the cosmological analogy
discussed in Sec.~\ref{section:effective_formalism}.\\

\emph{Black-box quantization (BBQ).\\}
Black-box quantization~\cite{Nigg2012} is a semiclassical method for
computing the effective low-energy quantum Hamiltonian of multi-mode
superconducting circuits containing weakly anharmonic Josephson
junctions. Its starting point is the classical linear-response
function (the admittance $Y(\omega)$ at the junction port), which
determines a complete basis of eigenmodes of the linearised circuit;
the Josephson nonlinearity is then incorporated perturbatively in
this basis, yielding a multimode Kerr Hamiltonian with dispersive
shifts computable directly from $Y(\omega)$. The comparison with MQM
reveals complementary strengths. BBQ is a spectral method: it
produces the low-energy spectrum and the qubit--cavity dispersive
shifts accurately and efficiently, but perturbatively in the
anharmonicity $E_C/E_J\ll 1$, so it inherits the limitations of the
Kerr approximation at moderate $E_J/E_C\sim 10$--$50$, and it does
not yield the dynamical equations of motion of the fluctuation
variables (e.g.\ $G^{2,0}(t)$, Sec.~\ref{sec:G20}) or their
back-reaction on the classical trajectory. MQM, by contrast, is a
dynamical framework, non-perturbative in $E_J/E_C$ within the
Gaussian approximation, but in its present formulation it is
restricted to a small number of modes. The two approaches are thus
not competing but complementary; applying the MQM Gaussian closure
to the multi-mode BBQ Hamiltonian is a natural extension of the
present work.\\

\emph{Gaussian variational methods.\\}
Within its Gaussian sector, the time-dependent variational framework
of Shi, Demler, and Cirac~\cite{ShiDemlerCirac2018} produces
equations of motion for the first and second moments that are
structurally identical to the MQM Gaussian-closure equations: both
implement the same Gaussian wave-packet ansatz, and applied to a
single JJ they would reproduce Eqs.~\eqref{eq:EOM-JJ} derived here.
The frameworks differ in scope and extension: Shi \emph{et al.}\
target quantum many-body systems and extend beyond Gaussianity via
entanglement and polaron transformations, whereas MQM targets
few-degree-of-freedom systems with strongly nonlinear potentials and
extends via the all-orders resummation
$\tfrac{1}{2}[V(\phi+s)+V(\phi-s)]$---a feature specific to the
MQM Hamiltonian formulation that the TDVP approach does not
produce. On the dissipative side the two frameworks are again
complementary: \cite{ShiDemlerCirac2018} employs a Lindblad
extension of the TDVP, while the present work additionally provides
the Bateman description with the explicit error bound of
Sec.~\ref{sec:discussion_dissipation}.\\

\emph{Summary.\\}
The distinctive feature of the MQM Gaussian closure, relevant
specifically to the Josephson junction, is the combination of
(i)~an unambiguous operator-ordering prescription (Weyl symmetrisation),
(ii)~a Hamiltonian formulation in the extended phase space that
allows the all-orders resummation of the cosine potential into the
closed form $\tfrac{1}{2}[V(\phi+s)+V(\phi-s)]$
(Eq.~\eqref{eq:Ham_all_orders})---a resummation that neither
second-order cumulant methods nor QHD-2 achieve, and that BBQ does
not target---and
(iii)~a natural connection to quantum cosmological models
studied with the same formalism~\cite{PhysRevD.110.043506,
doi:10.1142/S0129055X06002772}, which opens the possibility of
using superconducting circuits as analog simulators of
minisuperspace quantum gravity.

\subsection{Broader theoretical context}
\label{sec:broader}

The effective-method treatment developed in this work belongs
to a family of semiclassical approaches based on the
truncation and closure of the moment hierarchy of quantum
mechanics~\cite{doi:10.1142/S0129055X06002772,doi:10.1142/S0219887807001941,prezhdo2006quantized,PhysRevA.109.032209}.
In all of them the central challenge is the same: a nonlinear
potential generates an infinite coupled hierarchy of moment
equations, and one must find a physically motivated truncation that
captures the essential quantum effects while remaining
computationally tractable. The Gaussian closure employed here,
Eq.~\eqref{eq:Hierarchy}, is the simplest such truncation that goes
beyond the classical limit: it is exact for quadratic Hamiltonians
(harmonic oscillators, free particles) and provides the
leading-order quantum corrections for anharmonic systems, while for
the Josephson cosine it additionally resums the infinite Taylor
series into the closed-form effective potential of
Eq.~\eqref{eq:Ham_all_orders}.

The same mathematical structure has proven useful in other physical
contexts: quantum cosmological models with trigonometric or
exponential potentials~\cite{PhysRevD.110.043506}, tunneling-time
problems~\cite{PhysRevA.98.063417}, and the double-slit
experiment~\cite{HernandezHernandez_2023}. The present application to
circuit QED thus serves a dual purpose: it provides new theoretical
tools for the study of superconducting circuits, and it extends the
domain of applicability of the effective method to a system where
quantitative benchmarks exist (Secs.~\ref{sec:benchmark}
and~\ref{sec:comparison}), offering a testing ground for the
formalism itself.

An important open question is whether the Gaussian closure can be
systematically improved. One possibility is the extension to
fourth-order cumulant closures, which would capture non-Gaussian
features such as the Yurke--Stoler cat-like
states~\cite{PhysRevLett.57.13} that appear at the quantum revival
time. More broadly, the variance $G^{2,0}(t)$ computed in
Sec.~\ref{sec:G20} is the sharpest available diagnostic of the
Gaussian closure: it is more sensitive to variations in the parameters: $\alpha$ and the potential profile $E_{J}/E_{C}$. As a result it shows deviations from the exact dynamics from $t=0$ onward, and therefore precedes any departure of
$\langle\hat\theta\rangle(t)$. This makes $G^{2,0}(t)$ the natural
diagnostic for assessing the quality of any higher-order closure,
both in terms of the spectral accuracy of width oscillations and the
later onset of non-Gaussian cat-state features. Another is to
incorporate dissipation directly into the moment hierarchy, going
beyond the Bateman and Caldirola--Kanai models studied in
Sec.~\ref{sec:dissipative} and making contact with the Lindblad
formalism at the level of moments, not only at the level of
expectation values. These directions will be explored in future
work.
\section{Conclusions}
\label{sec:conclusions}

We have applied the momentous quantum mechanics (MQM) formalism to the
non-perturbative study of superconducting quantum circuits in the
transmon regime, deriving effective equations of motion that capture
the full cosine nonlinearity of the Josephson potential without
truncating its Taylor series at any finite order. The central results
of the paper are summarised below, followed by a discussion of their
implications and of the open problems they suggest.\\

\emph{All-orders resummation of the Josephson potential.\\}
The key theoretical contribution is the application of the Gaussian
closure (Eq.~\eqref{eq:Hierarchy}) to the infinite moment hierarchy
generated by the Josephson cosine. Under the second-order closure, the
infinite Taylor series of $-E_J\cos(\hat\theta)$ resums exactly
into the closed-form effective Hamiltonian
\begin{equation*}
\begin{split}
    \langle H\rangle &= 4E_{c}(n^{2}+n_{s}^{2})+\frac{4E_{c}U}{\theta_{s}^{2}}-E_{J}\text{Cos}(\theta)\text{Cos}(\theta_{s})
\end{split}
\end{equation*}
which is Eq.~\eqref{eq:All_orders_effective}. The resulting effective
potential $V_{\rm eff}(\phi)$ (Eq.~\eqref{eq:Veff_explicit}) preserves
the periodicity and boundedness of the original cosine at all phase
amplitudes, in contrast with the Kerr quartic approximation, which is
unbounded below for $|\theta|>\sqrt{6}$. The dressing factor
$\cos(\theta_{s})$ interpolates smoothly between the classical
limit ($\theta_s\to 0$, full Josephson depth) and the strong-fluctuation
limit ($\theta_{s}\gg 1$, potential washout), in analogy with the
Debye--Waller suppression in solid-state physics. The Kerr Hamiltonian
is recovered as the special case of a fourth-order Taylor truncation
of the cosine combined with the second-order moment closure,
confirming the hierarchy of approximations established in
Sec.~\ref{subsection: QJJ}.\\

\emph{Quantum-dressed frequency and anharmonicity.\\}
Linearising the all-orders equations of motion around the
minimum-uncertainty equilibrium point yields analytical
predictions:
\begin{equation*}
\omega_{\rm eff} = \Omega_p\sqrt{\cos\!\left(\theta_{s}\right)},
\end{equation*}
(Eq.~\eqref{eq:omega_dressed}), 
dressed by the zero-point factor $\cos(\theta_{zpf})<1$, which
deviates from unity by ${\sim}10\%$ at $E_J/E_C=50$ and ${\sim}22\%$
at $E_J/E_C=10$---the intermediate regime relevant to contemporary
transmon and fluxonium experiments. The two observables probe the
closure in opposite ways. For the small-oscillation frequency, the
dressed prediction coincides with the Kerr result at leading order and
reproduce the exact Mathieu value to better than $0.6\%$ over the
whole range studied (Fig.~\ref{fig:omega_eff}). The value of the non-perturbative
treatment thus lies not in the linear spectroscopy but in the bounded,
closed-form effective potential and in the nonlinear, large-excursion
dynamics summarised next.\\

\emph{Contrast against the exact Mathieu evolution.\\}
The dynamical accuracy of the Gaussian closure was assessed by
contrasting the all-orders equations of motion
(Eq.~\eqref{eq:EOM-JJ}) against the exact Schr\"odinger evolution
generated by the finite-dimensional coherent state
(Eq.~\eqref{eq:Appendix_Beta}) on the Mathieu energy basis. The
effective trajectory $\langle\hat\theta\rangle(t)$ reproduces the exact
carrier oscillation and its slow amplitude modulation, with an
agreement that improves systematically as $E_J/E_C$ increases
(Fig.~\ref{fig:JJ_dynamics}). Both the all-orders and the Kerr
descriptions eventually depart from the exact evolution near the
quantum revival time $t_{\rm rev}\sim 2\pi\hbar/E_C$, where the exact
state develops non-Gaussian (Yurke--Stoler cat-like)
features~\cite{PhysRevLett.57.13} that lie outside the Gaussian
closure. The variance $G^{2,0}(t)$ (Sec.~\ref{sec:G20},
Fig.~\ref{fig:G20}) is the most sensitive diagnostic of the closure's
limitations: it reveals a frequency mismatch present from $t=0$, and therefore
diverges from the exact value before any departure appears in
$\langle\hat\theta\rangle(t)$.\\

\emph{Cavity--Josephson junction: all-orders moment equations.\\}
Applying the same Gaussian closure to the cavity--JJ Hamiltonian
(Eq.~\eqref{eq:H_cav-JJ_classical}) generates the full set of
all-orders equations of motion
(Eqs.~\eqref{eq:classical_all_orders_cav-JJ}), which extend the
second-order (cavity--cavity) result of Appendix~\ref{appendix: II} to
include the complete cosine nonlinearity, without invoking a
rotating-wave or dispersive approximation. The comparison in
Fig.~\ref{fig:all_order_cavity_cavity_interaction} confirms that both
approximations agree at short times but diverge progressively in
amplitude as the cosine nonlinearity accumulates, with the JJ degree
of freedom showing the larger discrepancy---as expected, since it
carries the full periodicity of the Josephson potential.\\

\emph{Dissipation: Caldirola--Kanai, Bateman, and Lindblad.\\}
Three models of dissipation were compared within the MQM framework.
The Caldirola--Kanai Hamiltonian (Sec.~\ref{sec:CK}) yields exact
analytical solutions for the damped oscillator but violates
Heisenberg's uncertainty principle in the physical (kinetic) variables
for all $t>0$, as shown explicitly by Eq.~\eqref{eq:CK_violation}:
$\Uphys(t)=({\hbar^2}/{4})\,e^{-2\lambda t}\to 0$. The Bateman
dual-oscillator approach (Sec.~\ref{sec:bateman}) preserves the
Heisenberg bound under the modified commutation relations
Eq.~\eqref{eq:switch_commutation_Bateman}, and reproduces the Lindblad
expectation-value dynamics exactly under the identification
$\gamma=2\lambda$. The Lindblad master equation
(Sec.~\ref{sec:lindblad}) is valid for all $t>0$\footnote{In the sense that the solution obtained do not diverge or violates Heisenberg's uncertainty.}, drives the
oscillator to the quantum vacuum at $T=0$ (Eq.~\eqref{eq:Lindblad_ss}),
and provides the baseline against which the two Hamiltonian models
are assessed. The quantitative comparison
(Sec.~\ref{sec:discussion_dissipation},
Fig.~\ref{fig:Bateman_Lindblad}, and
Appendix~\ref{sec:quantitative_comparison}) shows that the two
Heisenberg-consistent descriptions share one and the same moment drift
matrix and differ structurally only in the inhomogeneous driving term:
a constant quantum-noise floor $\gamma\phizpf^2$ for Lindblad versus
the bounded, oscillatory auxiliary-oscillator contribution for
Bateman. For a vacuum-width initial state the Bateman physical moments
admit the exact closed forms Eq.~\eqref{eq:SBTH_vacuum_moments}, which
keep the uncertainty product at or marginally above the Heisenberg
floor (Eq.~\eqref{eq:SBTH_uncertainty}) and bound the deviation from
the Lindblad vacuum by $(\lambda/\omega_1)^2\phizpf^2$
(Eq.~\eqref{eq:SBTH_deviation_bound})---a $1\%$ effect at the weak
damping $\lambda/\omega_1=0.1$ used here, well below the $5\%$
threshold verified numerically out to $t\approx700/\gamma$ in
Appendix~\ref{sec:quantitative_comparison}. For the experimentally
relevant coherence window $t\lesssim 5\,T_1 = 10/\gamma$ of current transmon
devices ($T_1\sim 10$--$100\,\mu$s), the Bateman effective description
is therefore a reliable and computationally efficient alternative to
the full Lindblad master equation.\\

\emph{Position within the landscape of semiclassical methods.\\}
Section~\ref{sec:comparison_methods} situated MQM within the broader
family of related approaches. At the Gaussian level, MQM is
algebraically equivalent to cumulant-expansion
methods~\cite{Kubo1962,scholl2016control,bellac1992quantum},
to quantized Hamilton dynamics
(QHD-2)~\cite{Prezhdo2000,prezhdo2006quantized}, and to the Gaussian
sector of the variational framework of Shi, Demler, and
Cirac~\cite{ShiDemlerCirac2018}. Its distinctive advantage for the
Josephson problem is the combination of (i)~an unambiguous
Weyl-ordered operator prescription
(Eq.~\eqref{Generalized effective dynamical variables}),
(ii)~a Hamiltonian formulation in the extended phase space that allows
the all-orders resummation of the cosine potential without truncating
the Taylor series at any finite order---a resummation that QHD-2 and
second-order cumulant methods cannot achieve---and (iii)~a natural
mathematical connection to quantum cosmological models studied with
the same formalism
\cite{PhysRevD.110.043506,doi:10.1142/S0129055X06002772}, which opens
the possibility of using superconducting circuits as analog simulators
of minisuperspace quantum gravity. Relative to Nigg \textit{et al.}'s
black-box quantization (BBQ)~\cite{Nigg2012}, MQM is complementary:
BBQ provides accurate spectral data and dispersive shifts for
multi-mode circuits perturbatively in $E_C/E_J$, while MQM provides
non-perturbative dynamical equations valid for all $E_J/E_C$ at the
cost of a few-mode treatment.\\

\emph{Outlook.\\}
The present work opens several directions for future research. First,
the Gaussian closure can be systematically improved by extending to
fourth-order cumulant closures, which would capture non-Gaussian
features---in particular the Yurke--Stoler cat-like
superpositions~\cite{PhysRevLett.57.13} that emerge near the revival
time $t_{\rm rev}\sim 2\pi\hbar/E_C$,
at the cost of a larger but still finite
ODE system. The quantum width $G^{2,0}(t)$ provides the natural
diagnostic: the frequency mismatch between the all-orders and exact
variances is present from $t=0$ and signals immediately that a
higher-order closure is required for second-moment accuracy. Second, the
all-orders cavity--JJ equations of motion
(Eqs.~\eqref{eq:classical_all_orders_cav-JJ}) derived here contain the
full JJ nonlinearity in the moment sector and represent a
non-perturbative starting point for studying dispersive readout,
multi-photon resonances, and state preparation beyond the
rotating-wave approximation---phenomena that are relevant to current
cQED experiments~\cite{RevModPhys.93.025005} but are not accessible
within the Jaynes--Cummings (Kerr) approximation. Third,
incorporating the Lindblad dissipation directly into the MQM moment
hierarchy---rather than through the Bateman auxiliary
oscillator---would yield a dissipative effective phase-space
description that is completely positive for all times. Fourth,
combining the multi-mode spectral basis of BBQ with the
non-perturbative dynamical framework of MQM would produce a method
with the environmental accuracy of the former and the all-orders
capability of the latter---a natural next step for realistic
multi-junction circuit simulations. Finally, the same Gaussian closure
that succeeds here for the Josephson cosine was previously applied to
quantum cosmological minisuperspace
models~\cite{PhysRevD.110.043506,PhysRevD.84.043514}, where the
potential is also periodic or exponential. The formal analogy between
the effective JJ Hamiltonian and minisuperspace cosmology suggests
that superconducting circuits may serve as accessible quantum
simulators of early-universe phenomena, and that insights from quantum
gravity about non-perturbative moment closures may conversely inform
the design of more accurate effective theories for quantum hardware.

\section*{CRediT author statement}
{\bf G. Chacon-Acosta:} Conceptualization, Methodology, Formal analysis, Writing - Original Draft, Writing - Review and Editing,Supervision.
{\bf H. Hernandez-Hernandez:} Conceptualization, Methodology, Formal analysis, Writing - Original Draft, Writing - Review and Editing,Supervision.
{\bf C. Javier-Valdez:} Methodology, Software, Formal analysis, Resources, Writing - Original Draft,  Visualization.

\section*{Declaration of competing interests}
The authors declare that they have no known competing financial interests or personal relationships that could have appeared to influence the work reported in this paper.

\section*{Declaration of AI-Assisted Technologies}

The authors used Claude AI exclusively to improve English grammatical accuracy, clarity, and readability of the manuscript. It was not used for scientific content generation, data analysis, or interpretation of results. All scientific aspects of the work remain the sole responsibility of the authors, who have fully reviewed and approved the final manuscript.

\section*{Acknowledgments}
H.H. acknowledges support from SECIHTI grant CBF-2023-2024-1937 and Sabbatical Grant 2025. 

\appendix

\section{Expectation value for trigonometric operators}
\label{sec:Expect_val_Trig}
In Sec.~\ref{subsection: Effective_Josephson} we dealt with functions of
operators. Although this is formally addressed
in~\cite{doi:10.1142/S0129055X06002772}, let us review it in more depth.

The main problem is that
\begin{equation}
    \langle\text{Sin}[\hat{\phi}]\rangle\neq\text{Sin}[\langle\hat{\phi}\rangle], \quad \langle\text{Cos}[\hat{\phi}]\rangle\neq\text{Cos}[\langle\hat{\phi}\rangle].
\end{equation}
To gain insight into how to address this, let us represent the operators as
matrices. The exponential of a matrix is defined as~\cite{Horn2013}
\begin{equation}
    \text{exp}[A] = \sum_{k=0}^{\infty} \frac{1}{k!}A^{k},
\end{equation}
from which we can establish the cosine and sine functions as
\begin{equation}
    \text{Cos}[A] = \frac{1}{2}\left(\text{exp}[iA]+\text{exp}[-iA]\right), \quad \text{Sin}[A] = \frac{1}{2i}\left(\text{exp}[iA]-\text{exp}[-iA]\right).
\end{equation}

Now that we are furnished with the proper tools, we can tackle the original
problem. First, let us define $A = a\mathbb{I} +(A-a\mathbb{I}) = a\mathbb{I} +\delta A$, where $a$ is a scalar. Then
\begin{equation}
\begin{split}
    \text{Cos}[a+\delta A] &= \frac{1}{2}\left(\text{exp}[ia]\text{exp}[i\delta A]+\text{exp}[-ia]\text{exp}[-i\delta A]\right)\\
    &=\text{Cos}(a)\left[\frac{\text{exp}[i\delta A]+\text{exp}[-i\delta A]}{2}\right]-\text{Sin}(a)\left[\frac{\text{exp}[i\delta A]-\text{exp}[-i\delta A]}{2i}\right]\\
    &=\text{Cos}(a)\text{Cos}[\delta A]-\text{Sin}(a)\text{Sin}[\delta A]\\
    &=\text{Cos}(a)\sum_{n=0}^{\infty}\frac{(-1)^{n}}{(2n)!}(\delta A)^{2n}-\text{Sin}(a)\sum_{n=0}^{\infty}\frac{(-1)^{n}}{(2n+1)!}(\delta A)^{2n+1},
\end{split}
\end{equation}
and similarly
\begin{equation}
    \text{Sin}(a+\delta A) = \text{Sin}(a)\text{Cos}[\delta A]+\text{Cos}(a)\text{Sin}[\delta A],
\end{equation}
recovering the well-known trigonometric addition identities. Returning to the
operator representation and constraining the scalar as
\begin{equation}
    a = \langle\hat{A}\rangle\longrightarrow \delta A = \hat{A}-\langle\hat{A}\rangle,
\end{equation}
we see that $\delta A$ becomes a central moment. We thus obtain the expectation
value of $\text{Sin}(\hat{A})$,
\begin{equation}
\begin{split}
    \langle\text{Sin}(\hat{A})\rangle &= \text{Sin}(a)\sum_{n=0}^{\infty}\frac{(-1)^{n}}{(2n)!}\langle(\delta A)^{2n}\rangle+\text{Cos}(a)\sum_{n=0}^{\infty}\frac{(-1)^{n}}{(2n+1)!}\langle(\delta A)^{2n+1}\rangle\\
    &=\text{Sin}(a)\sum_{n=0}^{\infty}\frac{(-1)^{n}}{(2n)!}G^{2n,0}+\text{Cos}(a)\sum_{n=0}^{\infty}\frac{(-1)^{n}}{(2n+1)!}G^{2n+1,0},
\end{split}
\label{eq:App_sinA}
\end{equation}
and $\text{Cos}(\hat{A})$,
\begin{equation}
    \langle\text{Cos}(\hat{A})\rangle=\text{Cos}(a)\sum_{n=0}^{\infty}\frac{(-1)^{n}}{(2n)!}G^{2n,0}-\text{Sin}(a)\sum_{n=0}^{\infty}\frac{(-1)^{n}}{(2n+1)!}G^{2n+1,0}.
    \label{eq:App_cosA}
\end{equation}
These two results are important: they show that even if we truncate the series,
we get modifications coming from all orders. For example, taking the first term,
\begin{equation}
    \langle\text{Cos}(\hat{A})\rangle \approx \text{Cos}(a)-\text{Cos}(a)\frac{G^{2,0}}{2!},
\end{equation}
and by expanding the cosine in a Taylor series,
\begin{equation}
\label{eq:cos_operator}
\begin{split}
\langle\text{Cos}(\hat{A})\rangle &\approx \left(1-\frac{1}{2!}a^{2}\right)-\left(1-\frac{1}{2!}a^{2}\right)\frac{G^{2,0}}{2!}\\
&\approx \left(1-\frac{1}{2!}a^{2}\right)-\frac{G^{2,0}}{2!} = 1-\frac{\langle\hat{A}^{2}\rangle}{2!}.
\end{split}
\end{equation}
Therefore, the standard approximation involves a double truncation: first the
moment series Eqs.~\eqref{eq:App_sinA}--\eqref{eq:App_cosA} is cut at a finite
order (typically $G^{2,0}$ only), and then the cosine itself is expanded to a
finite Taylor order (typically fourth). Each truncation introduces uncontrolled
errors. The Gaussian closure of Sec.~\ref{section:effective_formalism}
eliminates the second truncation entirely by resumming the moment series into
the closed-form effective potential $\tfrac{1}{2}[V(x+s) + V(x-s)]$
(Eq.~\eqref{eq:Ham_all_orders}), leaving only the first truncation (Gaussianity
of the state) as the sole approximation. This is the key theoretical advantage
of the all-orders treatment.

\section{Effective dynamics of the cavity--cavity coupling}
\label{appendix: II}
A first approach to the dynamics of the cavity--Josephson-junction coupling was
addressed in Sec.~\ref{section: SQC}. As a preliminary step toward an effective
description, it is useful to work with the simpler model obtained by keeping the
second-order term in Eq.~\eqref{eq:H_cav-JJ_classical},
\begin{equation}
\label{eq:H_cav-cav_classical}
    H_{cav-cav} = \frac{1}{2C_{J1}}Q_{1}^{2}+\frac{1}{2C_{J2}}Q_{2}^{2}+\frac{C_{0}}{C_{T}}Q_{1}Q_{2}+\frac{1}{2L_{1}}\phi_{1}^{2}+\frac{1}{2L_{2}}\phi_{2}^{2},
\end{equation}
where $L_{2} = E_{J}/\phi_{0}^{2}$ is the linear part of the JJ inductance. This
describes the interaction between two resonant cavities and serves as a first
approximation to the dynamics of
Eq.~\eqref{eq:H_cav-JJ_classical}\footnote{In both the cavity--cavity and the
cavity--JJ Hamiltonians, if we work with annihilation and creation operators the
coupling term $Q_{1}Q_{2} = -Q_{{\rm zpf},1}Q_{{\rm zpf},2}(\hat{a}-\hat{a}^{\dagger})(\hat{b}-\hat{b}^{\dagger})$
contains counter-rotating elements of the type $\hat{a}\hat{b}$,
$\hat{a}^{\dagger}\hat{b}^{\dagger}$. In the effective formalism we find it more
practical to keep the system's canonical variables, since we then obtain a
closed system of EOM for Eq.~\eqref{eq:H_cav-cav_classical} and for its quantum
analog; this approach does not invoke the rotating-wave approximation.}. Writing
the Hamiltonian in terms of effective quantum variables gives
\begin{equation}
\label{eq:Appendix_Ham_second_order_cav-cav}
    H_Q = H_{cav-cav}+\frac{1}{2C_{J1}}G^{0,2,0,0}+\frac{1}{2L_{1}}G^{2,0,0,0}+\frac{1}{2C_{J2}}G^{0,0,0,2}+\frac{C_{0}}{C_{T}}G^{0,1,0,1}+\frac{1}{2L_{2}}G^{0,0,2,0},
\end{equation}
where the diagonal corrections carry the factor $1/2$ from
$\tfrac12\partial^2 H/\partial z^2$, whereas the mixed term
$(C_0/C_T)G^{0,1,0,1}$ does not (the mixed partial appears twice). The quantum
and classical degrees of freedom are uncoupled, e.g.\ there are no terms of the
type $\phi_{1}G^{a,b,c,d}$ or $Q_{1}G^{a,b,c,d}$, so there is no quantum
back-reaction. This is most easily seen from the EOM obtained through the
classical Hamiltonian formalism $\dot{f} = \{f,H\}$. For the classical part,
\begin{equation}
\label{eq:Appendix_EOM_second_order_cav-cav}
    \begin{split}
    \dot{\phi}_{1} &= \frac{Q_{1}}{C_{J1}} +\frac{C_{0}}{C_{T}}Q_{2},\quad \dot{Q}_{1} = -\frac{1}{L_{1}}\phi_{1},\\
    \dot{\phi}_{2} &= \frac{Q_{2}}{C_{J2}} +\frac{C_{0}}{C_{T}}Q_{1},\quad
    \dot{Q}_{2} = -\frac{1}{L_{2}}\phi_{2},
    \end{split}
\end{equation}
and for the quantum part,
\begin{equation}
\label{eq:Appendix_Hamiltonian_cav_cav_quan}
    \begin{split}
    \dot{G}^{2,0,0,0} &= \frac{2}{C_{J1}}G^{1,1,0,0}+\frac{2C_0}{C_T}G^{1,0,0,1}\\
    \dot{G}^{0,2,0,0} &= -\frac{2}{L_{1}}G^{1,1,0,0}\\
    \dot{G}^{0,0,2,0} &= \frac{2}{C_{J2}}G^{0,0,1,1}+\frac{2C_{0}}{C_{T}}G^{0,1,1,0}\\
    \dot{G}^{0,0,0,2} &= -\frac{2}{L_2}G^{0,0,1,1}\\
    \dot{G}^{1,0,0,1} &= \frac{1}{C_{J1}}G^{0,1,0,1}+\frac{C_{0}}{C_{T}}G^{0,0,0,2}-\frac{1}{L_2}G^{1,0,1,0}\\
    \dot{G}^{0,1,1,0} &= \frac{1}{C_{J2}}G^{0,1,0,1}+\frac{C_{0}}{C_{T}}G^{0,2,0,0}-\frac{1}{L_{1}}G^{1,0,1,0}\\
    \dot{G}^{1,0,1,0} &=\frac{1}{C_{J1}}G^{0,1,1,0}+\frac{1}{C_{J2}}G^{1,0,0,1}+\frac{C_{0}}{C_{T}}\left(G^{1,1,0,0}+G^{0,0,1,1}\right)\\
    \dot{G}^{0,1,0,1} &=-\frac{1}{L_2}G^{0,1,1,0}-\frac{1}{L_{1}}G^{1,0,0,1}\\
    \dot{G}^{1,1,0,0} &= \frac{1}{C_{J1}}G^{0,2,0,0}-\frac{1}{L_{1}}G^{2,0,0,0}+\frac{C_{0}}{C_{T}}G^{0,1,0,1}\\
    \dot{G}^{0,0,1,1} &=\frac{1}{C_{J2}}G^{0,0,0,2}-\frac{1}{L_2}G^{0,0,2,0}+\frac{C_{0}}{C_{T}}G^{0,1,0,1}
    \end{split}
\end{equation}
where we have used $1/L_2=E_J/\phi_0^2$ consistently. As the quantum dynamics is
independent of the (linear) classical one, it can be solved analytically; the
classical normal-mode frequencies are
\begin{equation}
\label{eq:cavcav_eigenfreq}
    \omega^{2}_{\pm} = \frac{1}{2}\left[(\omega^{2}_{1}+\omega^{2}_{2})\pm\sqrt{(\omega^{2}_{1}-\omega^{2}_{2})^{2}+\frac{4C_{0}^{2}}{C_{T}^{2}L_{1}L_{2}}}\,\right],
\end{equation}
with $\omega_i^2=1/(L_iC_{J_i})$.

\begin{figure}[H]
\centering
\includegraphics[width=0.72\linewidth]{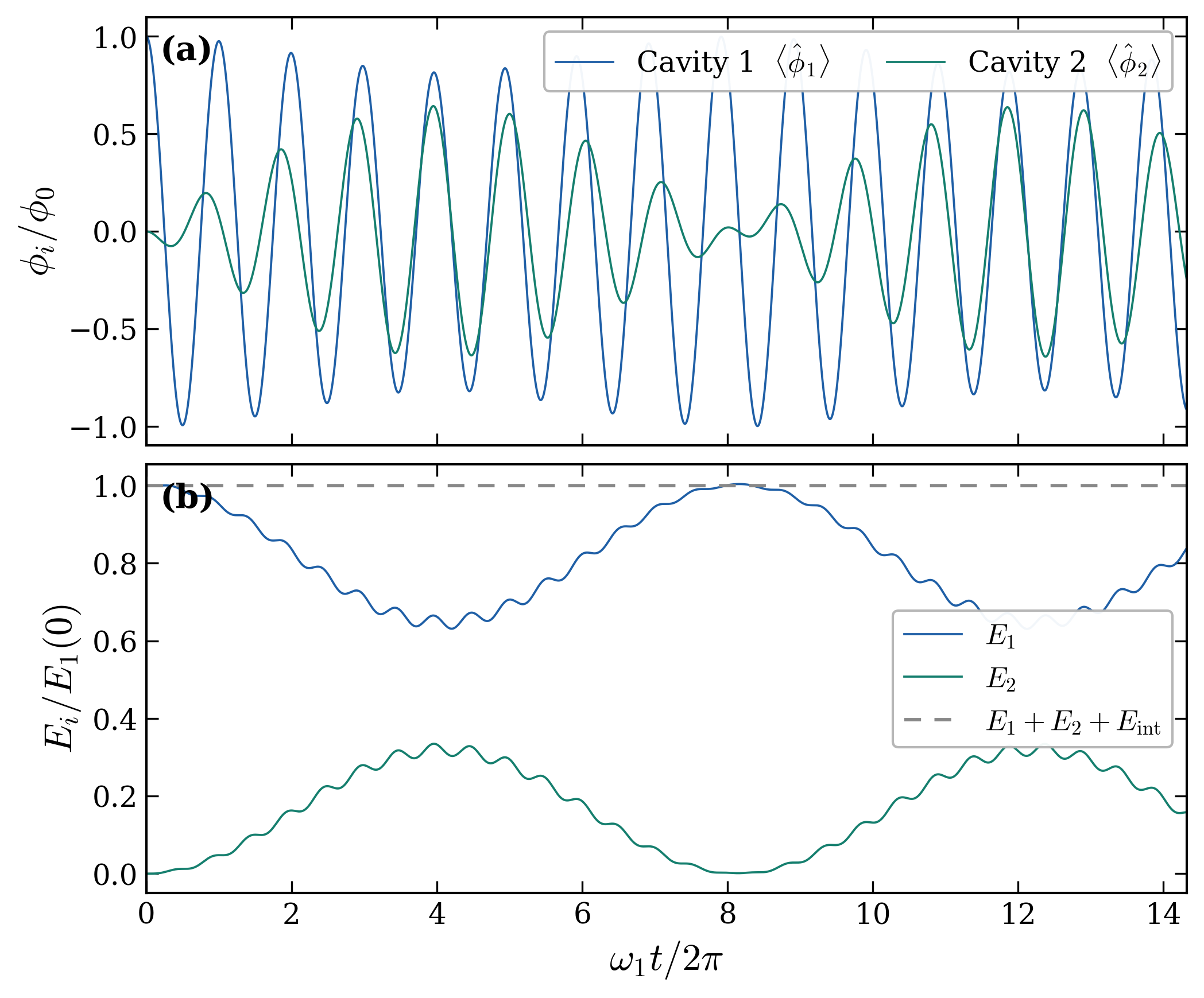}
\caption{Conservative dynamics of the cavity--cavity coupling, with frequency
    ratio $\omega_2/\omega_1 = 0.90$ and coupling strength $C_0/C_T = 0.15$.
    \textbf{(a)}~Flux expectation values $\phi_1(t)$ (cavity~1) and $\phi_2(t)$
    (cavity~2) versus the dimensionless time $\omega_1 t/2\pi$. Cavity~1 is
    initialised with a non-zero flux amplitude while cavity~2 starts in its
    ground state; the oscillatory exchange of flux is the classical analogue of
    quantum state transfer. \textbf{(b)}~Cavity energies $E_1(t)$, $E_2(t)$
    normalised by $E_1(0)$, together with the total $E_1+E_2+E_{\rm int}$
    (dashed). The total energy is conserved to numerical precision, confirming
    a purely conservative coupling.}
\label{fig:cavity_cavity_interaction}
\end{figure}

The dynamics obtained from the effective equations of motion is illustrated in
Fig.~\ref{fig:cavity_cavity_interaction}. Panel~(a) shows the flux expectation
values $\phi_1(t)$ and $\phi_2(t)$: starting from a state in which all excitation
resides in cavity~1, the flux amplitude is progressively transferred to cavity~2
and then returns, realising a coherent quantum-state-transfer protocol between
the two modes. Panel~(b) displays the individual cavity energies $E_1(t)$ and
$E_2(t)$, normalised by $E_1(0)$. Their exchange follows a Rabi-like envelope,
set by the effective coupling, superimposed on faster oscillations at the bare
cavity frequencies. Crucially, the total energy $E_1+E_2+E_{\rm int}$ (including
the interaction term) remains constant throughout, confirming that the coupling
is purely conservative and that the effective equations of motion preserve the
Hamiltonian structure of the original circuit.

\section{Caldirola--Kanai: analytical solutions}
\label{app:CK}

\subsection{Single-oscillator CK}
\label{app:CK_single}

The CK Hamiltonian Eq.~\eqref{eq:CK_ham} is quadratic; its classical solution in
the underdamped case ($\lambda < \omega_1$, relevant for transmon devices) is
\begin{equation}
  \phi_1(t) = e^{-\lambda t/2}
    \Bigl[A\cos(\omega_d t) + B\sin(\omega_d t)\Bigr],\quad
  \omega_d = \sqrt{\omega_1^2 - \lambda^2/4}.
  \label{eq:CK_phi_sol}
\end{equation}
The physical energy decays as $E_{\rm phys}(t) = E_0\,e^{-\lambda t}$.

\begin{figure}[H]
\centering
\includegraphics[width=0.72\linewidth]{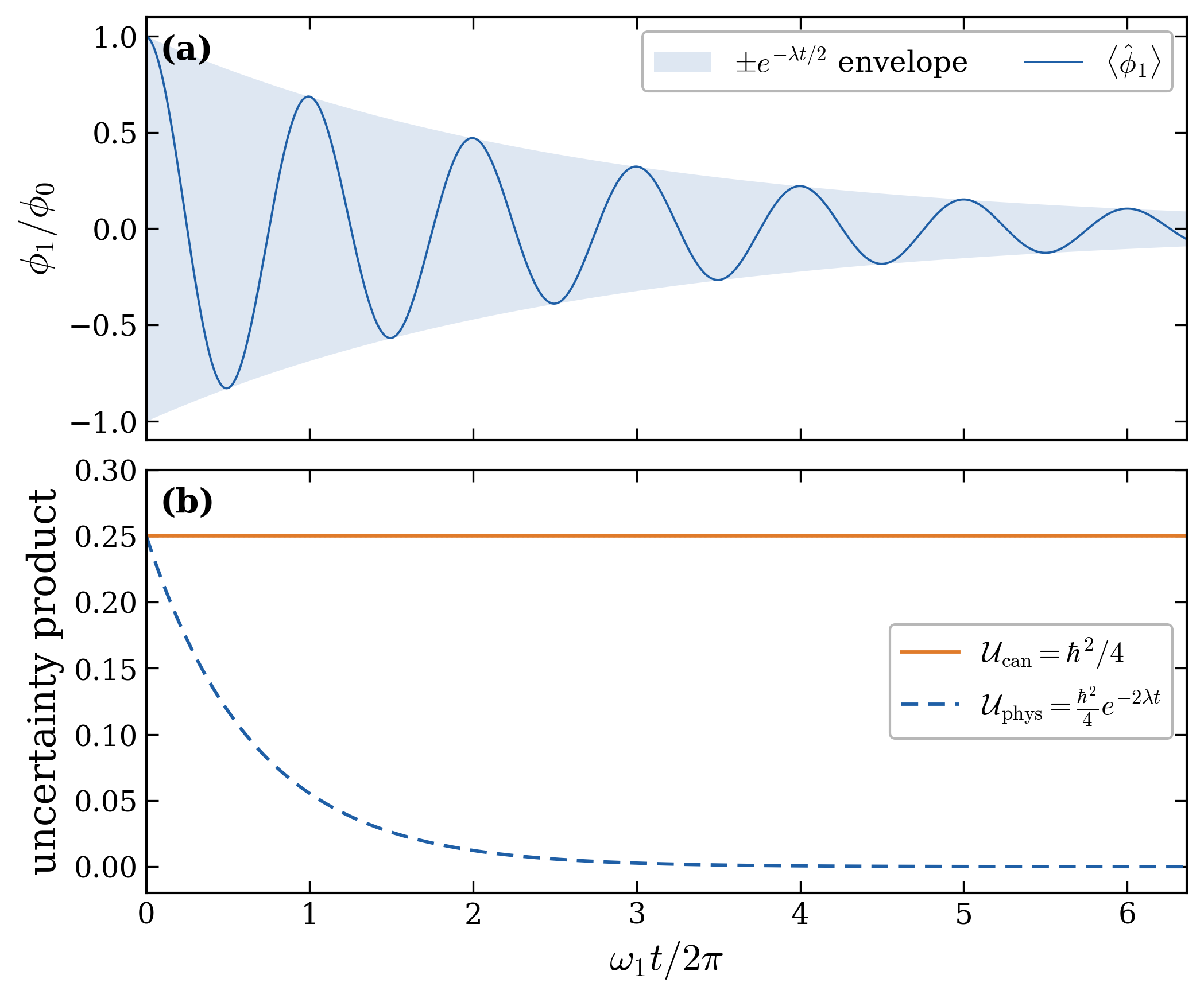}
\caption{Caldirola--Kanai single oscillator ($\lambda/\omega_1 = 0.12$).
    \textbf{(a)}~Flux expectation value $\langle\hat\phi_1\rangle$ and its
    amplitude envelope $\pm e^{-\lambda t/2}$. \textbf{(b)}~Canonical uncertainty
    product $\mathcal{U}_{\rm can}=\hbar^2/4$ (conserved) and physical
    uncertainty product $\mathcal{U}_{\rm phys}=(\hbar^2/4)e^{-2\lambda t}$,
    which decays to zero and so violates the Heisenberg bound in the physical
    variables.}
\label{fig:CK_violation}
\end{figure}

For the second-order moments, the characteristic equation of the linear system
Eq.~\eqref{eq:CK_Gs} factors as
\begin{equation}
  (\mu + \lambda)\,[\mu^2 + 2\lambda\mu + 4\omega_1^2] = 0,
  \label{eq:CK_char_poly}
\end{equation}
giving eigenvalues $\mu_1 = -\lambda$ and
$\mu_{2,3} = -\lambda \pm \sqrt{\lambda^2 - 4\omega_1^2}$. In the underdamped
case $\mu_{2,3} = -\lambda \pm 2i\omega_d$, and the general solution is
\begin{equation}
  \GT{2}{0}(t) = e^{-\lambda t}
    \Bigl[c_1 + c_2\cos(2\omega_d t) + c_3\sin(2\omega_d t)\Bigr],
  \label{eq:CK_G20_sol_underdamped}
\end{equation}
so that $\GT{2}{0}(t)\to 0$ as $t\to\infty$. The canonical uncertainty is
conserved,
\begin{equation}
  \Ucan(t) = \GT{2}{0}(t)\,\GT{0}{2}(t)-\bigl[\GT{1}{1}(t)\bigr]^2 = \frac{\hbar^2}{4} = \mathrm{const},
\end{equation}
whereas the physical uncertainty, built from the kinetic momentum
$p_{\rm mech}=Q_1e^{-\lambda t}$, decays,
\begin{equation}
  \Uphys(t) = e^{-2\lambda t}\,\Ucan(0) = \frac{\hbar^2}{4}\,e^{-2\lambda t}
            < \frac{\hbar^2}{4} \quad \forall\,t>0,
\end{equation}
in violation of the Heisenberg bound in the physical variables. The two
behaviours are illustrated in Fig.~\ref{fig:CK_violation}.

\subsection{CK coupled to a second oscillator}
\label{app:CK_coupled}

The coupled Hamiltonian Eq.~\eqref{eq:CK_coupled} yields the classical EOM
\begin{align}
  \dot\phi_1 &= \frac{Q_1}{\CJ}\,e^{-\lambda t} + \frac{C_0}{C_T}\,Q_2,
  &\dot Q_1 &= -\CJ\omega_1^2\,\phi_1\,e^{\lambda t},
  \label{eq:CK_coupled_eom_a}\\
  \dot\phi_2 &= \frac{Q_2}{\CJtwo} + \frac{C_0}{C_T}\,e^{-\lambda t}\,Q_1,
  &\dot Q_2 &= -\frac{\phi_2}{L_2},
  \label{eq:CK_coupled_eom_b}
\end{align}
and the moment equations
\begin{align}
    \dot{G}^{2,0,0,0} &= \frac{2}{C_{J_1}}e^{-\lambda t}G^{1,1,0,0}+\frac{2C_{0}}{C_{T}}e^{-\lambda t}G^{1,0,0,1} & \dot{G}^{0,2,0,0} &=-2C_{J_1}\omega^{2}e^{\lambda t}G^{1,1,0,0}\\
    \dot{G}^{0,0,2,0} &= \frac{2}{C_{J_2}}G^{0,0,1,1}+\frac{2C_{0}}{C_{T}}e^{-\lambda t}G^{0,1,1,0} & \dot{G}^{0,0,0,2} &=-\frac{2}{L_{2}}G^{0,0,1,1}
\end{align}
\begin{equation}
    \begin{split}
        \dot{G}^{1,1,0,0} &= \frac{e^{-\lambda t}}{C_{J_1}}G^{0,2,0,0}-C_{J_1}\omega^{2}e^{\lambda t}G^{2,0,0,0}+\frac{C_{0}}{C_{T}}e^{-\lambda t}G^{0,1,0,1}\\
        \dot{G}^{1,0,0,1} &=\frac{e^{-\lambda t}}{C_{J_1}}G^{0,1,0,1}+\frac{C_{0}}{C_{T}}e^{-\lambda t}G^{0,0,0,2}-\frac{1}{L_{2}}G^{1,0,1,0}\\
        \dot{G}^{1,0,1,0} &=\frac{e^{-\lambda t}}{C_{J_1}}G^{0,1,1,0}+\frac{1}{C_{J_2}}G^{1,0,0,1}+\frac{C_{0}}{C_{T}}e^{-\lambda t}\left(G^{1,1,0,0}+G^{0,0,1,1}\right)\\
        \dot{G}^{0,1,0,1} &= -C_{J_1}\omega^{2}e^{\lambda t} G^{1,0,0,1}-\frac{1}{L_{2}}G^{0,1,1,0}\\
        \dot{G}^{0,0,1,1} &= -\frac{1}{L_{2}}G^{0,0,2,0}+\frac{1}{C_{J_2}}G^{0,0,0,2}+\frac{C_{0}}{C_{T}}e^{-\lambda t}G^{0,1,0,1}\\
        \dot{G}^{0,1,1,0} &=-C_{J_1}\omega^{2}e^{\lambda t} G^{1,0,1,0}+\frac{1}{C_{J_2}}G^{0,1,0,1}+\frac{C_{0}}{C_{T}}e^{-\lambda t}G^{0,2,0,0}
    \end{split}
\end{equation}
The second-order moment for $\phi_1$ has the same structure as in
Eq.~\eqref{eq:CK_Gs}, with an additional coupling term to oscillator~2. The
physical uncertainty of oscillator~1 continues to violate the Heisenberg bound,
\begin{equation}
  \Uphys^{(1)}(t) = e^{-2\lambda t}\,\Ucan^{(1)}(0),
\end{equation}
irrespective of the coupling to oscillator~2, because the coupling preserves the
CK time-dependent exponents in the kinetic-momentum definition.

\section{Quantitative Bateman vs.\ Lindblad comparison}
\label{sec:quantitative_comparison}
Following~\cite{JavierValdez_2025}, we compare, for the single damped harmonic
oscillator, the second-order moment dynamics arising from the Lindblad master
equation with that of the Semiclassical Bateman--Tikochinsky Hamiltonian (SBTH).
The Lindblad system, in matrix form, reads
\begin{equation}
\label{eq:Lindblad_sde}
        \dot{\mathbf{G}}_{L} = \mathbf{M}\mathbf{G}_{L}+\mathbf{d}=
        \begin{pmatrix}
    -\gamma & 0 & 2/C_{J_1}\\
    0 & -\gamma & -2C_{J_1}\omega^{2}\\
    -C_{J_1}\omega^{2} & 1/C_{J_1} & -\gamma
    \end{pmatrix}
    \begin{pmatrix}
        G^{2,0,0,0}\\
        G^{0,2,0,0}\\
        G^{1,1,0,0}
    \end{pmatrix}
    + \begin{pmatrix}
        \dfrac{\gamma\hbar}{2C_{J_1}\omega}\\[4pt]
        \dfrac{\gamma\hbar C_{J_1}\omega}{2}\\[4pt]
        0
    \end{pmatrix}.
\end{equation}
If this dynamics reaches a stationary state ($\dot{\mathbf{G}}=0$),
\begin{equation}
    \mathbf{G}_{L}(\infty) = -\mathbf{M}^{-1}\mathbf{d} = \begin{pmatrix}
        \frac{\hbar}{2C_{J_1}\omega}\\[3pt]
        \frac{\hbar C_{J_1}\omega}{2}\\[3pt] 0
    \end{pmatrix} = \begin{pmatrix}
        \phi_{\rm zpf}^{2}\\
        Q_{\rm zpf}^{2}\\
        0
    \end{pmatrix},
\end{equation}
i.e.\ the vacuum, with uncertainty product $\phi_{\rm zpf}^2 Q_{\rm zpf}^2 =
\hbar^2/4$.

\begin{figure}[H]
    \centering
    \includegraphics[width=0.72\linewidth]{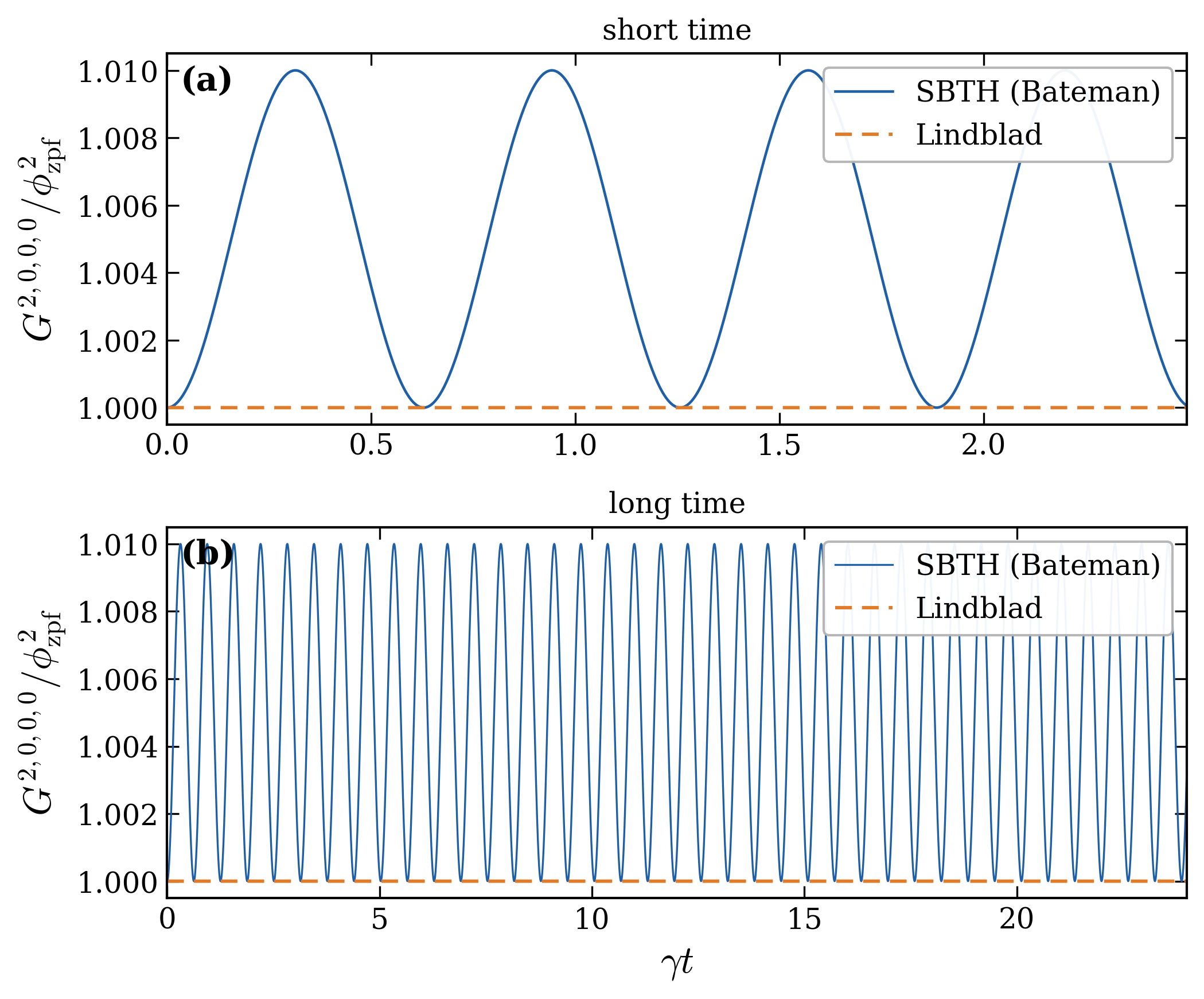}
    \caption{Second-order moment $G^{2,0,0,0}(t)$ of the single damped harmonic
    oscillator for the SBTH (Bateman) and Lindblad descriptions, normalised by
    $\phi_{\rm zpf}^2$, with $\lambda/\omega=0.1$ ($\gamma=2\lambda$).
    \textbf{(a)}~Short-time and \textbf{(b)}~long-time behaviour versus the
    dimensionless time $\gamma t$. The Lindblad moment sits at the vacuum value,
    while the SBTH moment oscillates around it with a small, bounded amplitude
    that persists at long times.}
    \label{fig:SBTH_vs_Lindblad}
\end{figure}

\begin{figure}[H]
    \centering
    \includegraphics[width=0.72\linewidth]{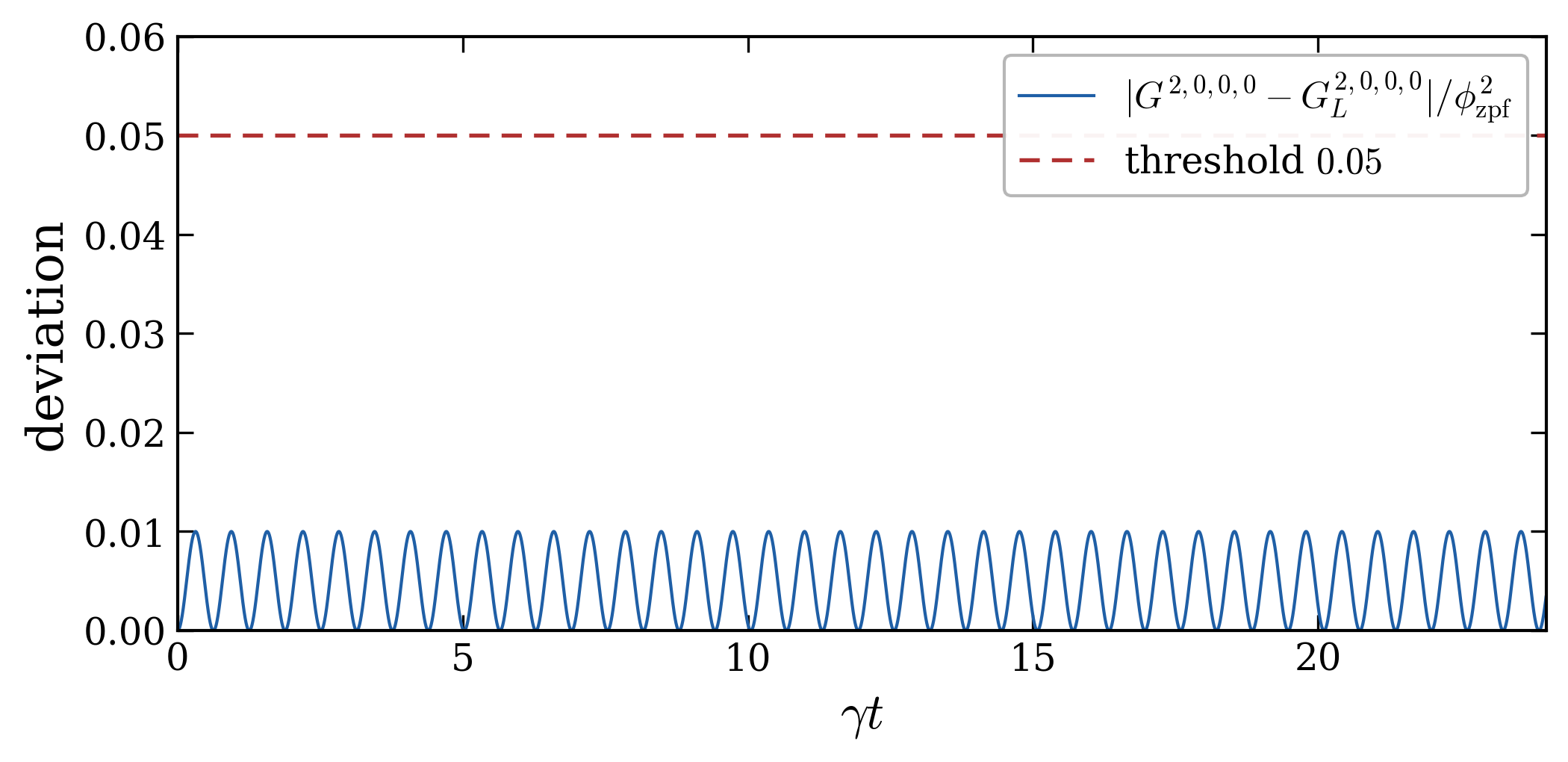}
    \caption{Deviation $|G^{2,0,0,0}-G^{2,0,0,0}_{L}|/\phi_{\rm zpf}^2$ between
    the SBTH and Lindblad moments as a function of $\gamma t$. It remains well
    below the threshold $0.05$ (dashed), confirming that the SBTH provides a
    faithful semiclassical description of the dissipative quantum moments.}
    \label{fig:threshold}
\end{figure}

For the SBTH, the doubled (Bateman) system is solved with the switched
commutation relations of Eq.~\eqref{eq:switch_commutation_Bateman} and projected
onto the physical moment $G^{2,0,0,0}=\mathrm{Var}(\phi_1)$. The comparison is
shown in Fig.~\ref{fig:SBTH_vs_Lindblad}: although the SBTH solution does not
converge to the Lindblad one, it oscillates around it and remains bounded. The
deviation is quantified in Fig.~\ref{fig:threshold}, where it stays below the
established threshold $|G^{2,0,0,0}-G^{2,0,0,0}_{L}|<0.05\,\phi_{\rm zpf}^{2}$;
this was verified out to $t\approx 700/\gamma$.

\begin{figure}[H]
\centering
\includegraphics[width=0.72\linewidth]{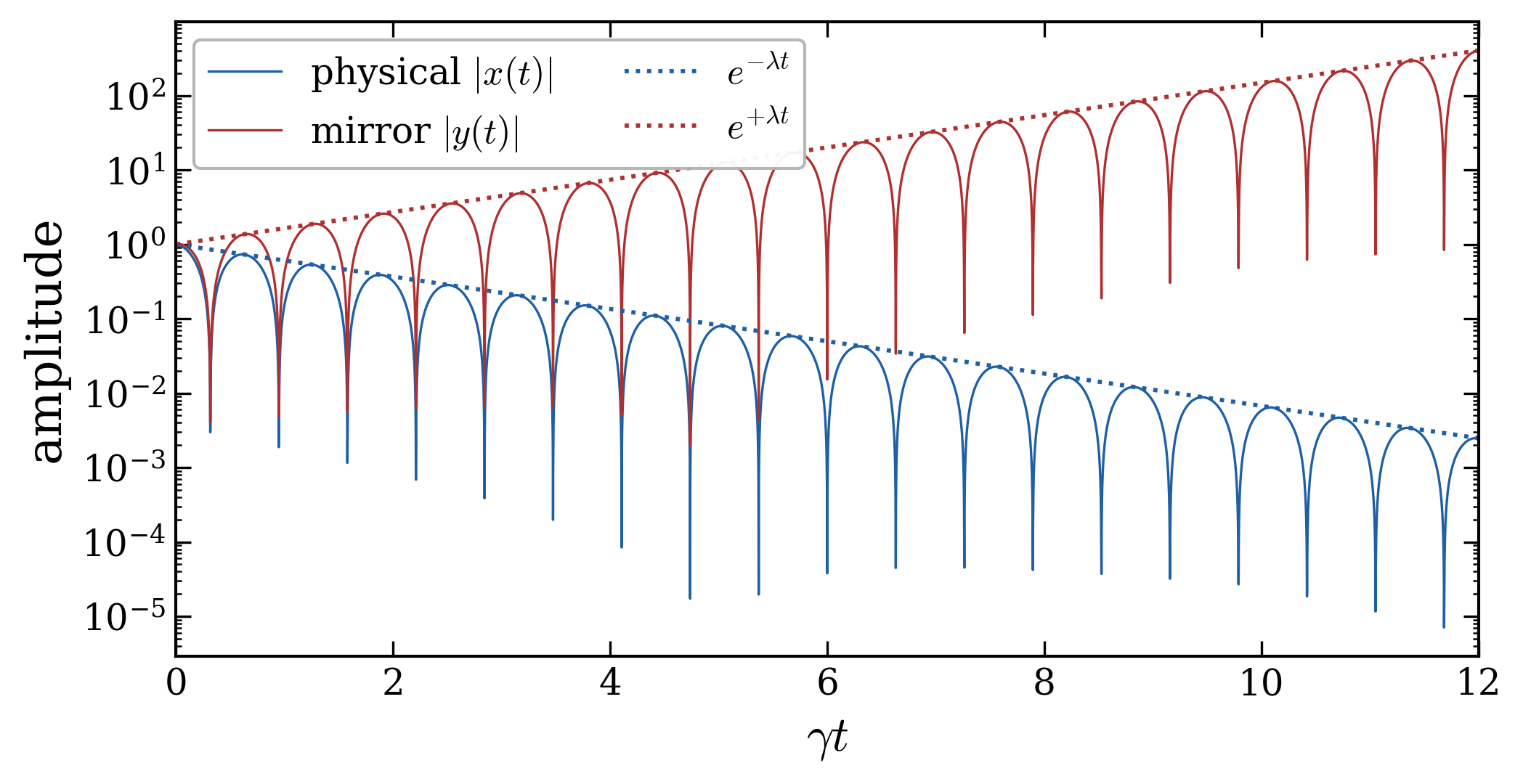}
\caption{Classical Bateman constraint. Amplitudes of the physical coordinate
    $|x(t)|$ and the mirror coordinate $|y(t)|$ (log scale) versus $\gamma t$.
    The physical mode decays as $e^{-\lambda t}$ while the mirror grows as
    $e^{+\lambda t}$; the latter eventually dominates and sets the finite
    validity window for the numerical integration of the classical equations of
    motion.}
\label{fig:classical_constr}
\end{figure}

The agreement at the level of the quantum moments holds only within a finite
validity window set by the classical dynamics. To keep the Bateman
Hamiltonian conservative, the mirror variable $y(t)$ grows in amplitude as the
physical variable $x(t)$ decays; numerically this exponential growth eventually
dominates and limits the integration of the classical EOM, as shown in
Fig.~\ref{fig:classical_constr}.

\newpage
\printbibliography

\end{document}